\documentclass{SciPost}

\hypersetup{
    colorlinks,
    linkcolor={red!50!black},
    citecolor={blue!50!black},
    urlcolor={blue!80!black}
}

\usepackage[bitstream-charter]{mathdesign}
\DeclareSymbolFont{usualmathcal}{OMS}{cmsy}{m}{n}
\DeclareSymbolFontAlphabet{\mathcal}{usualmathcal}

\usepackage{bm}
\usepackage{tikz}
\usetikzlibrary{arrows.meta,calc}
\usepackage{pgfplots}
\pgfplotsset{compat=1.18}
\usepackage{booktabs}
\usepackage[section]{placeins}  
\usepackage[most]{tcolorbox}

\newcommand{\dif}{\mathrm{d}}                 
\newcommand{\ca}{\mathrm{ca}}                 
\newcommand{\cc}{\mathrm{c.c.}}               
\newcommand{\ad}[1]{\hat{a}_{#1}^{\dagger}}   
\newcommand{\an}[1]{\hat{a}_{#1}}             
\newcommand{\ket}[1]{|#1\rangle}
\newcommand{\bra}[1]{\langle #1|}
\newcommand{\ue}{\hat{\bm{e}}}                
\newcommand{\Km}{\mathsf{K}}                  
\newcommand{\Cm}{\mathsf{C}}                  
\newcommand{\Gm}{\mathsf{G}}                  
\newcommand{\Id}{\mathsf{I}}                  
\newcommand{\tr}{\mathrm{tr}}
\newcommand{\Real}{\operatorname{Re}}
\newcommand{\Imag}{\operatorname{Im}}

\newtcolorbox{keybox}[1][]{enhanced, breakable, colback=scipostblue!3!white,
  colframe=scipostdeepblue, boxrule=0.6pt, arc=2pt, left=6pt, right=6pt,
  top=4pt, bottom=4pt, fonttitle=\bfseries, title={#1}}

\fancypagestyle{SPstyle}{
\fancyhf{}
\lhead{\colorbox{scipostblue}{\bf \color{white} ~SciPost Physics Lecture Notes }}
\rhead{{\bf \color{scipostdeepblue} ~Submission }}

\fancyfoot[C]{\textbf{\thepage}}
}

\begin{document}

\pagestyle{SPstyle}

\begin{center}{\Large \textbf{\color{scipostdeepblue}{
Why do single photons produce interference fringes?\\
The phase of a mode and the phase of a state
}}}\end{center}

\begin{center}\textbf{
Andrea Aiello\textsuperscript{$\star$}
}\end{center}

\begin{center}
{Max Planck Institute for the Science of Light, \\Staudtstrasse 2, 91058 Erlangen, Germany}
\\[\baselineskip]
$\star$ \href{mailto:andrea.aiello@mpl.mpg.de}{\small andrea.aiello@mpl.mpg.de}
\end{center}

\section*{\color{scipostdeepblue}{Abstract}}
\textbf{\boldmath{%
If the phase of a photon-number state is completely undetermined, how can single photons, sent one at a time through a two-slit apparatus, build up interference fringes? These lecture notes answer this question from first principles. Starting from Maxwell's equations, we expand the field in vector modes, quantize it, and introduce the quantity measured by a photodetector. We then show that the word ``phase'' refers to two different things: the phase of a mode, fixed by the optical setup, and the phase of the quantum state that excites the mode. Only the first determines the fringes. All calculations are given in full, and problems with solutions are included.
}}

\vspace{\baselineskip}

\noindent\textcolor{white!90!black}{%
\fbox{\parbox{0.975\linewidth}{%
\textcolor{white!40!black}{\begin{tabular}{lr}%
  \begin{minipage}{0.6\textwidth}%
    {\small Copyright attribution to authors. \newline
    This work is a submission to SciPost Physics Lecture Notes. \newline
    License information to appear upon publication. \newline
    Publication information to appear upon publication.}
  \end{minipage} & \begin{minipage}{0.4\textwidth}
    {\small Received Date \newline Accepted Date \newline Published Date}%
  \end{minipage}
\end{tabular}}
}}
}


\vspace{10pt}
\noindent\rule{\textwidth}{1pt}
\tableofcontents
\noindent\rule{\textwidth}{1pt}
\vspace{10pt}

\section{Introduction}
\label{sec:intro}

\subsection{A question from students}

These notes grew out of a question asked by students in a course on quantum optics. Photon-number states, also called Fock states, had just been introduced, together with the fact that their phase is completely undetermined: a number state is unchanged by a rotation of the phase of the field, and all values of the phase between $0$ and $2\pi$ are equally likely. The students recalled from another course the single-photon version of Young's two-slit experiment, in which photons are sent through the apparatus one at a time and the interference pattern builds up dot by dot on a screen~\cite{Grangier1986,Galvez2005,Aspden2016}. Their question was:
\begin{quote}
\emph{If the phase of a photon is completely undetermined, how can single photons produce interference fringes?}
\end{quote}
Classical optics teaches that fringes come from phase differences, so the question is a natural one, and not a naive one. Answering it properly requires almost everything that a first course on the quantum theory of light contains: modes of the electromagnetic field, quantization, photodetection, coherent and number states, and coherence functions. These notes collect all of this material in one place and derive every result from Maxwell's equations, without skipping steps.

\subsection{The short answer}

The answer rests on a distinction that textbooks rarely make explicit: the word ``phase'' refers to two different things.
\begin{itemize}
\item The phase of the \emph{quantum state} of a mode is, loosely speaking, the variable conjugate to the number of photons in that mode. It is completely undetermined for a number state, and well defined for an intense coherent state~\cite{Loudon}.
\item The phase of a \emph{mode} is a property of its mode function, which is the classical field configuration that the photons occupy. Its position-dependent phase is fixed by the optical setup: the positions of the slits, the path lengths, the beam splitters and the polarizers.
\end{itemize}
Interference fringes are produced by the second kind of phase only. A single photon prepared in a mode that is a superposition of the waves leaving the two slits has a completely undetermined state phase, but the relative phase of the two waves \emph{within the mode} is perfectly well defined, and it is this relative phase that the fringes reveal. This is the field-theoretic content of Dirac's dictum that ``each photon then interferes only with itself''~\cite{Dirac}.

The general statement behind this answer is a theorem of Titulaer and Glauber~\cite{TitulaerGlauber}: a field has full first-order coherence, and can therefore produce fringes of the maximal visibility allowed by the geometry, if and only if it is a \emph{single-mode excitation}, that is, if all its photons occupy one and the same mode, whatever their quantum state (coherent, number, squeezed or thermal). Walls illustrated this with examples drawn from Young's experiment~\cite{Walls}, several of which we use below, and the meaning of Dirac's dictum has been discussed many times since~\cite{Glauber1995,Paul1986,Mandel1999}.

\subsection{What these notes contain}

We treat classical and quantized fields with the same formalism. The fields are expanded in transverse \emph{vector} modes, so that the polarization of the light is part of the mode function and, as we shall see, contributes to the relative phase that fixes the position of the fringes. As the observable we use the quantity to which a photodetector responds: the cycle-averaged square of the electric field for a classical field~\cite{MSW1964,LambScully,MandelWolf}, and Glauber's normally ordered photodetection rate for a quantized one~\cite{Glauber1963}. The prefactors are chosen so that the two coincide exactly for coherent states. The mean intensity then becomes a quadratic form whose kernel is built from the mode functions and whose coefficients describe the state. This separation makes the answer to the students' question explicit: the fringe pattern of any single-mode excitation is the squared modulus of the excited mode function, and the phase of the state does not appear in it at all.

The notes are organized as follows.
\begin{itemize}
\item Section~\ref{sec:modes} solves Maxwell's equations in a box with periodic boundary conditions and constructs the transverse vector modes of the field, including modes that are not orthogonal to each other.
\item Section~\ref{sec:energy} computes the energy of a field made of several modes of the same frequency.
\item Section~\ref{sec:quantization} quantizes this field, first for orthonormal modes and then for nonorthogonal ones.
\item Section~\ref{sec:detection} explains what a photodetector measures and introduces the intensity.
\item Section~\ref{sec:intensity} shows that, for a classical field, the interference fringes are written in the mode functions, and illustrates this with two crossing plane waves, including the role of polarization.
\item Section~\ref{sec:glauber} introduces Glauber's intensity operator and the coherence matrix.
\item Section~\ref{sec:phase} contains the main result: it distinguishes the phase of a mode from the phase of a state, and shows that only the former determines the fringes.
\item Section~\ref{sec:states} works out examples, Sec.~\ref{sec:problems} proposes problems, and Sec.~\ref{sec:conclusions} summarizes.
\item The appendices collect the mathematical tools (Appendix~\ref{app:math}), the properties of number and coherent states that we use (Appendix~\ref{app:oscillator}), an example of modes that are not plane waves (Appendix~\ref{app:pinholes}), a discussion of how sharply the phase of a state is defined (Appendix~\ref{app:sharp}), and the solutions of the problems (Appendix~\ref{app:solutions}).
\end{itemize}

\subsection{Prerequisites and how to read these notes}

We assume that the reader knows Maxwell's equations in vacuum, complex numbers and Fourier series, the basic linear algebra of Hermitian matrices, and the quantum mechanics of the harmonic oscillator in Dirac notation. Everything else is derived here, and the less familiar mathematical facts are proved in Appendix~\ref{app:math}. Readers who already know how the electromagnetic field is quantized may skim Secs.~\ref{sec:modes}--\ref{sec:quantization} and start from Sec.~\ref{sec:detection}; the main results are in Secs.~\ref{sec:intensity}--\ref{sec:phase}. At the end of each section, a box summarizes what has been established.

\subsection{Notation and conventions}

We use SI units. Vectors are written in boldface, $\bm{r} = (x,y,z)$, and unit vectors carry a hat, as in $\hat{\bm{x}}$ or $\ue$. A hat on a non-bold symbol, as in $\an{j}$ or $\hat{H}$, denotes a quantum-mechanical operator; there is no risk of confusion because unit vectors are always boldface. The vacuum permittivity and permeability are $\varepsilon_0$ and $\mu_0$, with $\varepsilon_0\mu_0 = 1/c^2$. The complex conjugate of $z$ is $z^*$, and the Hermitian conjugate of an operator $\hat{X}$ is $\hat{X}^\dagger$; the abbreviations c.c.\ and h.c.\ stand for ``complex conjugate'' and ``Hermitian conjugate'' of the preceding terms. For complex vectors $\bm{a}$ and $\bm{b}$ we write $\bm{a}\cdot\bm{b} = \sum_{i}a_ib_i$, \emph{without} complex conjugation, so that the squared length of $\bm{a}$ is $|\bm{a}|^2 = \bm{a}^*\cdot\bm{a}$. Matrices are denoted by sans-serif letters, such as $\Gm$ or $\Km$, and their elements by $G_{jk}$ or $K_{jk}$. Finally, $\dif^3r = \dif x\,\dif y\,\dif z$, and the differential $\dif$ is upright.
\section{The classical field in a box: transverse vector modes}
\label{sec:modes}

In this section we solve Maxwell's equations in empty space, inside a cubic box with periodic boundary conditions, for a field of a single frequency $\omega$. The result is a precise definition of a \emph{mode}: a vector function of position that describes the spatial shape and the polarization of the field. Everything in this section is classical.

\subsection{From Maxwell's equations to the wave equation}
\label{sec:waveeq}

In empty space, with no charges and no currents, Maxwell's equations for the electric field $\bm{E}(\bm{r},t)$ and the magnetic field $\bm{B}(\bm{r},t)$ read
\begin{equation}
\nabla\cdot\bm{E} = 0, \qquad
\nabla\cdot\bm{B} = 0, \qquad
\nabla\times\bm{E} = -\frac{\partial\bm{B}}{\partial t}, \qquad
\nabla\times\bm{B} = \frac{1}{c^2}\frac{\partial\bm{E}}{\partial t}.
\label{eq:maxwell}
\end{equation}
The second and third equations are satisfied automatically if we express the fields through a vector potential $\bm{A}(\bm{r},t)$ and a scalar potential $\Phi(\bm{r},t)$,
\begin{equation}
\bm{B} = \nabla\times\bm{A}, \qquad \bm{E} = -\nabla\Phi - \frac{\partial\bm{A}}{\partial t}.
\label{eq:potentials}
\end{equation}
Indeed, $\nabla\cdot(\nabla\times\bm{A}) = 0$ for every vector field, and $\nabla\times\bm{E} = -\nabla\times\nabla\Phi - \partial_t(\nabla\times\bm{A}) = -\partial_t\bm{B}$, because the curl of a gradient vanishes. The potentials are not unique: the gauge transformation $\bm{A}\to\bm{A}+\nabla\chi$, $\Phi\to\Phi-\partial_t\chi$, with $\chi(\bm{r},t)$ an arbitrary function, leaves $\bm{E}$ and $\bm{B}$ unchanged. We use this freedom to impose the \emph{Coulomb gauge} condition
\begin{equation}
\nabla\cdot\bm{A} = 0 .
\label{eq:coulomb}
\end{equation}
With this choice, the first Maxwell equation becomes
\[
\nabla\cdot\bm{E} = -\nabla^2\Phi - \partial_t(\nabla\cdot\bm{A}) = -\nabla^2\Phi = 0 .
\]
In the periodic box introduced below, the only solution of $\nabla^2\Phi = 0$ is a constant (Appendix~\ref{app:laplace}), and a constant potential produces no field, so we can set $\Phi = 0$. Then
\begin{equation}
\bm{E} = -\frac{\partial\bm{A}}{\partial t}, \qquad \bm{B} = \nabla\times\bm{A}.
\label{eq:EBfromA}
\end{equation}
The last Maxwell equation now gives $\nabla\times(\nabla\times\bm{A}) = -c^{-2}\,\partial_t^2\bm{A}$. Using the identity $\nabla\times(\nabla\times\bm{A}) = \nabla(\nabla\cdot\bm{A}) - \nabla^2\bm{A}$, proved in Appendix~\ref{app:vector}, together with Eq.~\eqref{eq:coulomb}, we obtain the \emph{wave equation}
\begin{equation}
\nabla^2\bm{A} - \frac{1}{c^2}\frac{\partial^2\bm{A}}{\partial t^2} = 0, \qquad \nabla\cdot\bm{A} = 0 .
\label{eq:waveeq}
\end{equation}
Here $\nabla^2\bm{A}$ is the Laplacian applied to each Cartesian component of $\bm{A}$. The second condition, called the \emph{transversality condition}, will turn out to mean that each plane wave contained in the field oscillates perpendicularly to its own direction of propagation (Sec.~\ref{sec:polvectors}). A superposition of plane waves travelling in different directions, however, can have a field component along its mean direction of propagation (Sec.~\ref{sec:polarization}).

\subsection{The box and the periodic boundary conditions}

To make the mathematics simple, we enclose the field in a cube of side $L$ and require the vector potential to be periodic,
\begin{equation}
\bm{A}(x,y,z,t) = \bm{A}(x+L,y,z,t) = \bm{A}(x,y+L,z,t) = \bm{A}(x,y,z+L,t).
\label{eq:pbc}
\end{equation}
This is a mathematical device, not a physical box with walls: it turns integrals over all space into sums over a discrete set of plane waves. Physical results do not depend on $L$ as long as $L$ is much larger than the wavelength and than the size of the experimental setup, and at the end one may let $L\to\infty$. We write
\begin{equation}
\int_{L^3}\dif^3r \equiv \int_0^L\dif x\int_0^L\dif y\int_0^L\dif z
\label{eq:intbox}
\end{equation}
for the integral over the box.

\subsection{Fields of a single frequency: separation of variables}
\label{sec:separation}

We look for solutions of the wave equation \eqref{eq:waveeq} that factorize into a function of position and a function of time,
\begin{equation}
\bm{A}(\bm{r},t) = A\,\bm{u}(\bm{r})\,f(t),
\label{eq:sepansatz}
\end{equation}
where $A$ is a constant amplitude, $\bm{u}(\bm{r})$ is a vector function that describes the spatial pattern of the field, and $f(t)$ is a scalar function that describes its time dependence. Writing $u_n(\bm{r})$, $n = 1,2,3$, for the three Cartesian components of $\bm{u}$, and inserting Eq.~\eqref{eq:sepansatz} into Eq.~\eqref{eq:waveeq}, we obtain three equations, one for each component:
\begin{equation}
A\,f(t)\,\nabla^2u_n(\bm{r}) - \frac{1}{c^2}\,A\,u_n(\bm{r})\,\frac{\dif^2f}{\dif t^2} = 0, \qquad n = 1,2,3.
\label{eq:sepcomp}
\end{equation}
Dividing by $A\,u_n(\bm{r})\,f(t)$ at points where the components do not vanish, we find
\begin{equation}
\frac{\nabla^2u_n(\bm{r})}{u_n(\bm{r})} = \frac{1}{c^2}\,\frac{\ddot{f}(t)}{f(t)}, \qquad n = 1,2,3,
\label{eq:separated}
\end{equation}
where the dot denotes a time derivative. The left-hand side depends only on $\bm{r}$, and the right-hand side only on $t$. A function of $\bm{r}$ can be equal to a function of $t$ for all $\bm{r}$ and all $t$ only if both are equal to the same constant. Since the right-hand side is the same for $n = 1,2,3$, the constant is the same for all three components. We call it $-k^2$, with $k$ a real number (we show below that the constant cannot be positive):
\begin{equation}
\nabla^2u_n(\bm{r}) = -k^2\,u_n(\bm{r}), \qquad \ddot{f}(t) = -k^2c^2\,f(t).
\label{eq:sepode}
\end{equation}
Why must the constant be negative? If it were positive, say $+\kappa^2$, then $\ddot{f} = +\kappa^2c^2f$ would have exponentially growing solutions, and the energy of the field would grow without bound; moreover, in Sec.~\ref{sec:allowed} we shall see that the spatial equation has no periodic solution with a positive constant. This justifies taking $k$ real, and since only $k^2$ appears in Eq.~\eqref{eq:sepode}, we may choose $k \ge 0$. Defining the angular frequency
\begin{equation}
\omega \equiv kc \ge 0,
\label{eq:omega}
\end{equation}
the time equation becomes the equation of a harmonic oscillator, $\ddot{f} + \omega^2f = 0$, whose general solution is
\begin{equation}
f(t) = f_+\,e^{-i\omega t} + f_-\,e^{i\omega t}.
\label{eq:fsol}
\end{equation}
The constants $f_\pm$ are fixed by the initial conditions,
\[
f(0) = f_+ + f_-, \qquad \dot{f}(0) = -i\omega\,(f_+ - f_-),
\]
which give
\begin{equation}
f_+ = \frac{1}{2}\left[f(0) + \frac{i}{\omega}\,\dot{f}(0)\right], \qquad
f_- = \frac{1}{2}\left[f(0) - \frac{i}{\omega}\,\dot{f}(0)\right].
\label{eq:initial}
\end{equation}
We exclude $\omega = 0$, which describes static fields. The spatial part $\bm{u}(\bm{r})$ satisfies the vector \emph{Helmholtz equation}, the periodic boundary conditions, and the transversality condition inherited from Eq.~\eqref{eq:waveeq}:
\begin{gather}
\nabla^2\bm{u}(\bm{r}) + k^2\,\bm{u}(\bm{r}) = 0,
\nonumber\\
\bm{u}(\bm{r}+L\hat{\bm{x}}) = \bm{u}(\bm{r}+L\hat{\bm{y}}) = \bm{u}(\bm{r}+L\hat{\bm{z}}) = \bm{u}(\bm{r}), \qquad
\nabla\cdot\bm{u}(\bm{r}) = 0 .
\label{eq:helmholtz}
\end{gather}

\paragraph{Real fields.} The vector potential is a real physical quantity, whereas $\bm{u}$ and $f$ may be complex. Since the Helmholtz equation has real coefficients, if $\bm{u}$ is a solution then so is $\bm{u}^*$. A real field of frequency $\omega$ can therefore always be written as a term proportional to $e^{-i\omega t}$ plus its complex conjugate,
\begin{equation}
\bm{A}(\bm{r},t) = A\,\bm{u}(\bm{r})\,e^{-i\omega t} + A^*\,\bm{u}^*(\bm{r})\,e^{i\omega t},
\label{eq:Areal1}
\end{equation}
and this is the form we shall use. The term proportional to $e^{-i\omega t}$ is called the \emph{positive-frequency part} of the field, and its complex conjugate the \emph{negative-frequency part}.

\subsection{Fourier series in the box}
\label{sec:fourier}

Because of the periodic boundary conditions, $\bm{u}(\bm{r})$ can be expanded in a three-dimensional Fourier series,
\begin{equation}
\bm{u}(\bm{r}) = \sum_{l_1=-\infty}^{\infty}\sum_{l_2=-\infty}^{\infty}\sum_{l_3=-\infty}^{\infty}\bm{u}_l\,
\frac{e^{2\pi i l_1x/L}}{\sqrt{L}}\,\frac{e^{2\pi i l_2y/L}}{\sqrt{L}}\,\frac{e^{2\pi i l_3z/L}}{\sqrt{L}} ,
\label{eq:fourier_full}
\end{equation}
where the vector coefficients $\bm{u}_l$ are constant. Here and hereafter the symbol $l$ is a shorthand for the triple of integers
\begin{equation}
l \equiv (l_1,l_2,l_3), \qquad l_1,l_2,l_3 = 0,\pm1,\pm2,\ldots,
\label{eq:index_triple}
\end{equation}
and we use the shorthand
\begin{equation}
\sum_l \equiv \sum_{l_1=-\infty}^{\infty}\sum_{l_2=-\infty}^{\infty}\sum_{l_3=-\infty}^{\infty} .
\label{eq:sum_shorthand}
\end{equation}
For each triple $l$ we define the wave vector $\bm{k}_l$ and the frequency $\omega_l$,
\begin{equation}
\bm{k}_l \equiv \frac{2\pi}{L}\left(l_1\hat{\bm{x}} + l_2\hat{\bm{y}} + l_3\hat{\bm{z}}\right), \qquad
\omega_l \equiv c\,|\bm{k}_l| = \frac{2\pi c}{L}\sqrt{l_1^2+l_2^2+l_3^2} .
\label{eq:kl}
\end{equation}
For the opposite triple, $-l \equiv (-l_1,-l_2,-l_3)$, we have $\bm{k}_{-l} = -\bm{k}_l$ and $\omega_{-l} = \omega_l$. Since the scalar product $\bm{k}_l\cdot\bm{r}$ equals $2\pi(l_1x + l_2y + l_3z)/L$, the three exponentials in Eq.~\eqref{eq:fourier_full} combine into a single plane wave, and the Fourier series takes the compact form
\begin{equation}
\bm{u}(\bm{r}) = \sum_l \bm{u}_l\,\frac{e^{i\bm{k}_l\cdot\bm{r}}}{L^{3/2}} .
\label{eq:fourier_compact}
\end{equation}
Each term is a plane wave that is periodic in the box, because
\[
e^{i\bm{k}_l\cdot(\bm{r}+L\hat{\bm{x}})} = e^{i\bm{k}_l\cdot\bm{r}}\,e^{2\pi il_1} = e^{i\bm{k}_l\cdot\bm{r}} ,
\]
and similarly for $y$ and $z$.

The plane waves of the box are \emph{orthonormal}:
\begin{equation}
\int_{L^3}\frac{e^{-i\bm{k}_l\cdot\bm{r}}}{L^{3/2}}\,\frac{e^{i\bm{k}_{l'}\cdot\bm{r}}}{L^{3/2}}\,\dif^3r = \delta_{ll'} ,
\label{eq:orthonormal_pw}
\end{equation}
where $\delta_{ll'} = \delta_{l_1l_1'}\delta_{l_2l_2'}\delta_{l_3l_3'}$ is equal to one if the two triples coincide and to zero otherwise. The proof, which uses only the elementary integral $\int_0^L e^{2\pi inx/L}\,\dif x = L\,\delta_{n0}$ for integer $n$, is given in Appendix~\ref{app:orthogonality}. Multiplying Eq.~\eqref{eq:fourier_compact} by $e^{-i\bm{k}_{l'}\cdot\bm{r}}/L^{3/2}$, integrating over the box and using Eq.~\eqref{eq:orthonormal_pw}, we obtain the Fourier coefficients:
\begin{equation}
\bm{u}_l = \int_{L^3}\frac{e^{-i\bm{k}_l\cdot\bm{r}}}{L^{3/2}}\,\bm{u}(\bm{r})\,\dif^3r .
\label{eq:fourier_coef}
\end{equation}
An important consequence of Eq.~\eqref{eq:fourier_coef} is that the expansion is \emph{unique}: if a Fourier series vanishes identically, all its coefficients vanish. We shall use this fact twice below.

\subsection{Which plane waves are allowed}
\label{sec:allowed}

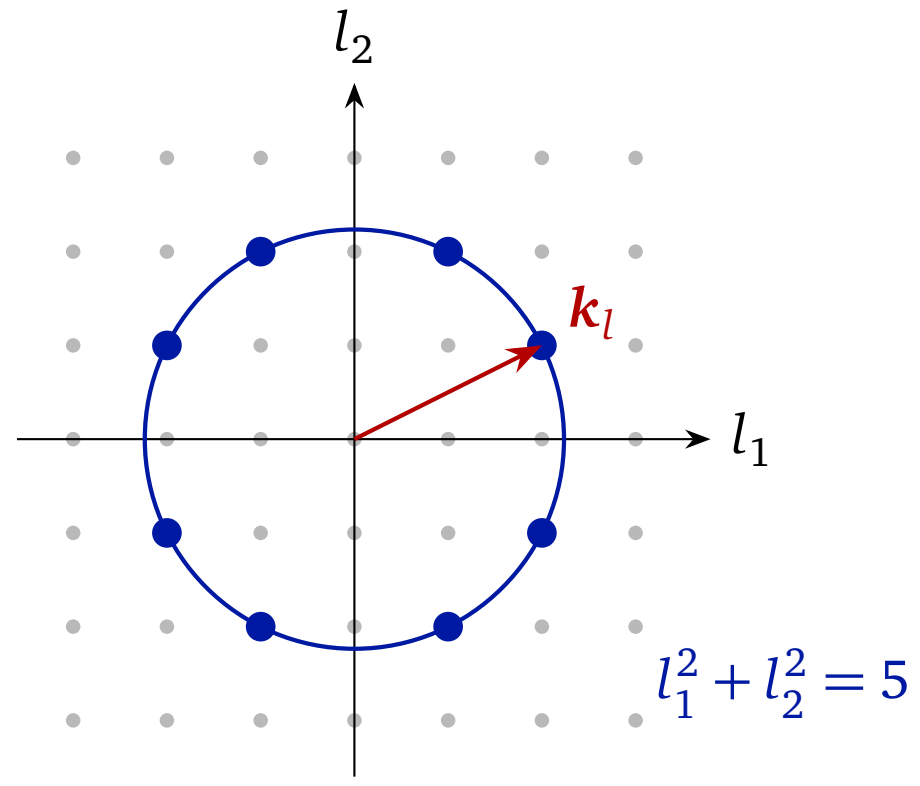
\begin{figure}[t]
\centering
\begin{tikzpicture}[scale=0.62,>=Stealth]
  \foreach \i in {-3,...,3}{\foreach \j in {-3,...,3}{\fill[gray!55] (\i,\j) circle (2.2pt);}}
  \draw[->] (-3.6,0) -- (3.8,0) node[right] {$l_1$};
  \draw[->] (0,-3.6) -- (0,3.8) node[above] {$l_2$};
  \draw[scipostblue,thick] (0,0) circle ({sqrt(5)});
  \foreach \p in {(1,2),(2,1),(-1,2),(-2,1),(1,-2),(2,-1),(-1,-2),(-2,-1)}{\fill[scipostblue] \p circle (4.5pt);}
  \draw[->,thick,red!70!black] (0,0) -- (2,1);
  \node[red!70!black] at (2.55,1.35) {$\bm{k}_l$};
  \node[scipostblue,align=left,anchor=west] at (3.0,-2.6) {$l_1^2+l_2^2 = 5$};
\end{tikzpicture}
\caption{Wave vectors of the box in the plane $l_3 = 0$, in units of $2\pi/L$ (gray dots). A field of frequency $\omega$ contains only the plane waves whose wave vectors lie on the circle (in three dimensions, the sphere) of radius $\omega/c$. For $(\omega/\omega_0)^2 = 5$ there are eight such wave vectors in this plane (blue dots), and 24 in total (Table~\ref{tab:shells}).}
\label{fig:lattice}
\end{figure}

We now insert the Fourier series \eqref{eq:fourier_compact} into the Helmholtz equation \eqref{eq:helmholtz}. The Laplacian of a plane wave is easy to compute:
\begin{align}
\nabla^2e^{i\bm{k}_l\cdot\bm{r}}
&= \left(\frac{\partial^2}{\partial x^2}+\frac{\partial^2}{\partial y^2}+\frac{\partial^2}{\partial z^2}\right)e^{i(k_{l,x}x+k_{l,y}y+k_{l,z}z)}
\nonumber\\
&= -\left(k_{l,x}^2+k_{l,y}^2+k_{l,z}^2\right)e^{i\bm{k}_l\cdot\bm{r}}
= -|\bm{k}_l|^2\,e^{i\bm{k}_l\cdot\bm{r}} ,
\label{eq:lap_pw}
\end{align}
where $k_{l,x}$, $k_{l,y}$, $k_{l,z}$ are the Cartesian components of $\bm{k}_l$. Hence, with $k = \omega/c$,
\begin{equation}
\nabla^2\bm{u} + \frac{\omega^2}{c^2}\,\bm{u} = \sum_l\left(\frac{\omega^2}{c^2} - |\bm{k}_l|^2\right)\bm{u}_l\,\frac{e^{i\bm{k}_l\cdot\bm{r}}}{L^{3/2}} = 0 .
\label{eq:helm_sub}
\end{equation}
This is a Fourier series that vanishes identically, so by uniqueness each coefficient must vanish: $(\omega^2/c^2 - |\bm{k}_l|^2)\,\bm{u}_l = 0$ for every $l$. Therefore $\bm{u}_l$ can be different from zero only if
\begin{equation}
|\bm{k}_l| = \frac{\omega}{c}, \qquad\text{or, explicitly,}\qquad \left(\frac{2\pi}{L}\right)^2\left(l_1^2+l_2^2+l_3^2\right) = \frac{\omega^2}{c^2} .
\label{eq:kcond}
\end{equation}
(This also shows why the separation constant in Eq.~\eqref{eq:sepode} could not be positive: a constant $+\kappa^2$ would require $-|\bm{k}_l|^2 = \kappa^2 > 0$, which is impossible.) A field of frequency $\omega$ is therefore a superposition of plane waves whose wave vectors all lie on the sphere of radius $\omega/c$; all these plane waves have the same frequency, but they travel in different directions. Introducing the \emph{fundamental frequency} of the box,
\begin{equation}
\omega_0 \equiv \frac{2\pi c}{L} ,
\label{eq:omega0}
\end{equation}
the condition \eqref{eq:kcond} reads
\begin{equation}
l_1^2 + l_2^2 + l_3^2 = \left(\frac{\omega}{\omega_0}\right)^2 .
\label{eq:mcond}
\end{equation}
Since the left-hand side is an integer, a nonzero field exists only if $(\omega/\omega_0)^2$ is an integer $m$ that can be written as a sum of three squares. The allowed frequencies are therefore
\begin{equation}
\omega = \sqrt{m}\,\omega_0 = \sqrt{m}\,\frac{2\pi c}{L}, \qquad m = l_1^2+l_2^2+l_3^2 \in \{1,2,3,4,5,6,8,9,\ldots\}.
\label{eq:freqs}
\end{equation}
Note that $\omega$ is in general \emph{not} an integer multiple of $\omega_0$: for instance, $l = (1,1,0)$ gives $\omega = \sqrt{2}\,\omega_0$. Moreover, not every integer is a sum of three squares. By Legendre's three-square theorem, the integers of the form $m = 4^a(8b+7)$, with $a,b = 0,1,2,\ldots$, are not; the smallest ones are $m = 7, 15, 23, 28, 31, 39,\ldots$, and these values never occur. Table~\ref{tab:shells} lists, for the first few values of $m$, the number of wave vectors on the corresponding sphere, and Fig.~\ref{fig:lattice} illustrates the construction in a two-dimensional cut.

\begin{table}[t]
\centering
\caption{Number of wave vectors $\bm{k}_l$ with $l_1^2+l_2^2+l_3^2 = m$, i.e., number of plane waves of frequency $\omega = \sqrt{m}\,\omega_0$ in the box. For $m = 7$ there are none.}
\label{tab:shells}
\begin{tabular}{lcccccccccc}
\toprule
$m$ & 1 & 2 & 3 & 4 & 5 & 6 & 7 & 8 & 9 & 10 \\
\midrule
number of $\bm{k}_l$ & 6 & 12 & 8 & 6 & 24 & 24 & 0 & 12 & 30 & 24 \\
\bottomrule
\end{tabular}
\end{table}

\subsection{Transversality and polarization vectors}
\label{sec:polvectors}

\begin{figure}[t]
\centering
\begin{tikzpicture}[>=Stealth,scale=1.0]
  \draw[thick,->] (0,0) -- (0,1.8) node[above] {$\ue_{l2}$};
  \draw[thick,->] (0,0) -- (1.5,1.15) node[above right] {$\ue_{l1}$};
  \draw[thick,->,scipostblue] (0,0) -- (3.2,0) node[right] {$\ue_{l3} = \bm{k}_l/|\bm{k}_l|$};
  \draw (0.2,0) -- (0.2,0.2) -- (0,0.2);
  \fill (0,0) circle (1.3pt);
\end{tikzpicture}
\caption{The right-handed orthonormal triad associated with the wave vector $\bm{k}_l$. The polarization vectors $\ue_{l1}$ and $\ue_{l2}$ are perpendicular to $\bm{k}_l$; the drawing is in perspective, with $\ue_{l1}$ pointing into the plane of the page, so that $\ue_{l1}\times\ue_{l2} = \ue_{l3}$.}
\label{fig:triad}
\end{figure}
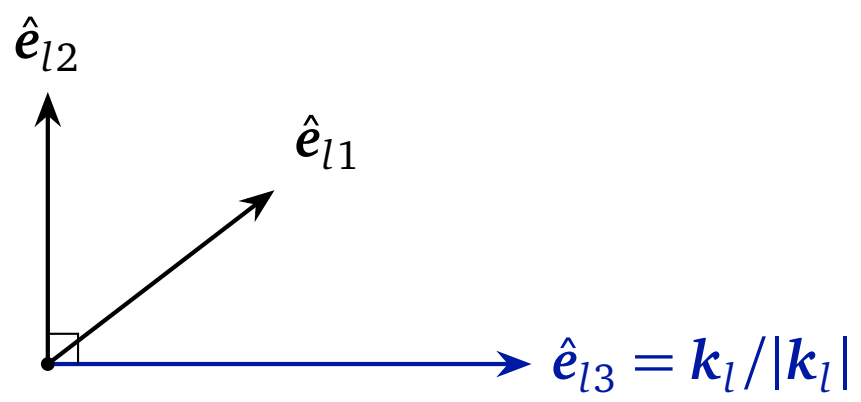

The transversality condition $\nabla\cdot\bm{u} = 0$ restricts the direction of the vector coefficients $\bm{u}_l$. Since $\nabla\cdot(\bm{u}_l\,e^{i\bm{k}_l\cdot\bm{r}}) = i\,\bm{k}_l\cdot\bm{u}_l\,e^{i\bm{k}_l\cdot\bm{r}}$ for a constant vector $\bm{u}_l$, we have
\begin{equation}
\nabla\cdot\bm{u} = \sum_l i\,\bm{k}_l\cdot\bm{u}_l\,\frac{e^{i\bm{k}_l\cdot\bm{r}}}{L^{3/2}} = 0
\qquad\Longrightarrow\qquad
\bm{k}_l\cdot\bm{u}_l = 0 \quad\text{for every } l,
\label{eq:ktransv}
\end{equation}
again by uniqueness of the Fourier expansion. Each plane wave in $\bm{u}$ is thus \emph{transverse}: its vector coefficient is perpendicular to its wave vector. To describe this, we associate with each wave vector three real unit vectors $\ue_{l1}$, $\ue_{l2}$, $\ue_{l3}$ (Fig.~\ref{fig:triad}). The third one points along the wave vector,
\begin{equation}
\ue_{l3} \equiv \frac{\bm{k}_l}{|\bm{k}_l|},
\label{eq:el3}
\end{equation}
and the three form a right-handed orthonormal triad:
\begin{equation}
\ue_{l\lambda}\cdot\ue_{l\lambda'} = \delta_{\lambda\lambda'}, \qquad
\ue_{l1}\times\ue_{l2} = \ue_{l3}, \quad
\ue_{l2}\times\ue_{l3} = \ue_{l1}, \quad
\ue_{l3}\times\ue_{l1} = \ue_{l2} .
\label{eq:triad}
\end{equation}
The vectors $\ue_{l1}$ and $\ue_{l2}$ span the plane perpendicular to $\bm{k}_l$ and are called \emph{polarization vectors}. For example, for $\bm{k}_l$ along $\hat{\bm{z}}$ one may take $\ue_{l1} = \hat{\bm{x}}$ and $\ue_{l2} = \hat{\bm{y}}$. Because of Eq.~\eqref{eq:ktransv}, each coefficient $\bm{u}_l$ has no component along $\ue_{l3}$, and can be written as
\begin{equation}
\bm{u}_l = \sum_{\lambda=1}^{2}u(l,\lambda)\,\ue_{l\lambda}, \qquad u(l,\lambda) = \ue_{l\lambda}\cdot\bm{u}_l \in \mathbb{C}.
\label{eq:ul_decomp}
\end{equation}

For each $l$, the choice of $\ue_{l1}$ and $\ue_{l2}$ is free up to a rotation about $\bm{k}_l$. For the opposite wave vector $\bm{k}_{-l} = -\bm{k}_l$ we fix this freedom by the convention
\begin{equation}
\ue_{-l1} = \ue_{l1}, \qquad \ue_{-l2} = -\ue_{l2}, \qquad \ue_{-l3} = -\ue_{l3},
\label{eq:inversion}
\end{equation}
i.e., $\ue_{-l\lambda} = -(-1)^{\lambda}\,\ue_{l\lambda}$ for $\lambda = 1,2$.
The third relation is forced by Eq.~\eqref{eq:el3}. The first two are a choice, and they keep the triad right-handed:
\[
\ue_{-l1}\times\ue_{-l2} = \ue_{l1}\times(-\ue_{l2}) = -\ue_{l3} = \ue_{-l3} .
\] This convention is used only in Sec.~\ref{sec:Delta}.

\subsection{The mode function and its normalization}
\label{sec:modefunction}

We can now write the general solution of Eq.~\eqref{eq:helmholtz} in a form that displays all the constraints at once. Let $\chi_l$ be the \emph{indicator} of the plane waves allowed at frequency $\omega$,
\begin{equation}
\chi_l \equiv
\begin{cases}
1, & \text{if } l_1^2+l_2^2+l_3^2 = (\omega/\omega_0)^2, \text{ i.e., } |\bm{k}_l| = \omega/c,\\[2pt]
0, & \text{otherwise.}
\end{cases}
\label{eq:chi}
\end{equation}
(The indicator depends on $\omega$, but we do not write this dependence explicitly, since all the modes we consider have the same frequency.) Combining Eqs.~\eqref{eq:fourier_compact}, \eqref{eq:kcond} and \eqref{eq:ul_decomp}, we obtain
\begin{equation}
\boxed{\;\bm{u}(\bm{r}) = \sum_l\sum_{\lambda=1}^{2}\chi_l\,u(l,\lambda)\,\ue_{l\lambda}\,\frac{e^{i\bm{k}_l\cdot\bm{r}}}{L^{3/2}}\;}
\label{eq:mode}
\end{equation}
The complex numbers $u(l,\lambda)$ can be chosen at will, and every choice defines a solution. We call such a solution a \emph{mode function}, or simply a \emph{mode}. A mode is a superposition of plane waves of the same frequency, each with its own direction of propagation and its own polarization. It can describe a single plane wave (only one coefficient different from zero), a standing wave, a focused beam, or the field radiated by a small aperture. In every case it is a \emph{vector} field, and its polarization is as much a part of it as its direction of propagation. A mode is an entirely classical object.

The scalar product of two vector functions in the box is defined as
\begin{equation}
(\bm{v},\bm{w}) \equiv \int_{L^3}\bm{v}^*(\bm{r})\cdot\bm{w}(\bm{r})\,\dif^3r .
\label{eq:scalarproduct}
\end{equation}
We normalize the modes to unity, $(\bm{u},\bm{u}) = 1$. In terms of the coefficients, this condition is computed by inserting Eq.~\eqref{eq:mode} twice, with independent summation indices, and using the orthonormality \eqref{eq:orthonormal_pw}:
\begin{align}
(\bm{u},\bm{u})
&= \sum_{l,\lambda}\sum_{l',\lambda'}\chi_l\chi_{l'}\,u^*(l,\lambda)\,u(l',\lambda')\;\ue_{l\lambda}\cdot\ue_{l'\lambda'}
\underbrace{\int_{L^3}\frac{e^{i(\bm{k}_{l'}-\bm{k}_l)\cdot\bm{r}}}{L^3}\,\dif^3r}_{=\,\delta_{ll'}}
\nonumber\\
&= \sum_{l}\sum_{\lambda,\lambda'}\chi_l\,u^*(l,\lambda)\,u(l,\lambda')\underbrace{\ue_{l\lambda}\cdot\ue_{l\lambda'}}_{=\,\delta_{\lambda\lambda'}}
= \sum_{l,\lambda}\chi_l\,|u(l,\lambda)|^2 .
\label{eq:norm}
\end{align}
In the first line we used that the polarization vectors are real, so that $\ue^*_{l\lambda} = \ue_{l\lambda}$, and that $\chi_l^2 = \chi_l$. The normalization condition is therefore
\begin{equation}
\sum_{l,\lambda}\chi_l\,|u(l,\lambda)|^2 = 1 ,
\label{eq:normresult}
\end{equation}
which says that the squared moduli of the plane-wave coefficients add up to one. Since $\int_{L^3}|\bm{u}|^2\dif^3r = 1$ and $\dif^3r$ has dimensions of a volume, the mode function has dimensions
\begin{equation}
[\bm{u}] = L^{-3/2}.
\label{eq:dimu}
\end{equation}

\subsection{Several modes: the Gram matrix}
\label{sec:gram}

Consider now $N$ normalized modes of the same frequency, $\bm{u}_1,\ldots,\bm{u}_N$, with coefficients $u_j(l,\lambda)$:
\begin{equation}
\bm{u}_j(\bm{r}) = \sum_l\sum_{\lambda=1}^{2}\chi_l\,u_j(l,\lambda)\,\ue_{l\lambda}\,\frac{e^{i\bm{k}_l\cdot\bm{r}}}{L^{3/2}}, \qquad j = 1,\ldots,N.
\label{eq:mode_j}
\end{equation}
Two different modes need not be orthogonal. Their scalar products form an $N\times N$ matrix $\Gm$, called the \emph{Gram matrix}:
\begin{equation}
G_{jk} \equiv (\bm{u}_j,\bm{u}_k) = \int_{L^3}\bm{u}_j^*(\bm{r})\cdot\bm{u}_k(\bm{r})\,\dif^3r .
\label{eq:gram_def}
\end{equation}
Repeating the calculation \eqref{eq:norm} with two different modes gives
\begin{equation}
G_{jk} = \sum_{l,\lambda}\sum_{l',\lambda'}\chi_l\chi_{l'}\,u_j^*(l,\lambda)\,u_k(l',\lambda')\;\ue_{l\lambda}\cdot\ue_{l'\lambda'}\,\delta_{ll'}
= \sum_{l,\lambda}\chi_l\,u_j^*(l,\lambda)\,u_k(l,\lambda) .
\label{eq:gram_coef}
\end{equation}
The diagonal elements are $G_{jj} = 1$, because the modes are normalized. The Gram matrix has two properties that we shall use repeatedly.
\begin{enumerate}
\item \emph{It is Hermitian}: $G_{kj} = G_{jk}^*$, as follows directly from Eq.~\eqref{eq:gram_def} by taking the complex conjugate.
\item \emph{It is positive semidefinite}: for every set of complex numbers $z_1,\ldots,z_N$,
\begin{equation}
\sum_{j,k=1}^{N}z_j^*\,G_{jk}\,z_k = \int_{L^3}\Big(\sum_j z_j\bm{u}_j\Big)^{\!*}\cdot\Big(\sum_k z_k\bm{u}_k\Big)\dif^3r = \int_{L^3}\Big|\sum_k z_k\,\bm{u}_k(\bm{r})\Big|^2\dif^3r \ge 0 .
\label{eq:gram_pos}
\end{equation}
The sum vanishes only if $\sum_kz_k\bm{u}_k = 0$. If the modes are \emph{linearly independent}, this happens only for $z_1 = \cdots = z_N = 0$, and $\Gm$ is \emph{positive definite}.\textbf{ We shall always assume linearly independent modes}.
\end{enumerate}
If $G_{jk} = \delta_{jk}$ the modes are called \emph{orthonormal}. A simple example of two modes that are \emph{not} orthogonal is provided by two plane waves travelling along $\hat{\bm{z}}$, one polarized along $\hat{\bm{x}}$ and one along a direction at an angle $\beta$ to the $x$ axis:
\begin{equation}
\bm{u}_1 = \hat{\bm{x}}\,\frac{e^{ikz}}{L^{3/2}}, \qquad
\bm{u}_2 = (\hat{\bm{x}}\cos\beta + \hat{\bm{y}}\sin\beta)\,\frac{e^{ikz}}{L^{3/2}}
\qquad\Longrightarrow\qquad G_{12} = \cos\beta .
\label{eq:gram_example}
\end{equation}
The two modes are orthogonal only if their polarizations are, $\beta = \pi/2$. In contrast, two plane waves with different wave vectors are always orthogonal, whatever their polarizations, because of Eq.~\eqref{eq:orthonormal_pw}.

\subsection{A second overlap integral}
\label{sec:Delta}

In the calculation of the energy we shall also meet the integral of $\bm{u}_j\cdot\bm{u}_k$ \emph{without} complex conjugation,
\begin{equation}
\Delta_{jk} \equiv \int_{L^3}\bm{u}_j(\bm{r})\cdot\bm{u}_k(\bm{r})\,\dif^3r = \Delta_{kj} .
\label{eq:Delta_def}
\end{equation}
The plane waves now appear as $e^{i\bm{k}_l\cdot\bm{r}}e^{i\bm{k}_{l'}\cdot\bm{r}}$, and the integral $\int_{L^3}e^{i(\bm{k}_l+\bm{k}_{l'})\cdot\bm{r}}\dif^3r/L^3$ is equal to $\delta_{l',-l}$: it pairs each plane wave with the \emph{opposite} one. Using also $\chi_{-l} = \chi_l$, we get
\begin{align}
\Delta_{jk}
&= \sum_{l,\lambda}\sum_{l',\lambda'}\chi_l\chi_{l'}\,u_j(l,\lambda)\,u_k(l',\lambda')\;\ue_{l\lambda}\cdot\ue_{l'\lambda'}\,\delta_{l',-l}
\nonumber\\
&= \sum_{l,\lambda}\chi_l\,u_j(l,\lambda)\sum_{\lambda'=1}^{2}u_k(-l,\lambda')\;\ue_{l\lambda}\cdot\ue_{-l\lambda'} .
\nonumber
\end{align}
By the convention \eqref{eq:inversion}, $\ue_{l\lambda}\cdot\ue_{-l\lambda'} = -(-1)^{\lambda'}\,\ue_{l\lambda}\cdot\ue_{l\lambda'} = -(-1)^{\lambda'}\delta_{\lambda\lambda'}$, so the sum over $\lambda'$ leaves only $\lambda' = \lambda$, and
\begin{equation}
\Delta_{jk} = -\sum_{l,\lambda}\chi_l\,(-1)^{\lambda}\,u_j(l,\lambda)\,u_k(-l,\lambda) .
\label{eq:Delta_final}
\end{equation}
For our purposes the explicit form of $\Delta_{jk}$ is not important. What matters is that it is in general different from zero, and that it will drop out of the energy.

\subsection{The field of \texorpdfstring{$N$}{N} modes}
\label{sec:Nmodes}

The most general real field of frequency $\omega$ built from the modes $\bm{u}_1,\ldots,\bm{u}_N$ generalizes Eq.~\eqref{eq:Areal1}:
\begin{equation}
\bm{A}(\bm{r},t) = \sum_{j=1}^{N}\left[A_j\,\bm{u}_j(\bm{r})\,e^{-i\omega t} + A_j^*\,\bm{u}_j^*(\bm{r})\,e^{i\omega t}\right],
\label{eq:Afield}
\end{equation}
where the amplitudes $A_j$ are arbitrary complex numbers. Since $[\bm{u}_j] = L^{-3/2}$ and the vector potential has units of V\,s\,m$^{-1}$, the amplitudes $A_j$ have units of V\,s\,m$^{1/2}$. From Eq.~\eqref{eq:EBfromA}, the electric and magnetic fields are
\begin{align}
\bm{E}(\bm{r},t) &= -\frac{\partial\bm{A}}{\partial t} = i\omega\sum_{j=1}^{N}\left[A_j\,\bm{u}_j(\bm{r})\,e^{-i\omega t} - A_j^*\,\bm{u}_j^*(\bm{r})\,e^{i\omega t}\right],
\label{eq:Efield}\\
\bm{B}(\bm{r},t) &= \nabla\times\bm{A} = \sum_{j=1}^{N}\left[A_j\,\nabla\times\bm{u}_j(\bm{r})\,e^{-i\omega t} + A_j^*\,\nabla\times\bm{u}_j^*(\bm{r})\,e^{i\omega t}\right].
\label{eq:Bfield}
\end{align}
For a single plane wave the curl is easy to compute: since $\nabla\times(\ue\,e^{i\bm{k}\cdot\bm{r}}) = i\,\bm{k}\times\ue\,e^{i\bm{k}\cdot\bm{r}}$ for a constant vector $\ue$, the magnetic field of a plane wave is perpendicular both to $\bm{k}$ and to $\bm{E}$, and $|\bm{B}| = |\bm{E}|/c$, as it should be.

\begin{keybox}[In a nutshell]
A monochromatic field in a periodic box is a superposition of plane waves of the same frequency $\omega$ whose wave vectors lie on the sphere $|\bm{k}_l| = \omega/c$ and whose polarizations are perpendicular to their wave vectors. A \emph{mode} $\bm{u}(\bm{r})$, Eq.~\eqref{eq:mode}, is one such superposition, normalized as in Eq.~\eqref{eq:normresult}. Several modes of the same frequency have a Hermitian, positive-definite Gram matrix $\Gm$, Eq.~\eqref{eq:gram_def}; they are orthonormal when $\Gm = \Id$. The general field is Eq.~\eqref{eq:Afield}.
\end{keybox}
\section{The energy of the field}
\label{sec:energy}

The energy stored in the electromagnetic field inside the box is
\begin{equation}
W = \frac{1}{2}\int_{L^3}\left[\varepsilon_0\,\bm{E}(\bm{r},t)\cdot\bm{E}(\bm{r},t) + \frac{1}{\mu_0}\,\bm{B}(\bm{r},t)\cdot\bm{B}(\bm{r},t)\right]\dif^3r \equiv W_E + W_B .
\label{eq:Wdef}
\end{equation}
In this section we compute $W$ for the field \eqref{eq:Afield} of $N$ modes. The result, Eq.~\eqref{eq:W}, is the starting point of the quantization in Sec.~\ref{sec:quantization}. We compute the electric and magnetic contributions separately.

\subsection{The electric energy}
\label{sec:WE}

Inserting Eq.~\eqref{eq:Efield} into $W_E$ and writing the product of the two sums as a double sum, we find
\begin{equation}
W_E = \frac{\varepsilon_0}{2}\,(i\omega)^2\sum_{j,k=1}^{N}\int_{L^3}\left(A_j\bm{u}_je^{-i\omega t} - A_j^*\bm{u}_j^*e^{i\omega t}\right)\cdot\left(A_k\bm{u}_ke^{-i\omega t} - A_k^*\bm{u}_k^*e^{i\omega t}\right)\dif^3r .
\label{eq:WE1}
\end{equation}
Multiplying out the four products, we meet four integrals. Two of them are elements of the Gram matrix \eqref{eq:gram_def}, and two are the overlaps \eqref{eq:Delta_def}:
\begin{align}
\int_{L^3}\bm{u}_j^*\cdot\bm{u}_k\,\dif^3r &= G_{jk}, &
\int_{L^3}\bm{u}_j\cdot\bm{u}_k^*\,\dif^3r &= G_{kj},
\nonumber\\
\int_{L^3}\bm{u}_j\cdot\bm{u}_k\,\dif^3r &= \Delta_{jk}, &
\int_{L^3}\bm{u}_j^*\cdot\bm{u}_k^*\,\dif^3r &= \Delta_{jk}^* .
\label{eq:four_integrals}
\end{align}
Using $(i\omega)^2 = -\omega^2$, Eq.~\eqref{eq:WE1} becomes
\begin{align}
W_E &= -\frac{\varepsilon_0\omega^2}{2}\sum_{j,k=1}^{N}\Big[A_jA_k\,\Delta_{jk}\,e^{-2i\omega t} - A_jA_k^*\,G_{kj} - A_j^*A_k\,G_{jk} + A_j^*A_k^*\,\Delta_{jk}^*\,e^{2i\omega t}\Big]
\nonumber\\
&= \frac{\varepsilon_0\omega^2}{2}\sum_{j,k=1}^{N}\Big[A_j^*G_{jk}A_k + A_k^*G_{kj}A_j\Big]
- \frac{\varepsilon_0\omega^2}{2}\Big[\sum_{j,k=1}^{N}A_jA_k\,\Delta_{jk}\,e^{-2i\omega t} + \cc\Big].
\label{eq:WE2}
\end{align}
The two terms in the first bracket are equal: exchanging the names of the summation indices $j\leftrightarrow k$ in the second one turns $\sum_{jk}A_k^*G_{kj}A_j$ into $\sum_{jk}A_j^*G_{jk}A_k$. Therefore
\begin{equation}
W_E = \varepsilon_0\omega^2\sum_{j,k=1}^{N}A_j^*\,G_{jk}\,A_k \;-\; \frac{\varepsilon_0\omega^2}{2}\Big[\sum_{j,k=1}^{N}A_jA_k\,\Delta_{jk}\,e^{-2i\omega t} + \cc\Big].
\label{eq:WE}
\end{equation}
The first term is constant in time. The second term oscillates at twice the optical frequency: the electric energy alone is not conserved.

\subsection{The magnetic energy}
\label{sec:WB}

Inserting Eq.~\eqref{eq:Bfield} into $W_B$ in the same way gives integrals of products of curls, such as $\int_{L^3}(\nabla\times\bm{u}_j^*)\cdot(\nabla\times\bm{u}_k)\,\dif^3r$. They can be reduced to the integrals \eqref{eq:four_integrals} with the help of three facts.
\begin{enumerate}
\item The vector identity, valid for any two differentiable vector fields $\bm{a}$ and $\bm{b}$ (Appendix~\ref{app:vector}),
\begin{equation}
(\nabla\times\bm{a})\cdot(\nabla\times\bm{b}) = \bm{b}\cdot\left[\nabla\times(\nabla\times\bm{a})\right] + \nabla\cdot\left[\bm{b}\times(\nabla\times\bm{a})\right].
\label{eq:vecid}
\end{equation}
\item The integral of the divergence of a \emph{periodic} vector field over the box vanishes, $\int_{L^3}\nabla\cdot\bm{F}\,\dif^3r = 0$. By the divergence theorem, this integral equals the flux of $\bm{F}$ through the surface of the cube; the contributions of two opposite faces cancel, because $\bm{F}$ takes the same values on them while their outward normals are opposite (Appendix~\ref{app:vector}).
\item Every mode function satisfies $\nabla\times(\nabla\times\bm{u}_j) = \nabla(\nabla\cdot\bm{u}_j) - \nabla^2\bm{u}_j = (\omega^2/c^2)\,\bm{u}_j$, because $\nabla\cdot\bm{u}_j = 0$ and $\nabla^2\bm{u}_j = -(\omega^2/c^2)\bm{u}_j$. The same holds for $\bm{u}_j^*$.
\end{enumerate}
Taking $\bm{a} = \bm{u}_j^*$ and $\bm{b} = \bm{u}_k$ in Eq.~\eqref{eq:vecid} and integrating over the box,
\begin{align}
\int_{L^3}(\nabla\times\bm{u}_j^*)\cdot(\nabla\times\bm{u}_k)\,\dif^3r
&= \frac{\omega^2}{c^2}\int_{L^3}\bm{u}_k\cdot\bm{u}_j^*\,\dif^3r + \underbrace{\int_{L^3}\nabla\cdot\left[\bm{u}_k\times(\nabla\times\bm{u}_j^*)\right]\dif^3r}_{=\,0}
\nonumber\\
&= \frac{\omega^2}{c^2}\,G_{jk} .
\label{eq:curlint1}
\end{align}
In exactly the same way, with or without complex conjugates,
\begin{equation}
\int_{L^3}(\nabla\times\bm{u}_j)\cdot(\nabla\times\bm{u}_k^*)\,\dif^3r = \frac{\omega^2}{c^2}\,G_{kj}, \qquad
\int_{L^3}(\nabla\times\bm{u}_j)\cdot(\nabla\times\bm{u}_k)\,\dif^3r = \frac{\omega^2}{c^2}\,\Delta_{jk},
\label{eq:curlint2}
\end{equation}
and the complex conjugate of the last one. Every curl integral is thus $\omega^2/c^2$ times the corresponding integral without curls. The magnetic energy is therefore
\begin{align}
W_B &= \frac{1}{2\mu_0}\,\frac{\omega^2}{c^2}\sum_{j,k=1}^{N}\Big[A_jA_k\,\Delta_{jk}\,e^{-2i\omega t} + A_jA_k^*\,G_{kj} + A_j^*A_k\,G_{jk} + A_j^*A_k^*\,\Delta_{jk}^*\,e^{2i\omega t}\Big]
\nonumber\\
&= \varepsilon_0\omega^2\sum_{j,k=1}^{N}A_j^*\,G_{jk}\,A_k \;+\; \frac{\varepsilon_0\omega^2}{2}\Big[\sum_{j,k=1}^{N}A_jA_k\,\Delta_{jk}\,e^{-2i\omega t} + \cc\Big],
\label{eq:WB}
\end{align}
where we used $1/(\mu_0c^2) = \varepsilon_0$ and, as before, the relabeling $j\leftrightarrow k$. Note the signs: in $W_E$ the products $\bm{E}\cdot\bm{E}$ contain $(i\omega)^2 = -\omega^2$, whereas in $W_B$ there is no such factor.

\subsection{The total energy}
\label{sec:Wtotal}

Adding Eqs.~\eqref{eq:WE} and \eqref{eq:WB}, the terms oscillating as $e^{\mp2i\omega t}$ appear with opposite signs and cancel exactly: energy flows back and forth between the electric and the magnetic field, but the total energy is conserved, as it must be for a free field. We obtain
\begin{equation}
\boxed{\;W = 2\varepsilon_0\omega^2\sum_{j,k=1}^{N}A_j^*\,G_{jk}\,A_k\;}
\label{eq:W}
\end{equation}
The energy is a quadratic form in the amplitudes, whose matrix is the Gram matrix of the modes. It is real, because $\Gm$ is Hermitian, and positive for any nonzero field, because $\Gm$ is positive definite. For orthonormal modes, $G_{jk} = \delta_{jk}$, it reduces to a sum of independent contributions, one for each mode:
\begin{equation}
W = 2\varepsilon_0\omega^2\sum_{j=1}^{N}|A_j|^2 \qquad\text{(orthonormal modes)}.
\label{eq:Worth}
\end{equation}

\paragraph{Check: a single plane wave.} For a single plane wave, $\bm{u} = \ue\,e^{i\bm{k}\cdot\bm{r}}/L^{3/2}$ with a real unit vector $\ue$ and $A = |A|e^{i\varphi}$, Eq.~\eqref{eq:Efield} gives the real field
\[
\bm{E} = i\omega\,(A\,\bm{u}\,e^{-i\omega t} - \cc) = -2\omega\,\Imag(A\,\bm{u}\,e^{-i\omega t}) = E_0\,\ue\,\sin(\omega t - \bm{k}\cdot\bm{r} - \varphi),
\]
with amplitude $E_0 = 2\omega|A|/L^{3/2}$. The cycle-averaged energy density of a plane wave is $\varepsilon_0E_0^2/2$, half electric and half magnetic, so the energy in the box is $\varepsilon_0E_0^2L^3/2 = 2\varepsilon_0\omega^2|A|^2$, in agreement with Eq.~\eqref{eq:Worth}.

\begin{keybox}[In a nutshell]
The energy of the field \eqref{eq:Afield} is $W = 2\varepsilon_0\omega^2\sum_{jk}A_j^*G_{jk}A_k$, Eq.~\eqref{eq:W}. It does not depend on time, because the oscillating parts of the electric and magnetic energies cancel. For orthonormal modes it is a sum of independent contributions, one per mode.
\end{keybox}
\section{Quantization of the field}
\label{sec:quantization}

\subsection{Each mode is a harmonic oscillator}
\label{sec:oscillator}

To see how the field should be quantized, recall the classical harmonic oscillator of mass $m$ and angular frequency $\omega$, with position $x$, momentum $p$ and energy $H = p^2/2m + m\omega^2x^2/2$. Define the complex, dimensionless variable
\begin{equation}
a \equiv \frac{m\omega\,x + i\,p}{\sqrt{2m\hbar\omega}} .
\label{eq:aosc}
\end{equation}
Then $|a|^2 = (m^2\omega^2x^2 + p^2)/(2m\hbar\omega)$, so that
\begin{equation}
H = \hbar\omega\,|a|^2 = \frac{\hbar\omega}{2}\left(a^*a + aa^*\right).
\label{eq:Hosc}
\end{equation}
(Classically $a^*a = aa^*$; we write the symmetric form for a reason that will become clear in a moment.) From the equations of motion $\dot{x} = p/m$ and $\dot{p} = -m\omega^2x$ we find $\dot{a} = (\omega p - im\omega^2x)/\sqrt{2m\hbar\omega} = -i\omega\,a$, so that $a(t) = a(0)\,e^{-i\omega t}$. In quantum mechanics $x$ and $p$ become operators with $[\hat{x},\hat{p}] = i\hbar$, and $a$ becomes the annihilation operator $\hat{a} = (m\omega\hat{x} + i\hat{p})/\sqrt{2m\hbar\omega}$, with
\begin{equation}
[\hat{a},\hat{a}^\dagger] = \frac{1}{2m\hbar\omega}\left[m\omega\hat{x}+i\hat{p},\,m\omega\hat{x}-i\hat{p}\right]
= \frac{1}{2m\hbar\omega}\left(-im\omega[\hat{x},\hat{p}] + im\omega[\hat{p},\hat{x}]\right) = 1 .
\label{eq:aadag}
\end{equation}
Since $\hat{a}^\dagger\hat{a}$ and $\hat{a}\hat{a}^\dagger$ differ by one, the classical expression $\hbar\omega|a|^2$ does not determine the quantum Hamiltonian uniquely; the symmetric form \eqref{eq:Hosc}, which is what one obtains from $p^2/2m + m\omega^2x^2/2$, gives $\hat{H} = \frac{\hbar\omega}{2}(\hat{a}^\dagger\hat{a} + \hat{a}\hat{a}^\dagger) = \hbar\omega(\hat{a}^\dagger\hat{a} + \frac{1}{2})$, including the zero-point energy. The properties of $\hat{a}$, of the number states $\ket{n}$ and of the coherent states $\ket{\alpha}$ are collected in Appendix~\ref{app:oscillator}.

The amplitudes of the field behave exactly like the variable $a$ of an oscillator: in Eq.~\eqref{eq:Afield} each amplitude appears multiplied by $e^{-i\omega t}$, and for orthonormal modes the energy \eqref{eq:Worth} is a sum of terms proportional to $|A_j|^2$. To make the analogy exact we only need to rescale the amplitudes.

\subsection{Orthonormal modes}
\label{sec:quant_orth}

For orthonormal modes, $G_{jk} = \delta_{jk}$, we introduce dimensionless amplitudes $a_j$ through
\begin{equation}
\boxed{\;A_j = \sqrt{\frac{\hbar}{2\varepsilon_0\omega}}\;a_j\;}
\label{eq:scaling}
\end{equation}
The factor has the right units: $\hbar/(\varepsilon_0\omega)$ has units of $\mathrm{J\,s}\cdot\mathrm{m\,F^{-1}}\cdot\mathrm{s} = \mathrm{V^2\,s^2\,m}$ (using $\mathrm{J/F} = \mathrm{V^2}$), so its square root has the units V\,s\,m$^{1/2}$ of $A_j$. Inserting Eq.~\eqref{eq:scaling} into Eq.~\eqref{eq:Worth},
\begin{equation}
W = 2\varepsilon_0\omega^2\,\frac{\hbar}{2\varepsilon_0\omega}\sum_{j=1}^{N}|a_j|^2 = \hbar\omega\sum_{j=1}^{N}|a_j|^2 = \frac{\hbar\omega}{2}\sum_{j=1}^{N}\left(a_j^*a_j + a_ja_j^*\right).
\label{eq:Wosc}
\end{equation}
This is the energy of $N$ independent harmonic oscillators of the same frequency, Eq.~\eqref{eq:Hosc}, and $|a_j|^2$ is the energy of mode $j$ in units of $\hbar\omega$. At the classical level, $\hbar$ in Eq.~\eqref{eq:scaling} is only bookkeeping: it fixes the units in which the amplitudes are measured.

The field is quantized by replacing the amplitudes with operators, $a_j\to\an{j}$ and $a_j^*\to\ad{j}$, with the canonical commutation relations
\begin{equation}
[\an{j},\ad{k}] = \delta_{jk}, \qquad [\an{j},\an{k}] = 0 = [\ad{j},\ad{k}] .
\label{eq:ccr}
\end{equation}
Operators of different modes commute: the modes are independent oscillators. The energy \eqref{eq:Wosc} becomes the Hamiltonian
\begin{equation}
\hat{H} = \frac{\hbar\omega}{2}\sum_{j=1}^{N}\left(\ad{j}\an{j} + \an{j}\ad{j}\right) = \hbar\omega\sum_{j=1}^{N}\left(\ad{j}\an{j} + \frac{1}{2}\right),
\label{eq:H}
\end{equation}
and the vector potential \eqref{eq:Afield} becomes the operator
\begin{equation}
\hat{\bm{A}}(\bm{r},t) = \sqrt{\frac{\hbar}{2\varepsilon_0\omega}}\sum_{j=1}^{N}\left[\an{j}\,\bm{u}_j(\bm{r})\,e^{-i\omega t} + \ad{j}\,\bm{u}_j^*(\bm{r})\,e^{i\omega t}\right].
\label{eq:Aop}
\end{equation}
The factors $e^{\mp i\omega t}$ express the time dependence of the operators in the Heisenberg picture: with the Hamiltonian \eqref{eq:H}, the Heisenberg equation $\dif\an{j}/\dif t = (i/\hbar)[\hat{H},\an{j}] = -i\omega\,\an{j}$ gives $\an{j}(t) = \an{j}\,e^{-i\omega t}$, exactly as for the classical amplitudes. From Eqs.~\eqref{eq:Efield} and \eqref{eq:Bfield} we obtain the field operators
\begin{equation}
\hat{\bm{E}} = \hat{\bm{E}}^{+} + \hat{\bm{E}}^{-}, \qquad \hat{\bm{B}} = \hat{\bm{B}}^{+} + \hat{\bm{B}}^{-},
\label{eq:EBsplit}
\end{equation}
with positive-frequency parts
\begin{equation}
\boxed{\;\begin{aligned}
\hat{\bm{E}}^{+}(\bm{r},t) &= i\sqrt{\frac{\hbar\omega}{2\varepsilon_0}}\sum_{j=1}^{N}\an{j}\,\bm{u}_j(\bm{r})\,e^{-i\omega t},\\
\hat{\bm{B}}^{+}(\bm{r},t) &= \sqrt{\frac{\hbar}{2\varepsilon_0\omega}}\sum_{j=1}^{N}\an{j}\,\nabla\times\bm{u}_j(\bm{r})\,e^{-i\omega t},
\end{aligned}\;}
\label{eq:fieldops}
\end{equation}
and negative-frequency parts $\hat{\bm{E}}^{-} = (\hat{\bm{E}}^{+})^\dagger$ and $\hat{\bm{B}}^{-} = (\hat{\bm{B}}^{+})^\dagger$. The prefactor of $\hat{\bm{E}}^+$ follows from $\omega\sqrt{\hbar/2\varepsilon_0\omega} = \sqrt{\hbar\omega/2\varepsilon_0}$. The positive-frequency part contains only annihilation operators, and the negative-frequency part only creation operators.

\paragraph{Check: units and comparison with textbooks.} For a single plane-wave mode,
\[
\bm{u}(\bm{r}) = \frac{\ue\,e^{i\bm{k}\cdot\bm{r}}}{L^{3/2}} ,
\]
and Eq.~\eqref{eq:fieldops} gives
\[
\hat{\bm{E}}^+ = i\sqrt{\frac{\hbar\omega}{2\varepsilon_0L^3}}\;\an{}\,\ue\,e^{i(\bm{k}\cdot\bm{r}-\omega t)},
\]
which is the familiar single-mode field of quantum-optics textbooks with quantization volume $V = L^3$~\cite{Loudon}. The magnetic field of the same mode contains $\nabla\times\bm{u} = i\bm{k}\times\bm{u}$, so that the ratio of the prefactors of $\hat{\bm{B}}^+$ and $\hat{\bm{E}}^+$ is $\sqrt{\hbar/2\varepsilon_0\omega}\,k/\sqrt{\hbar\omega/2\varepsilon_0} = k/\omega = 1/c$, as for a classical plane wave.

\subsection{Nonorthogonal modes}
\label{sec:nonorthogonal}

If the modes are not orthogonal, the energy \eqref{eq:W} contains products $A_j^*A_k$ with $j\ne k$: the amplitudes are coupled, and we cannot simply declare each of them to be an independent oscillator. The way out is to find new combinations of the modes that \emph{are} orthonormal. This is done by diagonalizing the Gram matrix.

\paragraph{Diagonalizing the Gram matrix.} Since $\Gm$ is Hermitian and positive definite, there is a unitary matrix $U$ ($U^\dagger U = UU^\dagger = \Id$) such that (Appendix~\ref{app:hermitian})
\begin{equation}
\Gm = U\,D\,U^\dagger, \qquad D = \mathrm{diag}(g_1,\ldots,g_N), \qquad g_l > 0 ,
\label{eq:Gdiag}
\end{equation}
i.e., $G_{jk} = \sum_{l}U_{jl}\,g_l\,U_{kl}^*$. The eigenvalues $g_l$ are strictly positive because $\Gm$ is positive definite; this will allow us to divide by $\sqrt{g_l}$ below. Inserting Eq.~\eqref{eq:Gdiag} into the energy \eqref{eq:W},
\begin{equation}
W = 2\varepsilon_0\omega^2\sum_{j,k=1}^{N}A_j^*\Big(\sum_{l=1}^{N}U_{jl}\,g_l\,U_{kl}^*\Big)A_k
= 2\varepsilon_0\omega^2\sum_{l=1}^{N}g_l\Big(\sum_{j=1}^{N}A_j^*U_{jl}\Big)\Big(\sum_{k=1}^{N}U_{kl}^*A_k\Big).
\label{eq:Wdiag1}
\end{equation}
This suggests defining new amplitudes
\begin{equation}
b_l \equiv \sum_{k=1}^{N}U_{kl}^*\,A_k, \qquad\text{so that}\qquad b_l^* = \sum_{j=1}^{N}A_j^*\,U_{jl},
\label{eq:bl}
\end{equation}
in terms of which the energy is diagonal:
\begin{equation}
W = 2\varepsilon_0\omega^2\sum_{l=1}^{N}g_l\,|b_l|^2 .
\label{eq:Wdiag}
\end{equation}
Since $U$ is unitary, Eq.~\eqref{eq:bl} can be inverted:
\[
\sum_lU_{jl}\,b_l = \sum_k\Big(\sum_lU_{jl}U^*_{kl}\Big)A_k = \sum_k(UU^\dagger)_{jk}A_k = A_j ,
\]
i.e.,
\begin{equation}
A_j = \sum_{l=1}^{N}U_{jl}\,b_l .
\label{eq:Ainv}
\end{equation}

\paragraph{Canonical amplitudes and new modes.} Comparing Eq.~\eqref{eq:Wdiag} with the oscillator energy $\hbar\omega\sum_l|a_l|^2$, we now rescale
\begin{equation}
b_l = \sqrt{\frac{\hbar}{2\varepsilon_0\omega\,g_l}}\;a_l
\qquad\Longrightarrow\qquad
W = \hbar\omega\sum_{l=1}^{N}|a_l|^2,
\qquad
A_j = \sqrt{\frac{\hbar}{2\varepsilon_0\omega}}\sum_{l=1}^{N}\frac{U_{jl}}{\sqrt{g_l}}\,a_l .
\label{eq:al}
\end{equation}
Inserting the last expression into the field \eqref{eq:Afield} and collecting the terms that multiply each $a_l$,
\begin{gather}
\bm{A}(\bm{r},t) = \sqrt{\frac{\hbar}{2\varepsilon_0\omega}}\sum_{l=1}^{N}\left[a_l\,\bm{v}_l(\bm{r})\,e^{-i\omega t} + a_l^*\,\bm{v}_l^*(\bm{r})\,e^{i\omega t}\right],
\qquad\text{with}
\nonumber\\
\boxed{\;\bm{v}_l(\bm{r}) \equiv \frac{1}{\sqrt{g_l}}\sum_{j=1}^{N}U_{jl}\,\bm{u}_j(\bm{r})\;}
\label{eq:vmodes}
\end{gather}
The new mode functions $\bm{v}_l$ are linear combinations of the old ones, and they are orthonormal:
\begin{align}
(\bm{v}_j,\bm{v}_k) &= \frac{1}{\sqrt{g_jg_k}}\sum_{m,n=1}^{N}U_{mj}^*\,U_{nk}\,(\bm{u}_m,\bm{u}_n)
= \frac{1}{\sqrt{g_jg_k}}\sum_{m,n=1}^{N}U_{mj}^*\,G_{mn}\,U_{nk}
\nonumber\\
&= \frac{(U^\dagger\Gm U)_{jk}}{\sqrt{g_jg_k}} = \frac{g_j\,\delta_{jk}}{\sqrt{g_jg_k}} = \delta_{jk} .
\label{eq:vorth}
\end{align}
The field \eqref{eq:vmodes} has exactly the form of the orthonormal case, with $\bm{u}_j$ replaced by $\bm{v}_l$ and with canonical amplitudes $a_l$. We quantize it as in Sec.~\ref{sec:quant_orth}: $a_l\to\an{l}$, $[\an{l},\ad{l'}] = \delta_{ll'}$, and Eq.~\eqref{eq:fieldops} holds with $\bm{u}_j\to\bm{v}_l$.

\paragraph{Quantizing the original amplitudes.} One could instead keep the original, nonorthogonal modes $\bm{u}_j$ and promote their amplitudes $A_j$ to operators $\hat{A}_j$. Their commutation relations then follow from Eq.~\eqref{eq:al} and from $[\an{l},\ad{m}] = \delta_{lm}$:
\begin{align}
[\hat{A}_j,\hat{A}_k^\dagger]
&= \frac{\hbar}{2\varepsilon_0\omega}\sum_{l,m=1}^{N}\frac{U_{jl}}{\sqrt{g_l}}\,\frac{U^*_{km}}{\sqrt{g_m}}\,[\an{l},\ad{m}]
= \frac{\hbar}{2\varepsilon_0\omega}\sum_{l=1}^{N}U_{jl}\,\frac{1}{g_l}\,U^*_{kl}
\nonumber\\
&= \frac{\hbar}{2\varepsilon_0\omega}\left(U\,D^{-1}\,U^\dagger\right)_{jk}
= \frac{\hbar}{2\varepsilon_0\omega}\left(\Gm^{-1}\right)_{jk},
\label{eq:Acomm}
\end{align}
where we used $\Gm^{-1} = UD^{-1}U^\dagger$, which follows from Eq.~\eqref{eq:Gdiag}. The scaled operators
\begin{equation}
\hat{c}_j \equiv \sqrt{\frac{2\varepsilon_0\omega}{\hbar}}\;\hat{A}_j
\qquad\text{satisfy}\qquad
[\hat{c}_j,\hat{c}_k^\dagger] = \left(\Gm^{-1}\right)_{jk}, \qquad [\hat{c}_j,\hat{c}_k] = 0,
\label{eq:ccomm}
\end{equation}
and the quantized field can be written with the original modes,
\begin{equation}
\hat{\bm{A}}(\bm{r},t) = \sqrt{\frac{\hbar}{2\varepsilon_0\omega}}\sum_{j=1}^{N}\left[\hat{c}_j\,\bm{u}_j(\bm{r})\,e^{-i\omega t} + \hat{c}_j^\dagger\,\bm{u}_j^*(\bm{r})\,e^{i\omega t}\right],
\label{eq:Aop_nonorth}
\end{equation}
provided the \emph{noncanonical} commutators \eqref{eq:ccomm} are used. For orthonormal modes, $\Gm = \Id$, the $\hat{c}_j$ are ordinary annihilation operators.

\paragraph{Example: two modes.} For two \textbf{normalized} modes ($G_{11} = 1 = G_{22}$), with overlap $G_{12} = g$, $0<|g|<1$, such as the modes \eqref{eq:gram_example} with $g = \cos\beta$, the Gram matrix and its inverse are
\begin{equation}
\Gm = \begin{pmatrix} 1 & g \\ g^* & 1\end{pmatrix}, \qquad
\Gm^{-1} = \frac{1}{1-|g|^2}\begin{pmatrix} 1 & -g \\ -g^* & 1\end{pmatrix}.
\label{eq:G2}
\end{equation}
The eigenvalues are $g_\pm = 1\pm|g|$, with eigenvectors proportional to $(1,\pm e^{-i\arg g})$, so the orthonormal modes are
\begin{equation}
\bm{v}_\pm = \frac{\bm{u}_1 \pm e^{-i\arg g}\,\bm{u}_2}{\sqrt{2(1\pm|g|)}} .
\label{eq:vpm}
\end{equation}
The amplitudes of the original modes satisfy $[\hat{c}_1,\hat{c}_2^\dagger] = -g/(1-|g|^2) \ne 0$. Two nonorthogonal modes are therefore \emph{not} independent degrees of freedom: they cannot be prepared in independent quantum states, such as a product of number states $\ket{n_1}_1\ket{n_2}_2$. This is why, from now on, we assume without loss of generality that the modes are orthonormal: any set of linearly independent modes can be replaced by the orthonormal set \eqref{eq:vmodes}.

\begin{keybox}[In a nutshell]
With the rescaling $A_j = \sqrt{\hbar/2\varepsilon_0\omega}\,a_j$, the field of $N$ orthonormal modes is a set of $N$ independent harmonic oscillators with energy $\hbar\omega\sum_j|a_j|^2$. Quantization replaces $a_j$ by $\an{j}$ with $[\an{j},\ad{k}] = \delta_{jk}$, and gives the field operators \eqref{eq:fieldops}. Nonorthogonal modes are first replaced by the orthonormal combinations \eqref{eq:vmodes}; if they are kept, their amplitudes obey $[\hat{c}_j,\hat{c}_k^\dagger] = (\Gm^{-1})_{jk}$. From now on all modes are orthonormal.
\end{keybox}
\section{What does a photodetector measure?}
\label{sec:detection}

Interference fringes are not seen directly: they are \emph{measured}, by a photographic plate, a camera, or a photon counter scanned across the screen. To discuss fringes we must therefore know which function of the field such a detector responds to. This section answers that question for a classical field; the quantum version is given in Sec.~\ref{sec:glauber}.

\subsection{Positive- and negative-frequency parts}

The electric field \eqref{eq:Efield} of a monochromatic field is the sum of a part oscillating as $e^{-i\omega t}$ and its complex conjugate,
\begin{equation}
\bm{E}(\bm{r},t) = \bm{E}^+(\bm{r},t) + \bm{E}^-(\bm{r},t), \qquad
\bm{E}^+ = i\omega\sum_{j=1}^{N}A_j\,\bm{u}_j\,e^{-i\omega t}, \qquad \bm{E}^- = \left(\bm{E}^+\right)^* .
\label{eq:Epm}
\end{equation}
With the dimensionless amplitudes \eqref{eq:scaling},
\begin{equation}
\bm{E}^+(\bm{r},t) = i\sqrt{\frac{\hbar\omega}{2\varepsilon_0}}\sum_{j=1}^{N}a_j\,\bm{u}_j(\bm{r})\,e^{-i\omega t},
\label{eq:Eplus_classical}
\end{equation}
which is the classical counterpart (a ``c-number'' version) of the operator \eqref{eq:fieldops}.

\subsection{The photodetection rate}

A photodetector registers light by \emph{absorbing} it. In the photoelectric effect, an electron bound in the photocathode is promoted to the continuum through the electric-dipole interaction $-\hat{\bm{d}}\cdot\bm{E}(\bm{r},t)$, where $\hat{\bm{d}}$ is the dipole operator of the electron. Treating the field classically and the electron quantum mechanically---the \emph{semiclassical theory of photodetection}~\cite{MSW1964,LambScully,MandelWolf}---one finds from Fermi's golden rule that the probability of a photoemission from a small area $\dif A$ around $\bm{r}$ during a short time $\dif t$ is
\begin{equation}
\dif P = \frac{\eta}{\hbar\omega}\,I(\bm{r},t)\,\dif A\,\dif t,
\label{eq:dP}
\end{equation}
where $\eta$, the \emph{quantum efficiency}, characterizes the detector, and
\begin{equation}
\boxed{\;I(\bm{r},t) = 2\varepsilon_0c\,\bm{E}^-(\bm{r},t)\cdot\bm{E}^+(\bm{r},t) = 2\varepsilon_0c\,|\bm{E}^+(\bm{r},t)|^2 = \varepsilon_0c\,\langle\bm{E}^2(\bm{r},t)\rangle_\ca\;}
\label{eq:Idef}
\end{equation}
is the \emph{intensity} of the light. The three expressions in Eq.~\eqref{eq:Idef} are equal; we now explain each of them.

\paragraph{Why only $\bm{E}^+$ appears.} In first-order perturbation theory, the transition amplitude contains the factor $e^{i(E_f-E_i)t/\hbar}$, which oscillates rapidly unless the energy $\hbar\omega$ supplied by the field matches the energy $E_f - E_i > 0$ gained by the electron. The term $\bm{E}^+\propto e^{-i\omega t}$ compensates this oscillation and gives a transition probability that grows steadily in time; the term $\bm{E}^-\propto e^{+i\omega t}$ does not, and averages out. The absorption rate is therefore proportional to the squared modulus of the positive-frequency part, $|\bm{E}^+|^2 = \bm{E}^-\cdot\bm{E}^+$, where we used $(\bm{E}^+)^* = \bm{E}^-$.

\paragraph{The cycle average.} The cycle average of a time-dependent quantity $f(t)$ is its mean over one optical period $T = 2\pi/\omega$ centered at $t$,
\begin{equation}
\langle f\rangle_\ca(t) \equiv \frac{1}{T}\int_{t-T/2}^{t+T/2}f(t')\,\dif t' .
\label{eq:ca}
\end{equation}
Squaring $\bm{E} = \bm{E}^+ + \bm{E}^-$ gives
\begin{equation}
\bm{E}^2 = \bm{E}^+\cdot\bm{E}^+ + \bm{E}^-\cdot\bm{E}^- + 2\,\bm{E}^-\cdot\bm{E}^+ .
\label{eq:Esq}
\end{equation}
The first term is proportional to $e^{-2i\omega t}$, and its cycle average vanishes:
\begin{equation}
\int_{t-T/2}^{t+T/2}e^{-2i\omega t'}\,\dif t' = \frac{e^{-2i\omega t}}{-2i\omega}\left(e^{-i\omega T} - e^{i\omega T}\right) = \frac{e^{-2i\omega t}}{\omega}\,\sin(\omega T) = \frac{e^{-2i\omega t}}{\omega}\,\sin(2\pi) = 0 .
\label{eq:ca_zero}
\end{equation}
The second term is its complex conjugate and also averages to zero. The third term does not depend on time at all, because the factors $e^{-i\omega t}$ in $\bm{E}^+$ and $e^{i\omega t}$ in $\bm{E}^-$ cancel. Hence $\langle\bm{E}^2\rangle_\ca = 2\,\bm{E}^-\cdot\bm{E}^+$, which is the last equality in Eq.~\eqref{eq:Idef}. For the strictly monochromatic fields considered here, $I$ does not depend on $t$; for quasi-monochromatic light, whose amplitude varies slowly on the scale of $T$, the same relations hold approximately and $I(\bm{r},t)$ follows the slow variations of the envelope.

\paragraph{The prefactor.} The factor $2\varepsilon_0c$ makes $I$ equal to the familiar intensity of a plane wave. For $\bm{E} = E_0\,\ue\cos(\bm{k}\cdot\bm{r}-\omega t)$ we have $\bm{E}^+ = \frac{1}{2}E_0\,\ue\,e^{i(\bm{k}\cdot\bm{r}-\omega t)}$, so $|\bm{E}^+|^2 = E_0^2/4$ and
\begin{equation}
I = 2\varepsilon_0c\,\frac{E_0^2}{4} = \frac{1}{2}\,\varepsilon_0cE_0^2 ,
\label{eq:Ipw}
\end{equation}
which is the magnitude of the cycle-averaged Poynting vector of the plane wave, i.e., the power per unit area that it carries. Then $I/\hbar\omega$ is the number of photons crossing a unit area per unit time, and Eq.~\eqref{eq:dP} says that $\eta$ is the probability that a photon impinging on the detector produces a photoelectron.

\begin{figure}[t]
\centering
\begin{tikzpicture}[x=1.15cm,y=1.0cm,>=Stealth]
  \fill[gray!35] (-0.25,-0.15) rectangle (0,2.45);
  \draw[thick] (0,-0.15) -- (0,2.45);
  \node[rotate=90,font=\small] at (-0.5,1.15) {mirror};
  \draw[very thick,scipostblue,domain=0:5.4,samples=200]
    plot (\x,{2*sin(deg(2*pi*\x/2.4))*sin(deg(2*pi*\x/2.4))});
  \draw[thick,red!70!black,dashed,domain=0:5.4,samples=200]
    plot (\x,{2*cos(deg(2*pi*\x/2.4))*cos(deg(2*pi*\x/2.4))});
  \draw[->] (0,0) -- (5.9,0) node[right] {$z$};
  \foreach \z in {0.6,1.8,3.0,4.2,5.4}{\fill[black] (\z-0.08,-0.38) rectangle (\z+0.08,-0.2);}
  \node[font=\small,anchor=north] at (2.7,-0.42) {blackened film};
  \draw[<->] (0,2.2) -- (0.6,2.2) node[midway,above,font=\small] {$\lambda/4$};
  \node[scipostblue,font=\small,anchor=west] at (1.3,3.05) {$\langle\bm{E}^2\rangle_\ca\propto\sin^2kz$};
  \node[red!70!black,font=\small,anchor=west] at (1.3,2.6) {$\langle\bm{B}^2\rangle_\ca\propto\cos^2kz$};
\end{tikzpicture}
\caption{Wiener's experiment. A plane wave reflected at normal incidence by a mirror at $z = 0$ forms a standing wave, whose cycle-averaged electric energy density (solid) has nodes on the mirror and antinodes at $z = \lambda/4, 3\lambda/4, \ldots$, whereas the magnetic energy density (dashed) has antinodes on the mirror. A photographic film is blackened at the antinodes of the electric field. The cycle-averaged Poynting vector of the standing wave vanishes everywhere.}
\label{fig:wiener}
\end{figure}
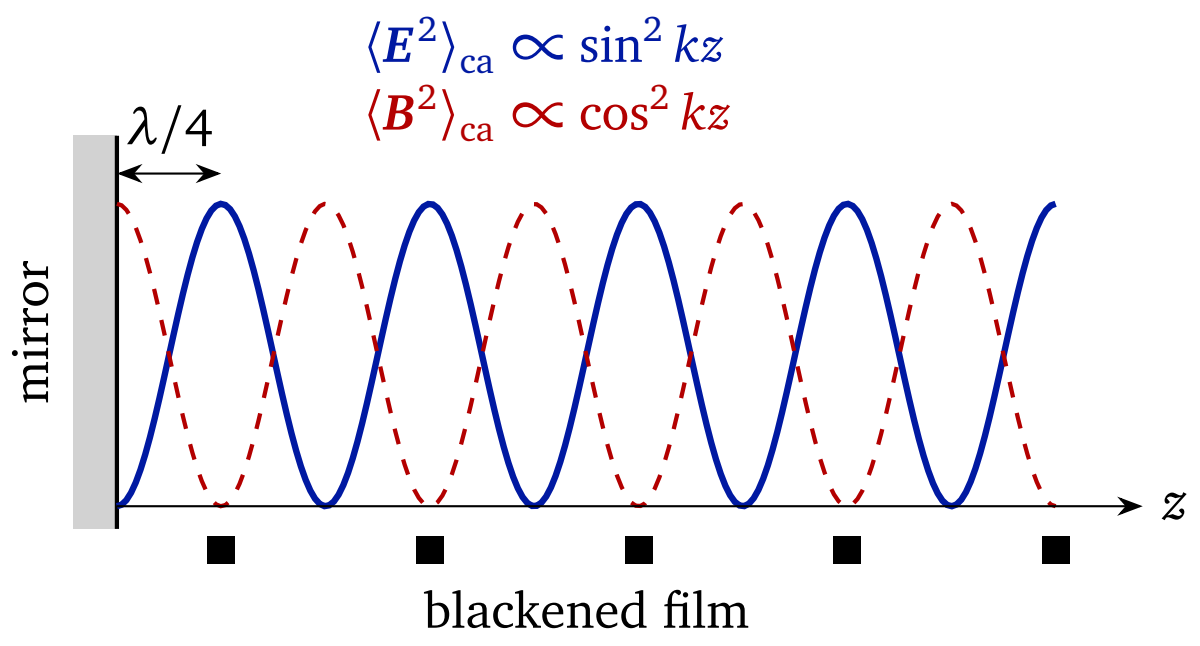

\paragraph{The magnetic field does not appear.} Its coupling to a bound electron is weaker than the electric-dipole coupling by a factor of the order of the fine-structure constant, $\alpha_{\rm fs}\approx 1/137$, and can be neglected. That photographic and photoelectric detectors respond to the electric field was established by Wiener's standing-wave experiment of 1890~\cite{Wiener1890,BornWolf}: a thin photographic film, slightly tilted with respect to a mirror in front of which a standing light wave had been formed, was blackened at the antinodes of the electric field, not at those of the magnetic field (Fig.~\ref{fig:wiener}).

\subsection{Intensity versus energy flux}
\label{sec:poynting}

Some authors, notably Loudon~\cite{Loudon}, define the intensity through the (normally ordered) Poynting vector, i.e., through the energy flux through the screen, instead of through Eq.~\eqref{eq:Idef}. For paraxial beams that hit the screen almost perpendicularly, the two definitions coincide. Beyond the paraxial regime they differ:
\begin{itemize}
\item the flux contains the obliquity factor of the screen and depends differently on polarization;
\item the flux is uniform in a standing wave, whose cycle-averaged Poynting vector vanishes everywhere, although Wiener's film recorded fringes (Fig.~\ref{fig:wiener});
\item the flux can even become negative in dark fringes, a phenomenon known as optical backflow~\cite{Eliezer2020}, whereas a detection probability cannot be negative.
\end{itemize}
Photodetectors record the intensity \eqref{eq:Idef}, not the energy flux, and this is the quantity we use throughout. We consider an ideal, point-like detector that responds equally to all directions of the electric field; an oriented or extended detector would replace $\bm{E}^-\cdot\bm{E}^+$ by $\sum_{mn}s_{mn}E^-_mE^+_n$ with a positive-semidefinite response tensor $s_{mn}$, which changes none of the conclusions below.

\begin{keybox}[In a nutshell]
A photodetector at $\bm{r}$ responds to the intensity $I(\bm{r},t) = 2\varepsilon_0c\,|\bm{E}^+(\bm{r},t)|^2$, Eq.~\eqref{eq:Idef}: the cycle average of $\varepsilon_0c\,\bm{E}^2(\bm{r},t) $. Only the positive-frequency part of the \emph{electric} field enters, because absorption requires energy $\hbar\omega$ to be taken from the field. The probability of a count is $\dif P = (\eta/\hbar\omega)\,I\,\dif A\,\dif t$.
\end{keybox}
\section{Interference of classical fields: fringes are written in the mode functions}
\label{sec:intensity}

\subsection{The intensity matrix}
\label{sec:intmatrix}

We now compute the intensity of the classical field of $N$ orthonormal modes. From Eq.~\eqref{eq:Eplus_classical} and its complex conjugate,
\begin{align}
\bm{E}^-\cdot\bm{E}^+ &= \left(-i\sqrt{\frac{\hbar\omega}{2\varepsilon_0}}\sum_{j}a_j^*\,\bm{u}_j^*\,e^{i\omega t}\right)\cdot\left(i\sqrt{\frac{\hbar\omega}{2\varepsilon_0}}\sum_{k}a_k\,\bm{u}_k\,e^{-i\omega t}\right)
\nonumber\\
&= \frac{\hbar\omega}{2\varepsilon_0}\sum_{j,k=1}^{N}a_j^*\,a_k\;\bm{u}_j^*\cdot\bm{u}_k ,
\label{eq:EmEp}
\end{align}
where we used $(-i)(i) = 1$ and $e^{i\omega t}e^{-i\omega t} = 1$: the time dependence drops out because all modes have the same frequency. Multiplying by $2\varepsilon_0c$ and using $2\varepsilon_0c\times\hbar\omega/(2\varepsilon_0) = \hbar\omega c$, Eq.~\eqref{eq:Idef} becomes
\begin{equation}
\boxed{\;I(\bm{r}) = \hbar\omega c\sum_{j,k=1}^{N}a_j^*\,K_{jk}(\bm{r})\,a_k, \qquad K_{jk}(\bm{r}) \equiv \bm{u}_j^*(\bm{r})\cdot\bm{u}_k(\bm{r}) .\;}
\label{eq:IK}
\end{equation}
We call $\Km(\bm{r}) = [K_{jk}(\bm{r})]$ the \emph{intensity matrix}. This equation is the backbone of everything that follows. The intensity is a quadratic form in the amplitudes: the matrix $\Km$ depends only on the \emph{modes}, and the coefficients $a_j^*a_k$ only on the \emph{excitation} of the modes. The intensity matrix has dimensions of $L^{-3}$ and three properties that we shall use repeatedly.

\paragraph{1. It is Hermitian.} $K_{kj} = \bm{u}_k^*\cdot\bm{u}_j = (\bm{u}_j^*\cdot\bm{u}_k)^* = K_{jk}^*$. Consequently $I$ is real: the complex conjugate of $\sum_{jk}a_j^*K_{jk}a_k$ is $\sum_{jk}a_jK_{jk}^*a_k^* = \sum_{jk}a_k^*K_{kj}a_j$, which is the same sum with the names of the indices exchanged.

\paragraph{2. It is positive semidefinite.} At every point $\bm{r}$, $K_{jk}(\bm{r})$ is the Gram matrix of the $N$ complex vectors $\bm{u}_1(\bm{r}),\ldots,\bm{u}_N(\bm{r})$ of $\mathbb{C}^3$. Hence, for all complex numbers $z_k$,
\begin{equation}
\sum_{j,k=1}^{N}z_j^*\,K_{jk}(\bm{r})\,z_k = \Big(\sum_jz_j\bm{u}_j(\bm{r})\Big)^{\!*}\cdot\Big(\sum_kz_k\bm{u}_k(\bm{r})\Big) = \Big|\sum_{k=1}^{N}z_k\,\bm{u}_k(\bm{r})\Big|^2 \ge 0 .
\label{eq:Kpos}
\end{equation}
Taking $z_k = a_k$ shows that $I(\bm{r}) = \hbar\omega c\,|\sum_ka_k\bm{u}_k(\bm{r})|^2 \ge 0$ for any amplitudes, as an intensity must be. Moreover, the Cauchy--Schwarz inequality in $\mathbb{C}^3$ (Appendix~\ref{app:cauchy}) gives
\begin{equation}
|K_{12}(\bm{r})|^2 = |\bm{u}_1^*\cdot\bm{u}_2|^2 \le |\bm{u}_1|^2\,|\bm{u}_2|^2 = K_{11}(\bm{r})\,K_{22}(\bm{r}),
\label{eq:Kcs}
\end{equation}
with equality if and only if $\bm{u}_1(\bm{r})$ and $\bm{u}_2(\bm{r})$ are parallel as complex vectors, i.e., if one is a complex multiple of the other.

\paragraph{3. Its integral is the Gram matrix.} For orthonormal modes, the integral of $K_{jk}$ over the box is $(\bm{u}_j,\bm{u}_k) = \delta_{jk}$. Hence
\begin{equation}
\int_{L^3}I(\bm{r})\,\dif^3r = \hbar\omega c\sum_{j=1}^{N}|a_j|^2 = c\,W ,
\label{eq:IW}
\end{equation}
with $W$ the field energy \eqref{eq:Wosc}: the intensity integrated over the box equals $c$ times the \emph{total} energy, electric plus magnetic. This may look surprising, because $I$ is built from the electric field alone. The explanation lies in the factor 2 of the prefactor $2\varepsilon_0c$ in Eq.~\eqref{eq:Idef}. Since $I = \varepsilon_0c\,\langle\bm{E}^2\rangle_\ca$, the cycle-averaged electric energy density is $\varepsilon_0\langle\bm{E}^2\rangle_\ca/2 = I/(2c)$: the intensity is $2c$ times the electric energy density, not $c$ times it. Integrating over the box, Eq.~\eqref{eq:IW} therefore says that the cycle-averaged electric energy is $\langle W_E\rangle_\ca = W/2$. This agrees with Sec.~\ref{sec:energy}: in Eqs.~\eqref{eq:WE} and \eqref{eq:WB} the terms oscillating as $e^{\mp2i\omega t}$ average to zero, so that $\langle W_E\rangle_\ca = \langle W_B\rangle_\ca = W/2$ (see also Problem~\ref{prob:halfenergy}). Hence
\[
\int_{L^3}I\,\dif^3r = 2c\,\langle W_E\rangle_\ca = c\left(\langle W_E\rangle_\ca + \langle W_B\rangle_\ca\right) = c\,W .
\]
The magnetic field does not enter $I$. Doubling the electric energy gives the total energy only because, on average and over the whole box, the magnetic energy equals the electric one.

This equality holds for the energy integrated over the box, not point by point. For a single traveling plane wave the electric and magnetic energy densities are equal at every point, and $I$ is $c$ times the local energy density, i.e., the magnitude of the cycle-averaged Poynting vector; the factor 2 in $2\varepsilon_0c$ was chosen precisely to make this true. In a standing wave, instead, the electric and magnetic energy densities are displaced by a quarter wavelength (Fig.~\ref{fig:wiener}). At the nodes of $\bm{E}$ the energy is entirely magnetic and $I = 0$, although the energy density is not zero; at the antinodes the energy is entirely electric and $I$ is $2c$ times the energy density. Only the average over a wavelength gives $c$ times the energy density.

\paragraph{Plane waves.} For plane-wave modes, $\bm{u}_j = \ue_j\,e^{i\bm{k}_j\cdot\bm{r}}/L^{3/2}$ with $|\bm{k}_j| = k = \omega/c$ and unit polarization vectors $\ue_j\perp\bm{k}_j$ (which may be complex, for circular or elliptical polarization),
\begin{equation}
K_{jk}(\bm{r}) = \frac{\ue_j^*\cdot\ue_k}{L^3}\;e^{i(\bm{k}_k-\bm{k}_j)\cdot\bm{r}} .
\label{eq:Kpw}
\end{equation}
For a single plane wave, $K_{11} = 1/L^3$ and $I = \hbar\omega c\,|a_1|^2/L^3$: the energy density $\hbar\omega|a_1|^2/L^3$ times the speed of light.

\subsection{Two modes}
\label{sec:twomodes}

For $N = 2$ the double sum in Eq.~\eqref{eq:IK} has four terms:
\begin{equation}
\sum_{j,k=1}^{2}a_j^*K_{jk}a_k = |a_1|^2K_{11} + |a_2|^2K_{22} + a_1^*a_2\,K_{12} + a_2^*a_1\,K_{21} .
\label{eq:four_terms}
\end{equation}
The diagonal elements $K_{jj} = |\bm{u}_j|^2$ are real and positive. The last two terms are complex conjugates of each other, because $K_{21} = K_{12}^*$; their sum is twice the real part of $z = a_1^*a_2K_{12}$. Writing the amplitudes and the off-diagonal element in polar form,
\begin{equation}
a_j = |a_j|\,e^{i\varphi_j}, \qquad K_{12}(\bm{r}) = |K_{12}(\bm{r})|\,e^{i\phi_{12}(\bm{r})},
\label{eq:polar}
\end{equation}
we get $z = |a_1||a_2||K_{12}|\,e^{i(\phi_{12}+\varphi_2-\varphi_1)}$ and $z + z^* = 2|a_1||a_2||K_{12}|\cos(\phi_{12}+\varphi_2-\varphi_1)$. Therefore
\begin{equation}
\boxed{\;I(\bm{r}) = \hbar\omega c\,\Big\{|a_1|^2K_{11}(\bm{r}) + |a_2|^2K_{22}(\bm{r}) + 2|a_1||a_2|\,|K_{12}(\bm{r})|\cos\vartheta(\bm{r})\Big\},\;}
\label{eq:I2}
\end{equation}
with
\begin{equation}
\vartheta(\bm{r}) = \underbrace{\phi_{12}(\bm{r})}_{\text{phase of the modes}} + \underbrace{\varphi_2 - \varphi_1}_{\text{phase of the amplitudes}}, \qquad
\phi_{12}(\bm{r}) = \arg\left[\bm{u}_1^*(\bm{r})\cdot\bm{u}_2(\bm{r})\right].
\label{eq:theta}
\end{equation}

Equation~\eqref{eq:I2} contains the main message of these notes in its simplest form. It has the structure
\begin{gather}
I(\bm{r}) = \hbar\omega c\,\big[A(\bm{r}) + B(\bm{r})\cos\vartheta(\bm{r})\big],
\nonumber\\
A = |a_1|^2K_{11} + |a_2|^2K_{22}, \qquad B = 2|a_1||a_2|\,|K_{12}| ,
\label{eq:AB}
\end{gather}
with a \emph{background} $A$ and a \emph{fringe amplitude} $B$. Let us read it carefully.
\begin{itemize}
\item The position dependence of $A$ and $B$ comes from the three real functions $K_{11}$, $K_{22}$ and $|K_{12}|$, which are determined by the mode functions alone.
\item The fringes themselves are produced by the cosine. Bright fringes ($\cos\vartheta = 1$) lie on the surfaces where
\begin{equation}
\phi_{12}(\bm{r}) = 2\pi m - (\varphi_2 - \varphi_1), \qquad m \in \mathbb{Z} .
\label{eq:brightsurf}
\end{equation}
The \emph{shape} of these surfaces, and their spacing, are fixed by the function $\phi_{12}(\bm{r})$, i.e., by the phase of the scalar product of the two vector mode functions. This is part of what we shall call the \emph{phase of the modes}.
\item The relative phase $\varphi_2 - \varphi_1$ of the amplitudes only relabels which surface of constant $\phi_{12}$ is bright: it translates the pattern rigidly. The global phase, $\varphi_1 + \varphi_2$, drops out altogether.
\item The moduli $|a_1|$ and $|a_2|$ affect only the contrast.
\end{itemize}

\subsection{Fringe visibility}
\label{sec:visibility}

The contrast of the fringes is measured by the \emph{visibility}. We define the local visibility as the ratio of fringe amplitude to background,
\begin{equation}
\mathcal{V}(\bm{r}) \equiv \frac{B(\bm{r})}{A(\bm{r})} = \frac{2|a_1||a_2|\,|K_{12}(\bm{r})|}{|a_1|^2K_{11}(\bm{r}) + |a_2|^2K_{22}(\bm{r})} .
\label{eq:visdef}
\end{equation}
To see that this agrees with the usual (Michelson) definition, keep $\bm{r}$ fixed and scan the relative phase $\varphi_2-\varphi_1$ over $2\pi$, as one does with a piezo-mounted mirror in an interferometer. Then $\cos\vartheta$ takes all values between $-1$ and $+1$, and $I$ oscillates between $I_{\max} = \hbar\omega c(A+B)$ and $I_{\min} = \hbar\omega c(A-B)$, so that
\begin{equation}
\frac{I_{\max}-I_{\min}}{I_{\max}+I_{\min}} = \frac{2B}{2A} = \mathcal{V} .
\label{eq:michelson}
\end{equation}
For fringes observed in space at fixed amplitudes, $\mathcal{V}$ gives the contrast of neighboring bright and dark fringes only if $A$ and $B$ vary little over one fringe period; for the plane waves of Sec.~\ref{sec:planewaves} they are constant, and the two definitions coincide exactly.

The visibility factorizes in an instructive way. Let $I_j = \hbar\omega c\,|a_j|^2K_{jj}$ be the intensity that beam $j$ would produce alone, and define the \emph{local overlap} of the two vector mode functions,
\begin{equation}
\mu_{12}(\bm{r}) \equiv \frac{K_{12}(\bm{r})}{\sqrt{K_{11}(\bm{r})\,K_{22}(\bm{r})}} = \frac{\bm{u}_1^*(\bm{r})\cdot\bm{u}_2(\bm{r})}{|\bm{u}_1(\bm{r})|\,|\bm{u}_2(\bm{r})|} .
\label{eq:mu}
\end{equation}
Writing $|K_{12}| = |\mu_{12}|\sqrt{K_{11}K_{22}}$ in the numerator of Eq.~\eqref{eq:visdef}, we have
\[
\hbar\omega c\,B = 2|\mu_{12}|\sqrt{I_1I_2} , \qquad \hbar\omega c\,A = I_1 + I_2 ,
\]
hence
\begin{equation}
\boxed{\;\mathcal{V}(\bm{r}) = |\mu_{12}(\bm{r})|\;\frac{2\sqrt{I_1(\bm{r})\,I_2(\bm{r})}}{I_1(\bm{r}) + I_2(\bm{r})} .\;}
\label{eq:visclass}
\end{equation}
Both factors lie between 0 and 1. For the first, this is the Cauchy--Schwarz inequality \eqref{eq:Kcs}. For the second, it follows from $(\sqrt{I_1}-\sqrt{I_2})^2 \ge 0$, i.e., $I_1 + I_2 \ge 2\sqrt{I_1I_2}$. Hence $0 \le \mathcal{V} \le 1$ automatically. Full contrast, $\mathcal{V} = 1$, requires \emph{both} parallel field vectors, $|\mu_{12}| = 1$, \emph{and} equal intensities, $I_1 = I_2$. The latter reduces to equal amplitudes only when $K_{11} = K_{22}$.

\paragraph{Random amplitudes.} Fringes disappear if the amplitudes fluctuate randomly and independently. Suppose that $a_1$ and $a_2$ are statistically independent random variables with zero mean, $\langle a_j\rangle = 0$, as for two independent sources whose phases drift. Averaging Eq.~\eqref{eq:IK} over the fluctuations, the cross terms contain $\langle a_1^*a_2\rangle = \langle a_1^*\rangle\langle a_2\rangle = 0$, and
\begin{equation}
\langle I(\bm{r})\rangle = \hbar\omega c\left(\langle|a_1|^2\rangle K_{11}(\bm{r}) + \langle|a_2|^2\rangle K_{22}(\bm{r})\right):
\label{eq:Irandom}
\end{equation}
the intensities of the two beams simply add, and there are no fringes. We shall meet the quantum counterpart of this situation in Sec.~\ref{sec:random}.

\subsection{Example: two crossing plane waves}
\label{sec:planewaves}

\begin{figure}[t]
\centering
\begin{tikzpicture}[scale=1.3, >=stealth]
  \draw[->, thick] (-1.2,0) -- (3.4,0) node[right] {$z$};
  \draw[->, thick] (0,-1.9) -- (0,1.9) node[above] {$y$};
  \draw (0,0) circle (0.09); \fill (0,0) circle (0.025);
  \node[below left] at (0,0) {$x$};
  \draw[->, red!75!black, ultra thick] (0,0) -- (1.97,1.01) node[above right] {$\hat{\bm{k}}_1$};
  \draw[->, blue!75!black, ultra thick] (0,0) -- (1.97,-1.01) node[below right] {$\hat{\bm{k}}_2$};
  \draw[dashed] (1.1,0) arc (0:27:1.1); \node at (1.32,0.25) {$\alpha$};
  \draw[dashed] (1.1,0) arc (0:-27:1.1); \node at (1.32,-0.27) {$\alpha$};
  \fill[gray!25] (2.9,-1.7) rectangle (3.02,1.7);
  \draw[gray!60!black] (2.9,-1.7) -- (2.9,1.7);
  \node[gray!40!black, rotate=90] at (3.25,-0.55) {\small screen};
\end{tikzpicture}
\caption{Geometry of the two-plane-wave example. The waves propagate in the $yz$ plane at angles $\pm\alpha$ to the $z$ axis and are observed on a screen parallel to the $xy$ plane. They are polarized either along $\hat{\bm{x}}$ (perpendicular to the plane of the figure) or in the plane of the figure.}
\label{fig:geometry}
\end{figure}
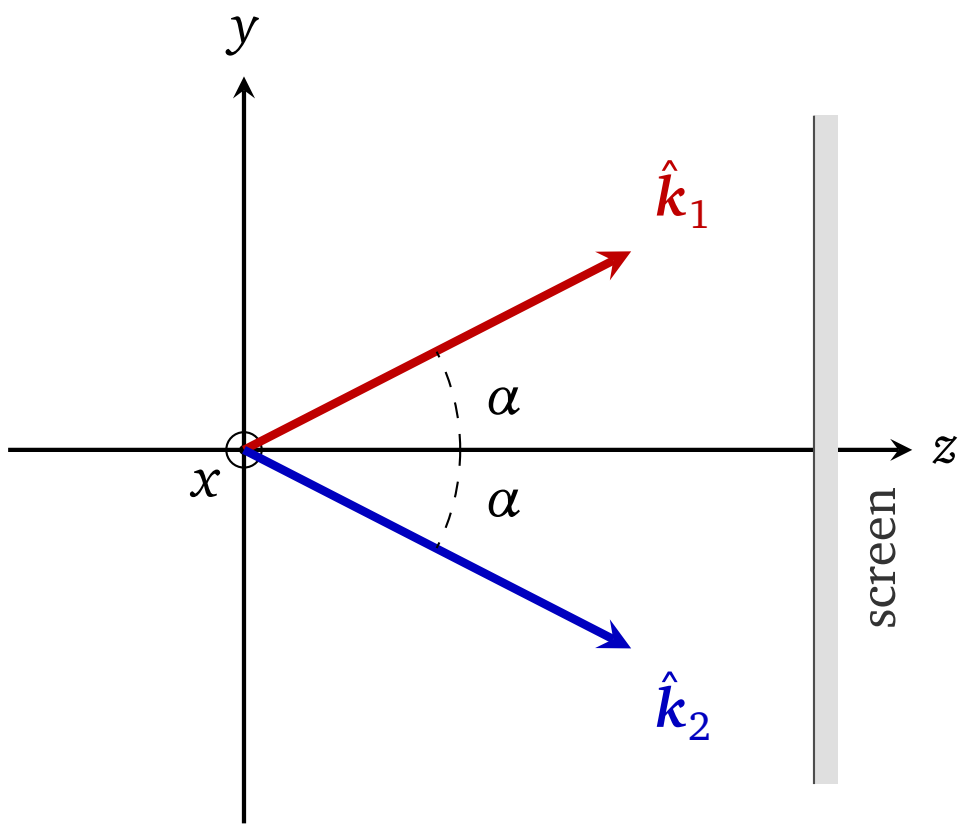

Consider two plane waves with propagation directions in the $yz$ plane,
\begin{equation}
\hat{\bm{k}}_j = \hat{\bm{y}}\sin\alpha_j + \hat{\bm{z}}\cos\alpha_j \qquad (j = 1,2),
\label{eq:pw}
\end{equation}
in the symmetric configuration $\alpha_1 = \alpha$, $\alpha_2 = -\alpha$, observed on a screen parallel to the $xy$ plane (Fig.~\ref{fig:geometry}). The difference of the wave vectors is
\begin{equation}
\bm{k}_2 - \bm{k}_1 = k\left[\hat{\bm{y}}\left(\sin\alpha_2 - \sin\alpha_1\right) + \hat{\bm{z}}\left(\cos\alpha_2 - \cos\alpha_1\right)\right] = -2k\sin\alpha\,\hat{\bm{y}} ,
\label{eq:kdiff}
\end{equation}
because $\sin(-\alpha) = -\sin\alpha$ and $\cos(-\alpha) = \cos\alpha$. By Eq.~\eqref{eq:Kpw}, the intensity matrix depends on position only through $(\bm{k}_2-\bm{k}_1)\cdot\bm{r} = -2ky\sin\alpha$.\footnote{With periodic boundary conditions the wave vectors $k\hat{\bm{k}}_j$ must belong to the lattice $\bm{k}_l$ of Eq.~\eqref{eq:kl}, so $\alpha$ can take only discrete values; for $L \gg \lambda$ this restriction is immaterial.}

\begin{figure}[t]
\centering
\begin{tikzpicture}
\begin{axis}[
    width=0.55\textwidth, height=5.6cm,
    xlabel={$y/\Delta y$}, ylabel={$I(y)/I_0$},
    xmin=-1.5, xmax=1.5, ymin=0, ymax=4.4,
    xtick={-1.5,-1,-0.5,0,0.5,1,1.5}, ytick={0,1,2,3,4},
    ymajorgrids=true, xmajorgrids=true, grid style={line width=0.2pt, draw=gray!40},
    legend columns=1, legend cell align={left}, legend style={font=\small, at={(1.03,0.5)}, anchor=west},
    samples=300, domain=-1.5:1.5]
\addplot[blue, very thick] {2 + 2*cos(deg(2*pi*x))};
\addlegendentry{$|a_1|/|a_2| = 1.0\ (\mathcal{V} = 1.00)$}
\addplot[red, thick, dashed] {1.25 + cos(deg(2*pi*x))};
\addlegendentry{$|a_1|/|a_2| = 0.5\ (\mathcal{V} = 0.80)$}
\addplot[black, thick, dotted] {1};
\addlegendentry{$|a_1|/|a_2| = 0.0\ (\mathcal{V} = 0.00)$}
\end{axis}
\end{tikzpicture}
\caption{Intensity \eqref{eq:Isym} of two crossing plane waves polarized along $\hat{\bm{x}}$, as a function of $y/\Delta y$, for $\varphi_1 = \varphi_2$ and three amplitude ratios; $I_0 = \hbar\omega c\,|a_2|^2/L^3$ and $\Delta y = \lambda/(2\sin\alpha)$. A nonzero phase difference $\varphi_1 - \varphi_2$ would translate all curves rigidly along $y$.}
\label{fig:fringes}
\end{figure}
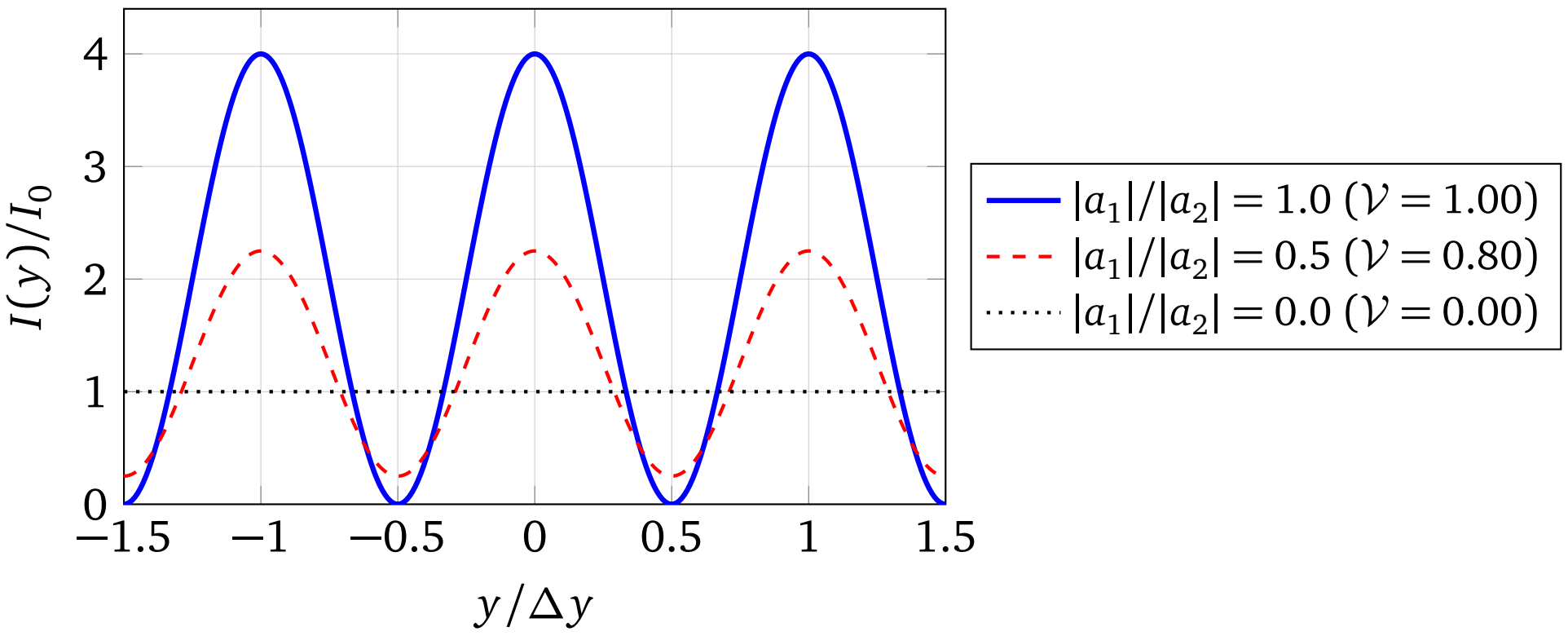

\paragraph{Polarization perpendicular to the plane of incidence.} For $\ue_1 = \ue_2 = \hat{\bm{x}}$ we have $\ue_1^*\cdot\ue_2 = 1$, and
\begin{equation}
\Km(\bm{r}) = \frac{1}{L^3}
\begin{pmatrix} 1 & e^{-2iky\sin\alpha} \\ e^{2iky\sin\alpha} & 1 \end{pmatrix},
\qquad \phi_{12}(\bm{r}) = -2ky\sin\alpha, \qquad |\mu_{12}| = 1 .
\label{eq:Ksym}
\end{equation}
Equation~\eqref{eq:I2} gives, using $\cos(-x) = \cos x$,
\begin{equation}
I(\bm{r}) = \frac{\hbar\omega c}{L^3}\Big\{|a_1|^2 + |a_2|^2 + 2|a_1||a_2|\cos\!\left(2ky\sin\alpha + \varphi_1 - \varphi_2\right)\Big\}.
\label{eq:Isym}
\end{equation}
The fringes are straight lines parallel to the $x$ axis. The cosine repeats when $y$ increases by $\Delta y$ such that $2k\sin\alpha\,\Delta y = 2\pi$, i.e., with $k = 2\pi/\lambda$,
\begin{equation}
\Delta y = \frac{\pi}{k\sin\alpha} = \frac{\lambda}{2\sin\alpha} .
\label{eq:spacing}
\end{equation}
The bright fringes lie at $y_m = [2\pi m - (\varphi_1-\varphi_2)]/(2k\sin\alpha) = m\,\Delta y + \Delta y\,(\varphi_2-\varphi_1)/2\pi$. Their orientation and spacing are fixed by the directions of the two modes; the phase difference of the amplitudes shifts them along $y$ by the fraction $(\varphi_2-\varphi_1)/2\pi$ of a period. The visibility $\mathcal{V} = 2|a_1||a_2|/(|a_1|^2+|a_2|^2)$ is uniform and reaches unity for $|a_1| = |a_2|$ (Fig.~\ref{fig:fringes}).

\subsection{Polarization is part of the mode}
\label{sec:polarization}

\begin{figure}[t]
\centering
\begin{tikzpicture}
\begin{axis}[
    width=0.55\textwidth, height=5.6cm,
    xlabel={$y/\Delta y$}, ylabel={$I(y)/(I_1+I_2)$},
    xmin=-1.5, xmax=1.5, ymin=0, ymax=2.1,
    xtick={-1.5,-1,-0.5,0,0.5,1,1.5}, ytick={0,0.5,1,1.5,2},
    ymajorgrids=true, xmajorgrids=true, grid style={line width=0.2pt, draw=gray!40},
    legend columns=1, legend cell align={left}, legend style={font=\small, at={(1.03,0.5)}, anchor=west},
    samples=300, domain=-1.5:1.5]
\addplot[blue, very thick] {1 + 0.70711*cos(deg(2*pi*x))};
\addlegendentry{$\alpha = \pi/8\ (\mathcal{V} = 0.71)$}
\addplot[black, thick, dotted] {1};
\addlegendentry{$\alpha = \pi/4\ (\mathcal{V} = 0)$}
\addplot[red, thick, dashed] {1 - 0.70711*cos(deg(2*pi*x))};
\addlegendentry{$\alpha = 3\pi/8\ (\mathcal{V} = 0.71)$}
\end{axis}
\end{tikzpicture}
\caption{Intensity \eqref{eq:Ip} of two crossing plane waves polarized in the plane of incidence, for $|a_1| = |a_2|$, $\varphi_1 = \varphi_2$ and three angles $\alpha$, normalized to the sum of the single-beam intensities, $I_1 + I_2 = 2\hbar\omega c\,|a_1|^2/L^3$. The visibility is reduced by the polarization overlap $|\cos 2\alpha|$; at $\alpha = \pi/4$ the fringes vanish, and for $\alpha > \pi/4$ bright and dark fringes are interchanged.}
\label{fig:ppol}
\end{figure}
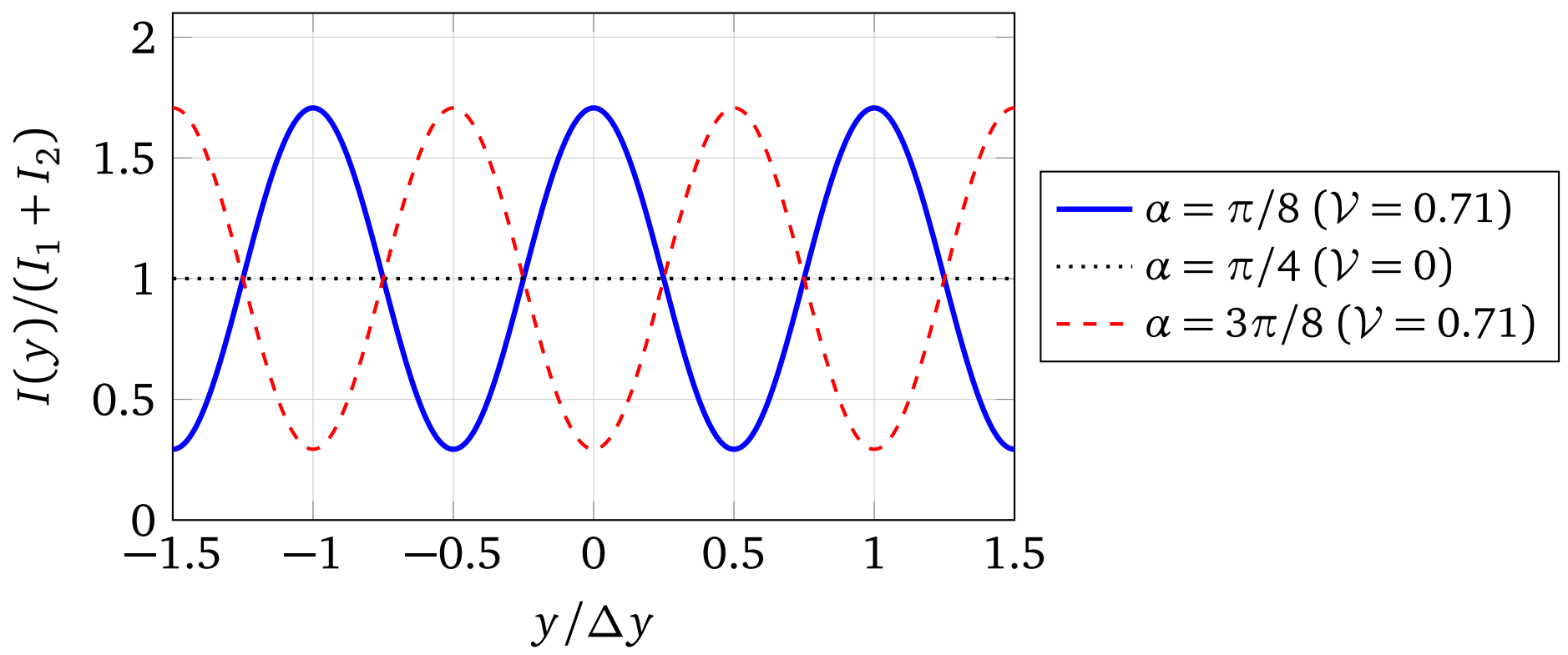

Suppose now that both waves are polarized \emph{in} the plane of incidence. The unit vector perpendicular to $\hat{\bm{k}}_j$ in the $yz$ plane is
\begin{equation}
\ue_j = \hat{\bm{x}}\times\hat{\bm{k}}_j = \hat{\bm{x}}\times\left(\hat{\bm{y}}\sin\alpha_j + \hat{\bm{z}}\cos\alpha_j\right) = \hat{\bm{z}}\sin\alpha_j - \hat{\bm{y}}\cos\alpha_j ,
\label{eq:ppol}
\end{equation}
where we used $\hat{\bm{x}}\times\hat{\bm{y}} = \hat{\bm{z}}$ and $\hat{\bm{x}}\times\hat{\bm{z}} = -\hat{\bm{y}}$. One checks that $\ue_j\cdot\hat{\bm{k}}_j = 0$ and $|\ue_j| = 1$. The polarization overlap is
\begin{equation}
\ue_1\cdot\ue_2 = \sin\alpha_1\sin\alpha_2 + \cos\alpha_1\cos\alpha_2 = \cos(\alpha_1-\alpha_2) = \cos2\alpha ,
\label{eq:cos2a}
\end{equation}
hence $K_{12} = \cos2\alpha\;e^{-2iky\sin\alpha}/L^3$ and
\begin{equation}
I(\bm{r}) = \frac{\hbar\omega c}{L^3}\Big\{|a_1|^2 + |a_2|^2 + 2|a_1||a_2|\cos2\alpha\,\cos\!\left(2ky\sin\alpha + \varphi_1 - \varphi_2\right)\Big\}.
\label{eq:Ip}
\end{equation}
Compared with Eq.~\eqref{eq:Isym}, the fringe spacing is unchanged, but the visibility is multiplied by $|\mu_{12}| = |\cos2\alpha|$ (Fig.~\ref{fig:ppol}).
\begin{itemize}
\item At $\alpha = \pi/4$ the electric fields of the two waves are orthogonal, $\ue_1\cdot\ue_2 = 0$, and the fringes disappear, although the two beams are perfectly coherent.
\item For $\alpha > \pi/4$, $\cos2\alpha < 0$. Writing $\cos2\alpha = |\cos2\alpha|\,e^{i\pi}$, the phase $\phi_{12}$ jumps by $\pi$: bright and dark fringes are interchanged, i.e., the pattern is shifted by half a period, although neither amplitude phase has changed.
\end{itemize}
The position of the fringes is thus a property of the vector mode functions, not only of the path difference.

\paragraph{A field that oscillates along its direction of propagation.} The same two waves show that transversality holds for each plane wave separately, $\ue_j\cdot\bm{k}_j = 0$, but not for their sum with respect to the mean direction of propagation, $\hat{\bm{z}}$. For equal amplitudes, $a_1 = a_2$, the electric field $\bm{E}^+\propto\sum_ja_j\bm{u}_j$ is proportional to
\begin{align*}
\ue_1\,e^{i\bm{k}_1\cdot\bm{r}} + \ue_2\,e^{i\bm{k}_2\cdot\bm{r}}
&= e^{ikz\cos\alpha}\left(\ue_1\,e^{i\theta} + \ue_2\,e^{-i\theta}\right)\\
&= -2\,e^{ikz\cos\alpha}\left(\hat{\bm{y}}\cos\alpha\,\cos\theta - i\,\hat{\bm{z}}\sin\alpha\,\sin\theta\right),
\end{align*}
where $\theta \equiv ky\sin\alpha$. Here we used $\ue_{1,2} = \pm\hat{\bm{z}}\sin\alpha - \hat{\bm{y}}\cos\alpha$ from Eq.~\eqref{eq:ppol} with $\alpha_{1,2} = \pm\alpha$, together with $e^{i\theta}+e^{-i\theta} = 2\cos\theta$ and $e^{i\theta}-e^{-i\theta} = 2i\sin\theta$. The field has a component along $\hat{\bm{z}}$, which vanishes only on the planes $\theta = n\pi$. Moreover, its $y$ and $z$ components are out of phase by $\pi/2$: at a fixed point, the real electric field $\bm{E} = \bm{E}^+ + \bm{E}^-$ runs around an ellipse in the $yz$ plane, which \emph{contains} the direction of propagation, like a spoke of a wheel rolling along $z$. For $\alpha = \pi/4$, where the intensity fringes vanish, the interference has not disappeared but has moved into the polarization: the field is linearly polarized along $\hat{\bm{y}}$ on the planes $\theta = n\pi$, along $\hat{\bm{z}}$ on the planes $\theta = \pi/2 + n\pi$, and circularly polarized in the $yz$ plane halfway between them~\cite{RodriguezLara2009}. Fields of this kind are called \emph{photonic wheels}. They carry spin angular momentum perpendicular to the direction of propagation and are common in tightly focused beams, evanescent waves and guided fields~\cite{Banzer2013,Aiello2015}. The familiar statement that light oscillates perpendicularly to its direction of propagation is thus true for each plane wave, but not for every field.

\paragraph{Arbitrary polarizations.} For two plane waves with arbitrary, possibly complex, polarization vectors, Eqs.~\eqref{eq:Kpw} and \eqref{eq:mu} give
\begin{equation}
\mu_{12}(\bm{r}) = \left(\ue_1^*\cdot\ue_2\right)e^{i(\bm{k}_2-\bm{k}_1)\cdot\bm{r}} .
\label{eq:mupw}
\end{equation}
The modulus of the polarization overlap reduces the contrast, and its phase, $\arg(\ue_1^*\cdot\ue_2)$, adds to $\phi_{12}$ and shifts the fringes rigidly. This is the phase that Pancharatnam identified as the natural definition of the phase difference between two beams in different states of polarization~\cite{Pancharatnam}. For classical light, these effects are the content of the Fresnel--Arago laws (for instance, two beams with orthogonal polarizations do not interfere)~\cite{Collett1971,Mujat2004}, which have been discussed and demonstrated many times in the teaching literature~\cite{HuntKarl1970,Andres1985,Kanseri2008,RodriguezLara2009}. Here they illustrate that the polarization of the modes contributes to the phase of the mode on the same footing as the directions of propagation. Problems~\ref{prob:pancharatnam} and~\ref{prob:helicity} work out two examples, and Appendix~\ref{app:pinholes} treats a case in which the modes are not plane waves: the spherical waves radiated by the two pinholes of Young's experiment, for which the factor $\cos2\alpha$ is replaced by the cosine of the angle between the two rays that reach the observation point.

\begin{keybox}[In a nutshell]
The classical intensity is $I(\bm{r}) = \hbar\omega c\sum_{jk}a_j^*K_{jk}(\bm{r})a_k$ with the intensity matrix $K_{jk} = \bm{u}_j^*\cdot\bm{u}_k$, which depends only on the modes. For two modes, the fringes are fixed by $\phi_{12}(\bm{r}) = \arg(\bm{u}_1^*\cdot\bm{u}_2)$, which includes the directions of propagation \emph{and} the polarizations; the phase difference of the amplitudes only translates the pattern, and their moduli only change the contrast. The visibility factorizes as $|\mu_{12}|\times(\text{balance of the two beams})$. Independent random amplitudes give no fringes.
\end{keybox}
\section{Quantized fields: Glauber's intensity operator}
\label{sec:glauber}

\subsection{The intensity operator}
\label{sec:Iop}

In Glauber's quantum theory of photodetection~\cite{Glauber1963,MandelWolf} the field is quantized as well. The derivation follows the semiclassical one. Since the electron can gain the energy $\hbar\omega$ only if the field loses it, only the annihilation part $\hat{\bm{E}}^+$ of the field acts on the initial state of the field; summing over all possible final states of the field replaces the classical product $E^-_mE^+_n$ by the expectation value $\langle\hat{E}^-_m\hat{E}^+_n\rangle$, in this order, with all atomic constants unchanged. The result is
\begin{equation}
\dif P = \frac{\eta}{\hbar\omega}\,\langle\hat{I}(\bm{r})\rangle\,\dif A\,\dif t, \qquad
\boxed{\;\hat{I}(\bm{r}) = 2\varepsilon_0c\;\hat{\bm{E}}^-(\bm{r},t)\cdot\hat{\bm{E}}^+(\bm{r},t),\;}
\label{eq:Iop}
\end{equation}
where the expectation value is taken in the state $\hat{\rho}$ of the field, $\langle\hat{I}\rangle = \tr(\hat{\rho}\,\hat{I})$; for a pure state $\hat{\rho} = \ket{\psi}\bra{\psi}$ this is $\bra{\psi}\hat{I}\ket{\psi}$. The prefactor $2\varepsilon_0c$ and the quantum efficiency $\eta$ are the same as in Eqs.~\eqref{eq:dP} and \eqref{eq:Idef}.

\paragraph{Normal order.} The operators in Eq.~\eqref{eq:Iop} are in \emph{normal order}: all creation operators (in $\hat{\bm{E}}^-$) stand to the left of all annihilation operators (in $\hat{\bm{E}}^+$). This reflects the fact that absorption removes a photon from the field, and it has two important consequences.
\begin{enumerate}
\item The vacuum produces no counts: since every $\an{k}$ annihilates the vacuum, $\hat{\bm{E}}^+\ket{0} = 0$ and therefore $\hat{I}\ket{0} = 0$.
\item The mean intensity is never negative. Writing $\hat{\bm{E}}^-\cdot\hat{\bm{E}}^+ = \sum_{n=x,y,z}(\hat{E}^+_n)^\dagger\hat{E}^+_n$,
\begin{equation}
\bra{\psi}\hat{I}\ket{\psi} = 2\varepsilon_0c\sum_{n}\bra{\psi}(\hat{E}^+_n)^\dagger\hat{E}^+_n\ket{\psi} = 2\varepsilon_0c\sum_{n}\big\|\hat{E}^+_n\ket{\psi}\big\|^2 \ge 0
\label{eq:Ipos}
\end{equation}
for every state, and therefore also for every mixture. Here $\|\ket{\chi}\| = \sqrt{\langle\chi|\chi\rangle}$ is the \emph{norm}, or length, of a state vector $\ket{\chi}$: it is never negative, and it vanishes only for the zero vector. We have used the fact that, for any operator $\hat{X}$, the bra associated with the ket $\ket{\chi} = \hat{X}\ket{\psi}$ is $\bra{\chi} = \bra{\psi}\hat{X}^\dagger$, so that
\[
\bra{\psi}\hat{X}^\dagger\hat{X}\ket{\psi} = \langle\chi|\chi\rangle = \big\|\hat{X}\ket{\psi}\big\|^2 \ge 0 .
\]
\end{enumerate}

\paragraph{The intensity operator in terms of modes.} Inserting the field operators \eqref{eq:fieldops}, with $\hat{\bm{E}}^- = -i\sqrt{\hbar\omega/2\varepsilon_0}\sum_j\ad{j}\bm{u}_j^*e^{i\omega t}$, the calculation is identical to the classical one, Eq.~\eqref{eq:EmEp}: the factors $e^{\pm i\omega t}$ cancel, $(-i)(i) = 1$, and
\begin{equation}
\boxed{\;\hat{I}(\bm{r}) = \hbar\omega c\sum_{j,k=1}^{N}\ad{j}\,K_{jk}(\bm{r})\,\an{k},\;}
\label{eq:IKop}
\end{equation}
with \emph{the same} intensity matrix and \emph{the same} prefactor $\hbar\omega c$ as the classical Eq.~\eqref{eq:IK}. The only change is $a_j^*\to\ad{j}$ and $a_k\to\an{k}$, in this order.

\paragraph{Why normal order matters.} A symmetrically ordered operator,
\[
\hat{I}_{\rm sym} = \frac{\hbar\omega c}{2}\sum_{jk}K_{jk}\left(\ad{j}\an{k} + \an{k}\ad{j}\right),
\]
would also reduce to the classical expression, but it is not what a photodetector measures. Using $\an{k}\ad{j} = \ad{j}\an{k} + \delta_{jk}$, we find $\hat{I}_{\rm sym} = \hat{I} + \frac{1}{2}\hbar\omega c\sum_jK_{jj}(\bm{r})$. The extra term is a nonzero signal in the vacuum, the contribution of the zero-point fluctuations of the field. A photodetector in the dark does not click, so this term must be absent.

\subsection{Coherent states reproduce the classical intensity}
\label{sec:coherent_classical}

The normalization of $\hat{I}$ can be checked with coherent states, whose properties are summarized in Appendix~\ref{app:oscillator}. For the product coherent state $\ket{\psi} = \ket{\alpha_1}_1\ket{\alpha_2}_2\cdots\ket{\alpha_N}_N$ we have $\an{k}\ket{\psi} = \alpha_k\ket{\psi}$, because $\an{k}$ acts only on the $k$th factor, and, taking the Hermitian conjugate, $\bra{\psi}\ad{j} = \alpha_j^*\bra{\psi}$. Therefore $\bra{\psi}\ad{j}\an{k}\ket{\psi} = \alpha_j^*\alpha_k$, and
\begin{equation}
\langle\hat{I}(\bm{r})\rangle = \hbar\omega c\sum_{j,k}\alpha_j^*\,K_{jk}(\bm{r})\,\alpha_k ,
\label{eq:Icoh}
\end{equation}
which is \emph{exactly} the classical intensity \eqref{eq:IK} with $a_j = \alpha_j$. The equality holds for every coherent state, bright or dim: the dimensionless classical amplitude $a_j$ introduced in Eq.~\eqref{eq:scaling} \emph{is} the coherent-state amplitude, and $|a_j|^2 = |\alpha_j|^2$ is the mean number of photons in mode $j$. This is the precise sense in which the prefactors of the classical and the quantum intensities have been ``chosen to agree''.

Two more checks. First, the quantum counterpart of Eq.~\eqref{eq:IW} is
\[
\int_{L^3}\langle\hat{I}\rangle\,\dif^3r = \hbar\omega c\sum_j\langle\ad{j}\an{j}\rangle ,
\]
which is $c$ times the energy of the field measured from the vacuum energy. Second, every state of the field can be written as a mixture of coherent states with a quasi-probability distribution $P(\{\alpha\})$, the Glauber--Sudarshan $P$ representation~\cite{Sudarshan1963,Glauber1963b},
\begin{equation}
\hat{\rho} = \int P(\{\alpha\})\,\ket{\{\alpha\}}\bra{\{\alpha\}}\,\dif^2\alpha_1\cdots\dif^2\alpha_N ,
\label{eq:Prep}
\end{equation}
where $\ket{\{\alpha\}}= \ket{\alpha_1}_1\ket{\alpha_2}_2\cdots\ket{\alpha_N}_N$ is the product coherent state. Since $\langle\hat{I}\rangle$ is linear in $\hat{\rho}$, Eq.~\eqref{eq:Icoh} gives $\langle\hat{I}\rangle = \int P(\{\alpha\})\,I(\{\alpha\})\,\dif^2\alpha_1\cdots\dif^2\alpha_N$: the quantum mean intensity is a weighted average of classical intensities. This is the \emph{optical equivalence theorem}. The function $P$ can be negative or very singular for nonclassical states, but the \emph{form} of the mean photodetection rate is the same in classical and quantum theory. What is specifically quantum shows up in the \emph{statistics} of the photocounts, not in their mean.

\subsection{The state enters only through the coherence matrix}
\label{sec:coherence}

Taking the expectation value of Eq.~\eqref{eq:IKop} in an arbitrary state $\hat{\rho}$, the matrix $K_{jk}(\bm{r})$, which is a number, comes out of the expectation value, and
\begin{equation}
\boxed{\;\langle\hat{I}(\bm{r})\rangle = \hbar\omega c\sum_{j,k=1}^{N}K_{jk}(\bm{r})\,C_{jk}, \qquad C_{jk} \equiv \langle\ad{j}\an{k}\rangle = \tr\!\left(\hat{\rho}\,\ad{j}\an{k}\right).\;}
\label{eq:IC}
\end{equation}
The intensity matrix $\Km$ carries the \emph{modes}; the \emph{coherence matrix} $\Cm = [C_{jk}]$ carries the \emph{state}. Comparing with Eq.~\eqref{eq:IK}, the dictionary between the classical and the quantum descriptions is simply
\begin{equation}
a_j^*\,a_k \quad\longleftrightarrow\quad C_{jk} = \langle\ad{j}\an{k}\rangle .
\label{eq:dictionary}
\end{equation}
The coherence matrix has the same two properties as the Gram matrix.
\begin{enumerate}
\item \emph{It is Hermitian}: $C_{jk}^* = \langle\ad{j}\an{k}\rangle^* = \langle(\ad{j}\an{k})^\dagger\rangle = \langle\ad{k}\an{j}\rangle = C_{kj}$. Its diagonal elements are the mean photon numbers, $C_{jj} = \langle\ad{j}\an{j}\rangle \equiv \bar{n}_j$.
\item \emph{It is positive semidefinite}: for all complex $z_k$, define $\hat{X} = \sum_kz_k\an{k}$; then
\begin{equation}
\sum_{j,k}z_j^*\,C_{jk}\,z_k = \langle\hat{X}^\dagger\hat{X}\rangle \ge 0 ,
\label{eq:Cpos}
\end{equation}
because $\langle\hat{X}^\dagger\hat{X}\rangle = \sum_ip_i\|\hat{X}\ket{\psi_i}\|^2$ for a mixture $\hat{\rho} = \sum_ip_i\ket{\psi_i}\bra{\psi_i}$ with probabilities $p_i\ge0$.
\end{enumerate}

\paragraph{Two modes.} For $N = 2$, let $\bar{n}_j = C_{jj}$ be the mean photon numbers, and define the \emph{degree of first-order coherence} between the two modes,
\begin{equation}
\gamma_{12} \equiv \frac{C_{12}}{\sqrt{\bar{n}_1\,\bar{n}_2}} = |\gamma_{12}|\,e^{i\psi} .
\label{eq:gamma}
\end{equation}
Since $\Cm$ is a positive-semidefinite $2\times2$ matrix, its two eigenvalues are nonnegative, and so is their product, the determinant: $\det\Cm = \bar{n}_1\bar{n}_2 - |C_{12}|^2 \ge 0$. Hence
\begin{equation}
|\gamma_{12}| \le 1 .
\label{eq:gamma_le1}
\end{equation}
Using $K_{21} = K_{12}^*$ and $C_{21} = C_{12}^*$, the same steps that led from Eq.~\eqref{eq:four_terms} to Eq.~\eqref{eq:I2} give
\begin{equation}
\langle\hat{I}(\bm{r})\rangle = \hbar\omega c\,\Big\{\bar{n}_1K_{11}(\bm{r}) + \bar{n}_2K_{22}(\bm{r}) + 2\sqrt{\bar{n}_1\bar{n}_2}\,|\gamma_{12}|\,|K_{12}(\bm{r})|\cos\big[\phi_{12}(\bm{r}) + \psi\big]\Big\},
\label{eq:IQ}
\end{equation}
and the visibility now factorizes into three factors, each between 0 and 1:
\begin{equation}
\boxed{\;\mathcal{V}(\bm{r}) = \underbrace{|\gamma_{12}|}_{\text{state}}\;\underbrace{|\mu_{12}(\bm{r})|}_{\text{modes}}\;\underbrace{\frac{2\sqrt{\bar{I}_1(\bm{r})\,\bar{I}_2(\bm{r})}}{\bar{I}_1(\bm{r}) + \bar{I}_2(\bm{r})}}_{\text{balance}},\qquad \bar{I}_j = \hbar\omega c\,\bar{n}_jK_{jj}.\;}
\label{eq:visQ}
\end{equation}
The first factor describes the state, the second the local overlap \eqref{eq:mu} of the two vector mode functions, and the third the balance of the two beams. This is a refinement of Walls's Eqs.~(3.14) and (3.16)~\cite{Walls}. The \emph{position} of the fringes is again set by $\phi_{12}(\bm{r})$, i.e., by the modes. The state can only shift them, through the phase $\psi$ of the coherence $C_{12}$, and reduce their contrast, through $|\gamma_{12}|$.

\paragraph{Computing $C_{12}$ in the number basis.} For a pure state written in the basis of number states of the two modes, $\ket{\psi} = \sum_{n_1,n_2}\Psi_{n_1n_2}\ket{n_1,n_2}$, the coherence can be computed directly. The operator $\an{2}$ lowers $n_2$ by one and $\ad{1}$ raises $n_1$ by one: $\ad{1}\an{2}\ket{m_1,m_2} = \sqrt{(m_1+1)\,m_2}\,\ket{m_1+1,m_2-1}$ (Appendix~\ref{app:oscillator}). Therefore
\begin{equation}
C_{12} = \bra{\psi}\ad{1}\an{2}\ket{\psi} = \sum_{m_1=0}^{\infty}\sum_{m_2=1}^{\infty}\Psi^*_{m_1+1,\,m_2-1}\,\Psi_{m_1,m_2}\,\sqrt{(m_1+1)\,m_2} .
\label{eq:Cfock}
\end{equation}
A nonzero coherence requires the state to contain pairs of components, $\ket{m_1,m_2}$ and $\ket{m_1+1,m_2-1}$, that differ by \emph{one photon moved from one mode to the other}, with a definite relative phase. We shall use this criterion repeatedly in Sec.~\ref{sec:states}.

\begin{keybox}[In a nutshell]
A photodetector measures $\langle\hat{I}\rangle$ with $\hat{I} = 2\varepsilon_0c\,\hat{\bm{E}}^-\cdot\hat{\bm{E}}^+ = \hbar\omega c\sum_{jk}\ad{j}K_{jk}\an{k}$: the same intensity matrix and prefactor as in the classical theory, but in normal order. For coherent states the quantum and classical intensities coincide exactly. In general $\langle\hat{I}\rangle = \hbar\omega c\sum_{jk}K_{jk}C_{jk}$: the modes enter through $\Km$, the state only through the coherence matrix $\Cm$. The visibility is $|\gamma_{12}|\times|\mu_{12}|\times(\text{balance})$.
\end{keybox}
\section{The phase of a mode and the phase of a state}
\label{sec:phase}

\subsection{Changing the mode basis}
\label{sec:newbasis}

The two orthonormal modes $\bm{u}_1$ and $\bm{u}_2$ are not the only natural pair: any two orthonormal combinations of them describe the same two-mode field. Consider the operator
\begin{equation}
\hat{b} = \beta_1\,\an{1} + \beta_2\,\an{2}, \qquad |\beta_1|^2 + |\beta_2|^2 = 1,
\label{eq:b}
\end{equation}
together with the orthogonal combination
\begin{equation}
\hat{c} = -\beta_2^*\,\an{1} + \beta_1^*\,\an{2} \equiv \zeta_1\,\an{1} + \zeta_2\,\an{2}, \qquad (\zeta_1,\zeta_2) = (-\beta_2^*,\beta_1^*) .
\label{eq:c}
\end{equation}

\paragraph{They are a new pair of independent modes.} Using $[\an{j},\ad{k}] = \delta_{jk}$, $\hat{b}^\dagger = \beta_1^*\ad{1} + \beta_2^*\ad{2}$ and $\hat{c}^\dagger = -\beta_2\ad{1} + \beta_1\ad{2}$,
\begin{align}
[\hat{b},\hat{b}^\dagger] &= |\beta_1|^2[\an{1},\ad{1}] + |\beta_2|^2[\an{2},\ad{2}] = |\beta_1|^2 + |\beta_2|^2 = 1, \nonumber\\
[\hat{c},\hat{c}^\dagger] &= |\beta_2|^2 + |\beta_1|^2 = 1, \nonumber\\
[\hat{b},\hat{c}^\dagger] &= \beta_1(-\beta_2)[\an{1},\ad{1}] + \beta_2\beta_1[\an{2},\ad{2}] = 0,
\label{eq:bccomm}
\end{align}
where the terms with $[\an{1},\ad{2}]$ and $[\an{2},\ad{1}]$ vanish. Also $[\hat{b},\hat{c}] = 0$. So $\hat{b}$ and $\hat{c}$ are the annihilation operators of two independent modes.

\paragraph{Inverse transformation.} In matrix form, $(\hat{b},\hat{c})^T = \mathsf{M}(\an{1},\an{2})^T$ with
\begin{equation}
\mathsf{M} = \begin{pmatrix}\beta_1 & \beta_2\\ -\beta_2^* & \beta_1^*\end{pmatrix}, \qquad
\mathsf{M}^\dagger\mathsf{M} = \Id ,
\label{eq:Mmatrix}
\end{equation}
as one checks by multiplying the matrices. The inverse is therefore $\mathsf{M}^{-1} = \mathsf{M}^\dagger$, which gives
\begin{equation}
\an{1} = \beta_1^*\,\hat{b} - \beta_2\,\hat{c}, \qquad \an{2} = \beta_2^*\,\hat{b} + \beta_1\,\hat{c}, \qquad\text{i.e.,}\qquad
\an{j} = \beta_j^*\,\hat{b} + \zeta_j^*\,\hat{c} .
\label{eq:ainv}
\end{equation}

\paragraph{The field in the new basis.} Substituting Eq.~\eqref{eq:ainv} into the field operator \eqref{eq:fieldops},
\begin{gather}
\sum_{j=1}^{2}\an{j}\,\bm{u}_j = \sum_{j=1}^{2}\left(\beta_j^*\hat{b} + \zeta_j^*\hat{c}\right)\bm{u}_j = \hat{b}\,\bm{w} + \hat{c}\,\bm{w}_\perp ,
\nonumber\\
\bm{w} \equiv \sum_{k=1}^{2}\beta_k^*\,\bm{u}_k, \qquad \bm{w}_\perp \equiv \sum_{k=1}^{2}\zeta_k^*\,\bm{u}_k ,
\label{eq:w}
\end{gather}
so that
\begin{equation}
\hat{\bm{E}}^+ = i\sqrt{\frac{\hbar\omega}{2\varepsilon_0}}\left[\hat{b}\,\bm{w}(\bm{r}) + \hat{c}\,\bm{w}_\perp(\bm{r})\right]e^{-i\omega t} .
\label{eq:Ebc}
\end{equation}
The function $\bm{w}(\bm{r})$ is the vector mode function associated with $\hat{b}$. It is normalized and orthogonal to $\bm{w}_\perp$:
\begin{align*}
(\bm{w},\bm{w}) &= \sum_{j,k}\beta_j\beta_k^*\,(\bm{u}_j,\bm{u}_k) = |\beta_1|^2+|\beta_2|^2 = 1,\\
(\bm{w},\bm{w}_\perp) &= \sum_j\beta_j\zeta_j^* = -\beta_1\beta_2 + \beta_2\beta_1 = 0 .
\end{align*}

\subsection{Single-mode excitations and the central result}
\label{sec:singlemode}

We call a state a \emph{single-mode excitation} if all its photons are in mode $\hat{b}$, i.e., if mode $\hat{c}$ is in the vacuum. For a pure state this means $\hat{c}\ket{\psi} = 0$; for a mixed state, $\hat{c}\,\hat{\rho} = 0$, and then also $\hat{\rho}\,\hat{c}^\dagger = (\hat{c}\hat{\rho})^\dagger = 0$. Examples are $\ket{n}_b\ket{0}_c$ (the $n$-photon state of mode $\hat{b}$), the coherent state $\ket{\alpha}_b\ket{0}_c$, and any mixture of such states.

Let us compute the coherence matrix of a single-mode excitation. From Eq.~\eqref{eq:ainv}, $\ad{j}\an{k} = (\beta_j\hat{b}^\dagger + \zeta_j\hat{c}^\dagger)(\beta_k^*\hat{b} + \zeta_k^*\hat{c})$, so
\begin{equation}
C_{jk} = \beta_j\beta_k^*\langle\hat{b}^\dagger\hat{b}\rangle + \beta_j\zeta_k^*\langle\hat{b}^\dagger\hat{c}\rangle + \zeta_j\beta_k^*\langle\hat{c}^\dagger\hat{b}\rangle + \zeta_j\zeta_k^*\langle\hat{c}^\dagger\hat{c}\rangle .
\label{eq:Cexpand}
\end{equation}
The last three expectation values vanish. Indeed, by the cyclic property of the trace, $\langle\hat{b}^\dagger\hat{c}\rangle = \tr(\hat{\rho}\,\hat{b}^\dagger\hat{c}) = \tr(\hat{c}\,\hat{\rho}\,\hat{b}^\dagger) = 0$ because $\hat{c}\hat{\rho} = 0$; $\langle\hat{c}^\dagger\hat{b}\rangle = \tr(\hat{\rho}\,\hat{c}^\dagger\hat{b}) = 0$ because $\hat{\rho}\hat{c}^\dagger = 0$; and $\langle\hat{c}^\dagger\hat{c}\rangle = \tr(\hat{\rho}\,\hat{c}^\dagger\hat{c}) = 0$ for the same reason. Hence
\begin{equation}
C_{jk} = \beta_j\,\beta_k^*\,\langle\hat{b}^\dagger\hat{b}\rangle \equiv \beta_j\,\beta_k^*\,\bar{n}_b ,
\label{eq:Csingle}
\end{equation}
where $\bar{n}_b$ is the mean number of photons in mode $\hat{b}$. Inserting this into Eq.~\eqref{eq:IC},
\begin{equation}
\langle\hat{I}(\bm{r})\rangle = \hbar\omega c\,\bar{n}_b\sum_{j,k}\beta_j\beta_k^*\,\bm{u}_j^*\cdot\bm{u}_k
= \hbar\omega c\,\bar{n}_b\Big(\sum_j\beta_j\bm{u}_j^*\Big)\cdot\Big(\sum_k\beta_k^*\bm{u}_k\Big)
= \hbar\omega c\,\bar{n}_b\,\bm{w}^*\cdot\bm{w} ,
\nonumber
\end{equation}
and we arrive at the central result of these notes:
\begin{equation}
\boxed{\;\langle\hat{I}(\bm{r})\rangle = \hbar\omega c\;\bar{n}_b\,|\bm{w}(\bm{r})|^2 .\;}
\label{eq:central}
\end{equation}

\begin{keybox}[The central result]
The fringe pattern of a single-mode excitation is the squared modulus of the single vector mode function $\bm{w}(\bm{r})$ of the excited mode. It is the same for \emph{every} state of mode $\hat{b}$---number, coherent, squeezed, thermal---and the same as for a classical field with amplitudes $a_j\propto\beta_j^*$. The state only fixes the overall scale, through the mean photon number $\bar{n}_b$.
\end{keybox}

For a single photon, $\bar{n}_b = 1$, and Eqs.~\eqref{eq:Iop} and \eqref{eq:central} give the detection probability
\begin{equation}
\dif P = \eta\,c\,|\bm{w}(\bm{r})|^2\,\dif A\,\dif t .
\label{eq:singlephoton}
\end{equation}
The vector mode function plays the role of a detection amplitude for the photon: its squared modulus gives the probability of detecting the photon at $\bm{r}$. (One checks that $c\,|\bm{w}|^2\,\dif A\,\dif t$ is dimensionless, since $[\bm{w}] = L^{-3/2}$.)

\subsection{The converse: the Titulaer--Glauber theorem}
\label{sec:TG}

Since $C_{jk} = \bar{n}_b\,\beta_j\beta_k^*$ is the product of a column vector and a row vector, the matrix $\Cm$ of a single-mode excitation has rank one, and $|C_{12}|^2 = \bar{n}_b^2|\beta_1|^2|\beta_2|^2 = C_{11}C_{22}$. Hence $|\gamma_{12}| = 1$: a single-mode excitation has full first-order coherence.

The converse also holds. Suppose that $|\gamma_{12}| = 1$, i.e., $\det\Cm = 0$. Then $\Cm$ has an eigenvalue equal to zero, with a normalized eigenvector $(\zeta_1,\zeta_2)$: $\sum_kC_{jk}\zeta_k = 0$. Define the mode $\hat{c} = \sum_k\zeta_k\an{k}$. Its mean photon number is
\begin{equation}
\langle\hat{c}^\dagger\hat{c}\rangle = \sum_{j,k}\zeta_j^*\,\langle\ad{j}\an{k}\rangle\,\zeta_k = \sum_{j,k}\zeta_j^*\,C_{jk}\,\zeta_k = 0 .
\label{eq:cempty}
\end{equation}
What does it mean that mode $\hat{c}$ has zero mean photon number? Consider first a pure state $\ket{\psi}$. Then $\langle\hat{c}^\dagger\hat{c}\rangle = \bra{\psi}\hat{c}^\dagger\hat{c}\ket{\psi} = \|\hat{c}\ket{\psi}\|^2$ is the squared norm of the vector $\hat{c}\ket{\psi}$ (see the remark after Eq.~\eqref{eq:Ipos}). A norm vanishes only for the zero vector, so Eq.~\eqref{eq:cempty} gives $\hat{c}\ket{\psi} = 0$: trying to remove a photon from mode $\hat{c}$ gives zero, because the mode contains no photons. For a mixture $\hat{\rho} = \sum_ip_i\ket{\psi_i}\bra{\psi_i}$, with probabilities $p_i > 0$, the mean value is the weighted average of the mean values in the pure states $\ket{\psi_i}$:
\[
\langle\hat{c}^\dagger\hat{c}\rangle = \tr\!\left(\hat{\rho}\,\hat{c}^\dagger\hat{c}\right) = \sum_ip_i\bra{\psi_i}\hat{c}^\dagger\hat{c}\ket{\psi_i} = \sum_ip_i\,\big\|\hat{c}\ket{\psi_i}\big\|^2 .
\]
This is a sum of nonnegative terms, and it vanishes only if every term vanishes: $\hat{c}\ket{\psi_i} = 0$ for every state $\ket{\psi_i}$ of the mixture. In every component, therefore, mode $\hat{c}$ is in the vacuum, and all the photons are in the orthogonal mode $\hat{b} = \zeta_2^*\,\an{1} - \zeta_1^*\,\an{2}$, related to $\hat{c}$ as in Eqs.~\eqref{eq:b} and \eqref{eq:c}: the state is a single-mode excitation of $\hat{b}$. We have proved:

\begin{keybox}[Titulaer--Glauber theorem (two-mode version)]
Two modes have full first-order coherence, $|\gamma_{12}| = 1$, if and only if the state is a single-mode excitation, i.e., all photons occupy one and the same mode, whatever their quantum state~\cite{TitulaerGlauber}.
\end{keybox}

\subsection{Fringes of a single-mode excitation}
\label{sec:fringes_single}

Without loss of generality we parametrize the coefficients of $\hat{b}$ as
\begin{gather}
\beta_1 = \cos\theta, \qquad \beta_2 = e^{i\xi}\sin\theta, \qquad 0\le\theta\le\pi/2,
\nonumber\\
\text{so that}\qquad \bm{w} = \cos\theta\,\bm{u}_1 + e^{-i\xi}\sin\theta\,\bm{u}_2 .
\label{eq:betapar}
\end{gather}
(An overall phase of $\hat{b}$ is irrelevant, because it does not change $\hat{b}^\dagger\hat{b}$.) Then
\begin{equation}
|\bm{w}|^2 = \cos^2\theta\,|\bm{u}_1|^2 + \sin^2\theta\,|\bm{u}_2|^2 + 2\cos\theta\sin\theta\,\Real\!\left(e^{-i\xi}\,\bm{u}_1^*\cdot\bm{u}_2\right),
\end{equation}
and with $\bm{u}_1^*\cdot\bm{u}_2 = |K_{12}|e^{i\phi_{12}}$ and $2\cos\theta\sin\theta = \sin2\theta$, Eq.~\eqref{eq:central} becomes
\begin{equation}
\langle\hat{I}(\bm{r})\rangle = \hbar\omega c\;\bar{n}_b\left\{\cos^2\theta\,K_{11}(\bm{r}) + \sin^2\theta\,K_{22}(\bm{r}) + \sin2\theta\,|K_{12}(\bm{r})|\cos\!\big[\phi_{12}(\bm{r}) - \xi\big]\right\},
\label{eq:Itheta}
\end{equation}
with visibility
\begin{equation}
\mathcal{V}(\bm{r}) = \frac{|K_{12}(\bm{r})|\sin2\theta}{K_{11}(\bm{r})\cos^2\theta + K_{22}(\bm{r})\sin^2\theta} .
\label{eq:vistheta}
\end{equation}
The visibility does not depend on $\bar{n}_b$, nor on the state of mode $\hat{b}$, and has the same form as the classical Eq.~\eqref{eq:visdef} with $|a_1|^2:|a_2|^2 = \cos^2\theta:\sin^2\theta$. The relative phase $\xi$ \emph{in the definition of the mode} shifts the fringes; the mixing angle $\theta$ sets the balance between the two beams.

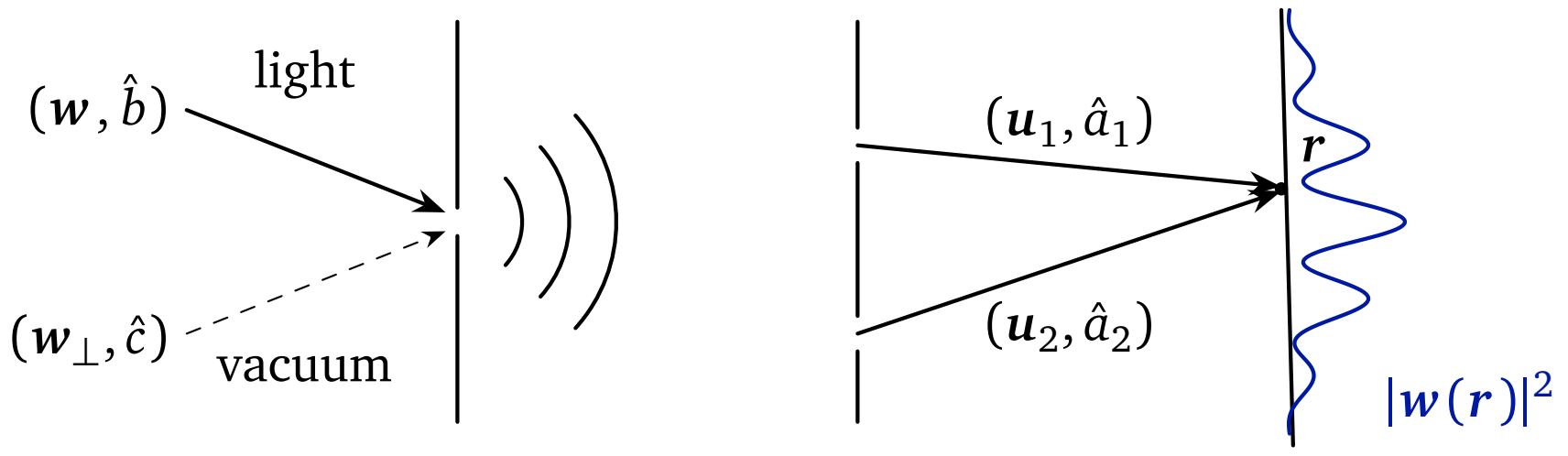
\begin{figure}[t]
\centering
\begin{tikzpicture}[>=Stealth,scale=0.85,line cap=round]
  \draw[thick] (0,-1.7) -- (0,-0.12);
  \draw[thick] (0,0.12) -- (0,1.7);
  \draw[->,thick] (-2.3,0.95) -- (-0.1,0.08);
  \draw[->,dashed] (-2.3,-0.95) -- (-0.1,-0.08);
  \node[anchor=east] at (-2.3,1.0) {$(\bm{w},\hat{b})$};          \node at (-1.3,1.25) {light};
  \node[anchor=east] at (-2.3,-1.0) {$(\bm{w}_\perp,\hat{c})$};    \node at (-1.3,-1.25) {vacuum};
  \foreach \r in {0.55,0.95,1.35}{\draw[thick] ([shift=(-42:\r)]0,0) arc (-42:42:\r);}
  \draw[thick] (3.4,1.7) -- (3.4,0.8);
  \draw[thick] (3.4,0.5) -- (3.4,-0.8);
  \draw[thick] (3.4,-1.1) -- (3.4,-1.7);
  \draw[thick] (7.0,1.8) -- (7.1,-1.9);
  \draw[->,thick] (3.4,0.65) -- (7.0,0.30) node[midway,above] {$(\bm{u}_1,\an{1})$};
  \draw[->,thick] (3.4,-0.95) -- (7.0,0.26) node[midway,below,yshift=-4pt] {$(\bm{u}_2,\an{2})$};
  \fill (7.0,0.28) circle (1.6pt);
  \node[right] at (7.0,0.62) {$\bm{r}$};
  \draw[thick,scipostblue,smooth,samples=140,domain=-1.8:1.8,variable=\y]
     plot ({7.05+1.0*exp(-\y*\y/1.2)*(0.15+0.85*cos(4.6*\y r)*cos(4.6*\y r))},{\y});
  \node[scipostblue] at (8.6,-1.5) {$|\bm{w}(\bm{r})|^2$};
\end{tikzpicture}
\caption{Young's experiment in the language of modes. Each arrow is labeled by a pair (mode function, annihilation operator). The pinhole and the two slits define the mode $(\bm{w},\hat{b})$, whose mode function $\bm{w} = \cos\theta\,\bm{u}_1 + e^{-i\xi}\sin\theta\,\bm{u}_2$, Eq.~\eqref{eq:w}, is a superposition of the waves $\bm{u}_1$ and $\bm{u}_2$ leaving the two slits. The orthogonal mode $(\bm{w}_\perp,\hat{c})$ is in the vacuum. For every state of mode $\hat{b}$ the mean intensity on the screen is proportional to $|\bm{w}(\bm{r})|^2$, Eq.~\eqref{eq:central}.}
\label{fig:young}
\end{figure}

\paragraph{The physical picture: Young's experiment.} In Young's experiment (Fig.~\ref{fig:young}), light illuminates a small pinhole and then two slits; $\bm{u}_1$ and $\bm{u}_2$ are the waves leaving the two slits (for two pinholes instead of slits, an explicit model of these waves is worked out in Appendix~\ref{app:pinholes}). The pinhole and the slits select a definite superposition of these two waves, which is the mode $\hat{b}$; the orthogonal combination $\hat{c}$ is not illuminated and stays in the vacuum. Whatever light is sent in---a laser beam, a thermal lamp filtered by the pinhole, or a source of single photons---the output is a single-mode excitation of $\hat{b}$, and Eq.~\eqref{eq:central} applies. The positions of the slits and of the pinhole fix $\phi_{12}(\bm{r})$ and $\xi$, i.e., the phase of the mode.

\subsection{Two different phases}
\label{sec:twophases}

Section~\ref{sec:fringes_single} has shown that the fringes of a single-mode excitation are written in the mode function $\bm{w}(\bm{r})$. Their position is fixed by the phase $\phi_{12}(\bm{r}) - \xi$, which belongs to the mode and is set by the optical setup, whereas the state of the mode enters only through the mean photon number $\bar{n}_b$. The students' question is about a different phase: the phase of the light itself, that is, of the quantum state that excites the mode. In this subsection we make this second notion precise, find out which states have such a phase, and show that the mean intensity is completely insensitive to it. How sharply the phase of a state is defined is discussed in Appendix~\ref{app:sharp}.

\subsubsection{Rotating the phase of a mode}
\label{sec:rotation}

Classically, a single mode with complex amplitude $a = |a|\,e^{i\varphi}$ has a well-defined phase $\varphi$. Rotating this phase by an angle $\phi$ means replacing $a$ with $a\,e^{-i\phi}$, i.e., multiplying the field of the mode by $e^{-i\phi}$. In quantum theory the same operation is performed by the unitary operator
\begin{equation}
\hat{U}_b(\phi) = e^{-i\phi\hat{N}_b} , \qquad \hat{N}_b = \hat{b}^\dagger\hat{b} ,
\label{eq:Ub}
\end{equation}
which transforms the annihilation and creation operators of the mode as
\begin{equation}
\hat{U}_b^\dagger(\phi)\,\hat{b}\,\hat{U}_b(\phi) = \hat{b}\,e^{-i\phi} ,
\qquad
\hat{U}_b^\dagger(\phi)\,\hat{b}^\dagger\,\hat{U}_b(\phi) = \hat{b}^\dagger\,e^{i\phi}
\label{eq:Ubrot}
\end{equation}
These relations follow from the commutator $[\hat{N}_b,\hat{b}] = [\hat{b}^\dagger,\hat{b}]\,\hat{b} = -\hat{b}$ and the operator expansion $e^{\hat{X}}\hat{Y}e^{-\hat{X}} = \hat{Y} + [\hat{X},\hat{Y}] + [\hat{X},[\hat{X},\hat{Y}]]/2! + \cdots$, a consequence of the Baker--Hausdorff theorem~\cite[Chap.~10]{MandelWolf}. With $\hat{X} = i\phi\hat{N}_b$ each commutator multiplies $\hat{b}$ by $-i\phi$, and the series sums to $\hat{b}\,e^{-i\phi}$. If the state $\hat{\rho}$ of the mode is replaced by the rotated state $\hat{U}_b\hat{\rho}\,\hat{U}_b^\dagger$, the mean field changes accordingly. By the cyclic property of the trace,
\[
\tr\!\left(\hat{U}_b\hat{\rho}\,\hat{U}_b^\dagger\,\hat{b}\right) = \tr\!\left(\hat{\rho}\;\hat{U}_b^\dagger\hat{b}\,\hat{U}_b\right) = \langle\hat{b}\rangle\,e^{-i\phi} ,
\]
exactly as for the classical amplitude $a$. Thus $\hat{U}_b(\phi)$ rotates the phase of the field of mode $\hat{b}$ by $-\phi$.
\subsubsection{When does a state have a phase?}
\label{sec:criterion}

A state has a phase if rotating it produces a physically different state: then there is something, a direction in phase space, that the rotation changes. If the rotated state is physically identical to the original one, the state has no phase at all. Whether two states are physically identical must be judged on their density operators, not on their state vectors, because two kets that differ only by a global phase factor describe the same physical state. Let $\hat{\rho}$ be the state of mode $\hat{b}$; for a pure state, $\hat{\rho} = \ket{\psi}\bra{\psi}$. The rotated state is
\begin{equation}
\hat{\rho}(\phi) = \hat{U}_b(\phi)\,\hat{\rho}\,\hat{U}_b^\dagger(\phi) = e^{-i\phi\hat{N}_b}\,\hat{\rho}\,e^{i\phi\hat{N}_b} ,
\label{eq:rhophi}
\end{equation}
and differentiating with respect to $\phi$, with $\hat{N}_b$ commuting with $e^{\pm i\phi\hat{N}_b}$, we obtain
\begin{equation}
\frac{\dif\hat{\rho}(\phi)}{\dif\phi} = -i\hat{N}_b\,\hat{\rho}(\phi) + i\,\hat{\rho}(\phi)\,\hat{N}_b = -i\,[\hat{N}_b,\hat{\rho}(\phi)] .
\label{eq:rhoode}
\end{equation}
This is the von Neumann equation of motion, $i\hbar\,\dif\hat{\rho}/\dif t = [\hat{H},\hat{\rho}]$, with $\hat{N}_b$ in place of $\hat{H}/\hbar$ and the angle $\phi$ in place of the time $t$: the photon-number operator is the \emph{generator} of phase rotations, just as the Hamiltonian is the generator of translations in time.

To see which states change, we take matrix elements of Eq.~\eqref{eq:rhophi} between number states of mode $\hat{b}$, writing $\ket{n}$ for $\ket{n}_b$. Since $e^{i\phi\hat{N}_b}\ket{m} = e^{im\phi}\ket{m}$ and $\bra{n}e^{-i\phi\hat{N}_b} = e^{-in\phi}\bra{n}$,
\begin{equation}
\bra{n}\hat{\rho}(\phi)\ket{m} = e^{-i(n-m)\phi}\,\bra{n}\hat{\rho}\ket{m} .
\label{eq:rhonm}
\end{equation}
The diagonal elements, $n = m$, are the photon-number probabilities, and they never change. The off-diagonal elements, $n\ne m$, are the \emph{coherences} between different photon numbers, and they rotate at a rate proportional to $n - m$. This gives a simple criterion.

\begin{keybox}[When does a state have a phase?]
A state $\hat{\rho}$ of mode $\hat{b}$ is left unchanged by every phase rotation if and only if it is diagonal in the number basis:
\begin{equation}
\hat{\rho}(\phi) = \hat{\rho}\ \text{for all } \phi
\quad\Longleftrightarrow\quad
[\hat{N}_b,\hat{\rho}] = 0
\quad\Longleftrightarrow\quad
\bra{n}\hat{\rho}\ket{m} = 0\ \text{for all } n\ne m .
\label{eq:nophase}
\end{equation}
In words: \emph{a state has a phase if and only if it contains coherences between different photon numbers}.
\end{keybox}

The first equivalence in Eq.~\eqref{eq:nophase} follows from Eq.~\eqref{eq:rhoode}. If $\hat{\rho}(\phi)$ never changes, its derivative at $\phi = 0$, which is $-i[\hat{N}_b,\hat{\rho}]$, vanishes. Conversely, if $\hat{\rho}$ commutes with $\hat{N}_b$, it also commutes with $e^{-i\phi\hat{N}_b}$, and Eq.~\eqref{eq:rhophi} gives $\hat{\rho}(\phi) = \hat{\rho}\,e^{-i\phi\hat{N}_b}e^{i\phi\hat{N}_b} = \hat{\rho}$. The second equivalence follows from $\bra{n}[\hat{N}_b,\hat{\rho}]\ket{m} = (n-m)\bra{n}\hat{\rho}\ket{m}$, which vanishes for all $n$ and $m$ if and only if every off-diagonal element vanishes.

\begin{figure}[t]
\centering
\begin{tikzpicture}[arr/.style={-{Stealth[length=3.2pt]},thick,scipostdeepblue},line cap=round]
\node[font=\small\bfseries,anchor=west] at (-0.6,1.35) {(a)};
\node[font=\small] at (4.460,1.35) {state $\ket{\psi}$};
\node[font=\small] at (9.210,1.35) {rotated state $\hat{U}_b(\phi)\ket{\psi}$};
\node[font=\small,align=center] at (0.7,0.150) {coherent\\ $\ket{\alpha}$, $\alpha=1.2$};
\draw[gray] (2.350,0.000) -- (6.570,0.000);
\draw[gray] (2.600,-0.060) -- (2.600,0.060);
\node[font=\scriptsize,gray!70!black] at (2.600,-0.420) {0};
\draw[gray] (3.220,-0.060) -- (3.220,0.060);
\node[font=\scriptsize,gray!70!black] at (3.220,-0.420) {1};
\draw[gray] (3.840,-0.060) -- (3.840,0.060);
\node[font=\scriptsize,gray!70!black] at (3.840,-0.420) {2};
\draw[gray] (4.460,-0.060) -- (4.460,0.060);
\node[font=\scriptsize,gray!70!black] at (4.460,-0.420) {3};
\draw[gray] (5.080,-0.060) -- (5.080,0.060);
\node[font=\scriptsize,gray!70!black] at (5.080,-0.420) {4};
\draw[gray] (5.700,-0.060) -- (5.700,0.060);
\node[font=\scriptsize,gray!70!black] at (5.700,-0.420) {5};
\draw[gray] (6.320,-0.060) -- (6.320,0.060);
\node[font=\scriptsize,gray!70!black] at (6.320,-0.420) {6};
\draw[arr] (2.600,0.000) -- (2.600,0.550);
\draw[arr] (3.220,0.000) -- (3.220,0.660);
\draw[arr] (3.840,0.000) -- (3.840,0.560);
\draw[arr] (4.460,0.000) -- (4.460,0.388);
\draw[arr] (5.080,0.000) -- (5.080,0.233);
\draw[arr] (5.700,0.000) -- (5.700,0.125);
\draw[arr] (6.320,0.000) -- (6.320,0.061);
\draw[gray] (7.100,0.000) -- (11.320,0.000);
\draw[gray] (7.350,-0.060) -- (7.350,0.060);
\node[font=\scriptsize,gray!70!black] at (7.350,-0.420) {0};
\draw[gray] (7.970,-0.060) -- (7.970,0.060);
\node[font=\scriptsize,gray!70!black] at (7.970,-0.420) {1};
\draw[gray] (8.590,-0.060) -- (8.590,0.060);
\node[font=\scriptsize,gray!70!black] at (8.590,-0.420) {2};
\draw[gray] (9.210,-0.060) -- (9.210,0.060);
\node[font=\scriptsize,gray!70!black] at (9.210,-0.420) {3};
\draw[gray] (9.830,-0.060) -- (9.830,0.060);
\node[font=\scriptsize,gray!70!black] at (9.830,-0.420) {4};
\draw[gray] (10.450,-0.060) -- (10.450,0.060);
\node[font=\scriptsize,gray!70!black] at (10.450,-0.420) {5};
\draw[gray] (11.070,-0.060) -- (11.070,0.060);
\node[font=\scriptsize,gray!70!black] at (11.070,-0.420) {6};
\draw[arr] (7.350,0.000) -- (7.350,0.550);
\draw[arr] (7.970,0.000) -- (8.300,0.572);
\draw[arr] (8.590,0.000) -- (9.075,0.280);
\draw[arr] (9.210,0.000) -- (9.598,0.000);
\draw[arr] (9.830,0.000) -- (10.032,-0.116);
\draw[arr] (10.450,0.000) -- (10.512,-0.108);
\draw[arr] (11.070,0.000) -- (11.070,-0.061);
\node[font=\scriptsize,red!70!black,anchor=west] at (11.420,0.450) {$\phi=\pi/6$};
\node[font=\small,align=center] at (0.7,-1.400) {number\\ $\ket{2}$};
\draw[gray] (2.350,-1.550) -- (6.570,-1.550);
\draw[gray] (2.600,-1.610) -- (2.600,-1.490);
\node[font=\scriptsize,gray!70!black] at (2.600,-1.970) {0};
\draw[gray] (3.220,-1.610) -- (3.220,-1.490);
\node[font=\scriptsize,gray!70!black] at (3.220,-1.970) {1};
\draw[gray] (3.840,-1.610) -- (3.840,-1.490);
\node[font=\scriptsize,gray!70!black] at (3.840,-1.970) {2};
\draw[gray] (4.460,-1.610) -- (4.460,-1.490);
\node[font=\scriptsize,gray!70!black] at (4.460,-1.970) {3};
\draw[gray] (5.080,-1.610) -- (5.080,-1.490);
\node[font=\scriptsize,gray!70!black] at (5.080,-1.970) {4};
\draw[gray] (5.700,-1.610) -- (5.700,-1.490);
\node[font=\scriptsize,gray!70!black] at (5.700,-1.970) {5};
\draw[gray] (6.320,-1.610) -- (6.320,-1.490);
\node[font=\scriptsize,gray!70!black] at (6.320,-1.970) {6};
\draw[arr] (3.840,-1.550) -- (3.840,-1.000);
\draw[gray] (7.100,-1.550) -- (11.320,-1.550);
\draw[gray] (7.350,-1.610) -- (7.350,-1.490);
\node[font=\scriptsize,gray!70!black] at (7.350,-1.970) {0};
\draw[gray] (7.970,-1.610) -- (7.970,-1.490);
\node[font=\scriptsize,gray!70!black] at (7.970,-1.970) {1};
\draw[gray] (8.590,-1.610) -- (8.590,-1.490);
\node[font=\scriptsize,gray!70!black] at (8.590,-1.970) {2};
\draw[gray] (9.210,-1.610) -- (9.210,-1.490);
\node[font=\scriptsize,gray!70!black] at (9.210,-1.970) {3};
\draw[gray] (9.830,-1.610) -- (9.830,-1.490);
\node[font=\scriptsize,gray!70!black] at (9.830,-1.970) {4};
\draw[gray] (10.450,-1.610) -- (10.450,-1.490);
\node[font=\scriptsize,gray!70!black] at (10.450,-1.970) {5};
\draw[gray] (11.070,-1.610) -- (11.070,-1.490);
\node[font=\scriptsize,gray!70!black] at (11.070,-1.970) {6};
\draw[arr] (8.590,-1.550) -- (8.590,-1.000);
\node[font=\scriptsize,red!70!black,anchor=west] at (11.420,-1.100) {any $\phi$};
\node[font=\small,align=center] at (0.7,-2.950) {squeezed\\ vacuum, $s=0.8$};
\draw[gray] (2.350,-3.100) -- (6.570,-3.100);
\draw[gray] (2.600,-3.160) -- (2.600,-3.040);
\node[font=\scriptsize,gray!70!black] at (2.600,-3.520) {0};
\draw[gray] (3.220,-3.160) -- (3.220,-3.040);
\node[font=\scriptsize,gray!70!black] at (3.220,-3.520) {1};
\draw[gray] (3.840,-3.160) -- (3.840,-3.040);
\node[font=\scriptsize,gray!70!black] at (3.840,-3.520) {2};
\draw[gray] (4.460,-3.160) -- (4.460,-3.040);
\node[font=\scriptsize,gray!70!black] at (4.460,-3.520) {3};
\draw[gray] (5.080,-3.160) -- (5.080,-3.040);
\node[font=\scriptsize,gray!70!black] at (5.080,-3.520) {4};
\draw[gray] (5.700,-3.160) -- (5.700,-3.040);
\node[font=\scriptsize,gray!70!black] at (5.700,-3.520) {5};
\draw[gray] (6.320,-3.160) -- (6.320,-3.040);
\node[font=\scriptsize,gray!70!black] at (6.320,-3.520) {6};
\draw[arr] (2.600,-3.100) -- (2.600,-2.550);
\draw[arr] (3.840,-3.100) -- (3.840,-3.358);
\draw[arr] (5.080,-3.100) -- (5.080,-2.951);
\draw[arr] (6.320,-3.100) -- (6.320,-3.190);
\draw[gray] (7.100,-3.100) -- (11.320,-3.100);
\draw[gray] (7.350,-3.160) -- (7.350,-3.040);
\node[font=\scriptsize,gray!70!black] at (7.350,-3.520) {0};
\draw[gray] (7.970,-3.160) -- (7.970,-3.040);
\node[font=\scriptsize,gray!70!black] at (7.970,-3.520) {1};
\draw[gray] (8.590,-3.160) -- (8.590,-3.040);
\node[font=\scriptsize,gray!70!black] at (8.590,-3.520) {2};
\draw[gray] (9.210,-3.160) -- (9.210,-3.040);
\node[font=\scriptsize,gray!70!black] at (9.210,-3.520) {3};
\draw[gray] (9.830,-3.160) -- (9.830,-3.040);
\node[font=\scriptsize,gray!70!black] at (9.830,-3.520) {4};
\draw[gray] (10.450,-3.160) -- (10.450,-3.040);
\node[font=\scriptsize,gray!70!black] at (10.450,-3.520) {5};
\draw[gray] (11.070,-3.160) -- (11.070,-3.040);
\node[font=\scriptsize,gray!70!black] at (11.070,-3.520) {6};
\draw[arr] (7.350,-3.100) -- (7.350,-2.550);
\draw[arr] (8.590,-3.100) -- (8.590,-2.842);
\draw[arr] (9.830,-3.100) -- (9.830,-2.951);
\draw[arr] (11.070,-3.100) -- (11.070,-3.010);
\node[font=\scriptsize,red!70!black,anchor=west] at (11.420,-2.650) {$\phi=\pi/2$};
\node[font=\scriptsize,gray!70!black,anchor=west] at (11.420,-3.520) {$n$};
\node[font=\small\bfseries,anchor=west] at (-0.6,-3.970) {(b)};
\node[font=\small] at (2.575,-3.970) {coherent, $\alpha=1.2$};
\fill[gray!60] (1.300,-5.350) circle (0.6pt);
\draw[arr] (1.300,-5.350) -- (1.300,-5.058);
\fill[gray!60] (2.150,-5.350) circle (0.6pt);
\draw[arr] (2.150,-5.350) -- (2.150,-5.000);
\fill[gray!60] (3.000,-5.350) circle (0.6pt);
\draw[arr] (3.000,-5.350) -- (3.000,-5.053);
\fill[gray!60] (3.850,-5.350) circle (0.6pt);
\draw[arr] (3.850,-5.350) -- (3.850,-5.144);
\fill[gray!60] (1.300,-6.200) circle (0.6pt);
\draw[arr] (1.300,-6.200) -- (1.300,-5.850);
\fill[gray!60] (2.150,-6.200) circle (0.6pt);
\draw[arr] (2.150,-6.200) -- (2.150,-5.780);
\fill[gray!60] (3.000,-6.200) circle (0.6pt);
\draw[arr] (3.000,-6.200) -- (3.000,-5.844);
\fill[gray!60] (3.850,-6.200) circle (0.6pt);
\draw[arr] (3.850,-6.200) -- (3.850,-5.953);
\fill[gray!60] (1.300,-7.050) circle (0.6pt);
\draw[arr] (1.300,-7.050) -- (1.300,-6.753);
\fill[gray!60] (2.150,-7.050) circle (0.6pt);
\draw[arr] (2.150,-7.050) -- (2.150,-6.694);
\fill[gray!60] (3.000,-7.050) circle (0.6pt);
\draw[arr] (3.000,-7.050) -- (3.000,-6.748);
\fill[gray!60] (3.850,-7.050) circle (0.6pt);
\draw[arr] (3.850,-7.050) -- (3.850,-6.840);
\fill[gray!60] (1.300,-7.900) circle (0.6pt);
\draw[arr] (1.300,-7.900) -- (1.300,-7.694);
\fill[gray!60] (2.150,-7.900) circle (0.6pt);
\draw[arr] (2.150,-7.900) -- (2.150,-7.653);
\fill[gray!60] (3.000,-7.900) circle (0.6pt);
\draw[arr] (3.000,-7.900) -- (3.000,-7.690);
\fill[gray!60] (3.850,-7.900) circle (0.6pt);
\draw[arr] (3.850,-7.900) -- (3.850,-7.755);
\node[font=\scriptsize,gray!70!black] at (1.300,-4.730) {0};
\node[font=\scriptsize,gray!70!black] at (2.150,-4.730) {1};
\node[font=\scriptsize,gray!70!black] at (3.000,-4.730) {2};
\node[font=\scriptsize,gray!70!black] at (3.850,-4.730) {3};
\node[font=\scriptsize,gray!70!black] at (0.850,-5.350) {0};
\node[font=\scriptsize,gray!70!black] at (0.850,-6.200) {1};
\node[font=\scriptsize,gray!70!black] at (0.850,-7.050) {2};
\node[font=\scriptsize,gray!70!black] at (0.850,-7.900) {3};
\node[font=\scriptsize] at (2.575,-4.400) {$m$};
\node[font=\scriptsize] at (0.450,-6.625) {$n$};
\node[font=\small] at (6.675,-3.970) {rotated by $\phi=\pi/6$};
\fill[gray!60] (5.400,-5.350) circle (0.6pt);
\draw[arr] (5.400,-5.350) -- (5.400,-5.058);
\fill[gray!60] (6.250,-5.350) circle (0.6pt);
\draw[arr] (6.250,-5.350) -- (6.075,-5.047);
\fill[gray!60] (7.100,-5.350) circle (0.6pt);
\draw[arr] (7.100,-5.350) -- (6.843,-5.202);
\fill[gray!60] (7.950,-5.350) circle (0.6pt);
\draw[arr] (7.950,-5.350) -- (7.744,-5.350);
\fill[gray!60] (5.400,-6.200) circle (0.6pt);
\draw[arr] (5.400,-6.200) -- (5.575,-5.897);
\fill[gray!60] (6.250,-6.200) circle (0.6pt);
\draw[arr] (6.250,-6.200) -- (6.250,-5.780);
\fill[gray!60] (7.100,-6.200) circle (0.6pt);
\draw[arr] (7.100,-6.200) -- (6.922,-5.891);
\fill[gray!60] (7.950,-6.200) circle (0.6pt);
\draw[arr] (7.950,-6.200) -- (7.736,-6.077);
\fill[gray!60] (5.400,-7.050) circle (0.6pt);
\draw[arr] (5.400,-7.050) -- (5.657,-6.902);
\fill[gray!60] (6.250,-7.050) circle (0.6pt);
\draw[arr] (6.250,-7.050) -- (6.428,-6.741);
\fill[gray!60] (7.100,-7.050) circle (0.6pt);
\draw[arr] (7.100,-7.050) -- (7.100,-6.748);
\fill[gray!60] (7.950,-7.050) circle (0.6pt);
\draw[arr] (7.950,-7.050) -- (7.845,-6.869);
\fill[gray!60] (5.400,-7.900) circle (0.6pt);
\draw[arr] (5.400,-7.900) -- (5.606,-7.900);
\fill[gray!60] (6.250,-7.900) circle (0.6pt);
\draw[arr] (6.250,-7.900) -- (6.464,-7.777);
\fill[gray!60] (7.100,-7.900) circle (0.6pt);
\draw[arr] (7.100,-7.900) -- (7.205,-7.719);
\fill[gray!60] (7.950,-7.900) circle (0.6pt);
\draw[arr] (7.950,-7.900) -- (7.950,-7.755);
\node[font=\scriptsize,gray!70!black] at (5.400,-4.730) {0};
\node[font=\scriptsize,gray!70!black] at (6.250,-4.730) {1};
\node[font=\scriptsize,gray!70!black] at (7.100,-4.730) {2};
\node[font=\scriptsize,gray!70!black] at (7.950,-4.730) {3};
\node[font=\scriptsize,gray!70!black] at (4.950,-5.350) {0};
\node[font=\scriptsize,gray!70!black] at (4.950,-6.200) {1};
\node[font=\scriptsize,gray!70!black] at (4.950,-7.050) {2};
\node[font=\scriptsize,gray!70!black] at (4.950,-7.900) {3};
\node[font=\scriptsize] at (6.675,-4.400) {$m$};
\node[font=\scriptsize] at (4.550,-6.625) {$n$};
\node[font=\small] at (10.775,-3.970) {thermal, $\bar{n}=2$};
\fill[gray!60] (9.500,-5.350) circle (0.6pt);
\draw[arr] (9.500,-5.350) -- (9.500,-4.930);
\fill[gray!60] (10.350,-5.350) circle (0.6pt);
\fill[gray!60] (11.200,-5.350) circle (0.6pt);
\fill[gray!60] (12.050,-5.350) circle (0.6pt);
\fill[gray!60] (9.500,-6.200) circle (0.6pt);
\fill[gray!60] (10.350,-6.200) circle (0.6pt);
\draw[arr] (10.350,-6.200) -- (10.350,-5.920);
\fill[gray!60] (11.200,-6.200) circle (0.6pt);
\fill[gray!60] (12.050,-6.200) circle (0.6pt);
\fill[gray!60] (9.500,-7.050) circle (0.6pt);
\fill[gray!60] (10.350,-7.050) circle (0.6pt);
\fill[gray!60] (11.200,-7.050) circle (0.6pt);
\draw[arr] (11.200,-7.050) -- (11.200,-6.863);
\fill[gray!60] (12.050,-7.050) circle (0.6pt);
\fill[gray!60] (9.500,-7.900) circle (0.6pt);
\fill[gray!60] (10.350,-7.900) circle (0.6pt);
\fill[gray!60] (11.200,-7.900) circle (0.6pt);
\fill[gray!60] (12.050,-7.900) circle (0.6pt);
\draw[arr] (12.050,-7.900) -- (12.050,-7.776);
\node[font=\scriptsize,gray!70!black] at (9.500,-4.730) {0};
\node[font=\scriptsize,gray!70!black] at (10.350,-4.730) {1};
\node[font=\scriptsize,gray!70!black] at (11.200,-4.730) {2};
\node[font=\scriptsize,gray!70!black] at (12.050,-4.730) {3};
\node[font=\scriptsize,gray!70!black] at (9.050,-5.350) {0};
\node[font=\scriptsize,gray!70!black] at (9.050,-6.200) {1};
\node[font=\scriptsize,gray!70!black] at (9.050,-7.050) {2};
\node[font=\scriptsize,gray!70!black] at (9.050,-7.900) {3};
\node[font=\scriptsize] at (10.775,-4.400) {$m$};
\node[font=\scriptsize] at (8.650,-6.625) {$n$};
\end{tikzpicture}
\caption{Number-state diagrams of some states of a mode, constructed as explained in the text. (a)~Pure states before (left) and after (right) a rotation $\hat{U}_b(\phi)$, which turns the arrow at $n$ clockwise by $(n-n_0)\,\phi$. The coherent state changes; the number state has a single arrow and cannot change; in the squeezed vacuum a rotation by $\pi/2$ reverses the arrows at $n = 2$ and $6$, whereas a rotation by $\pi$ would leave every arrow unchanged. (b)~Density matrices, with the longest arrow of each grid drawn with the same length. Rotating the coherent state (left) turns its off-diagonal arrows (center); the thermal state (right) has only diagonal arrows and no phase.}
\label{fig:arrows}
\end{figure}
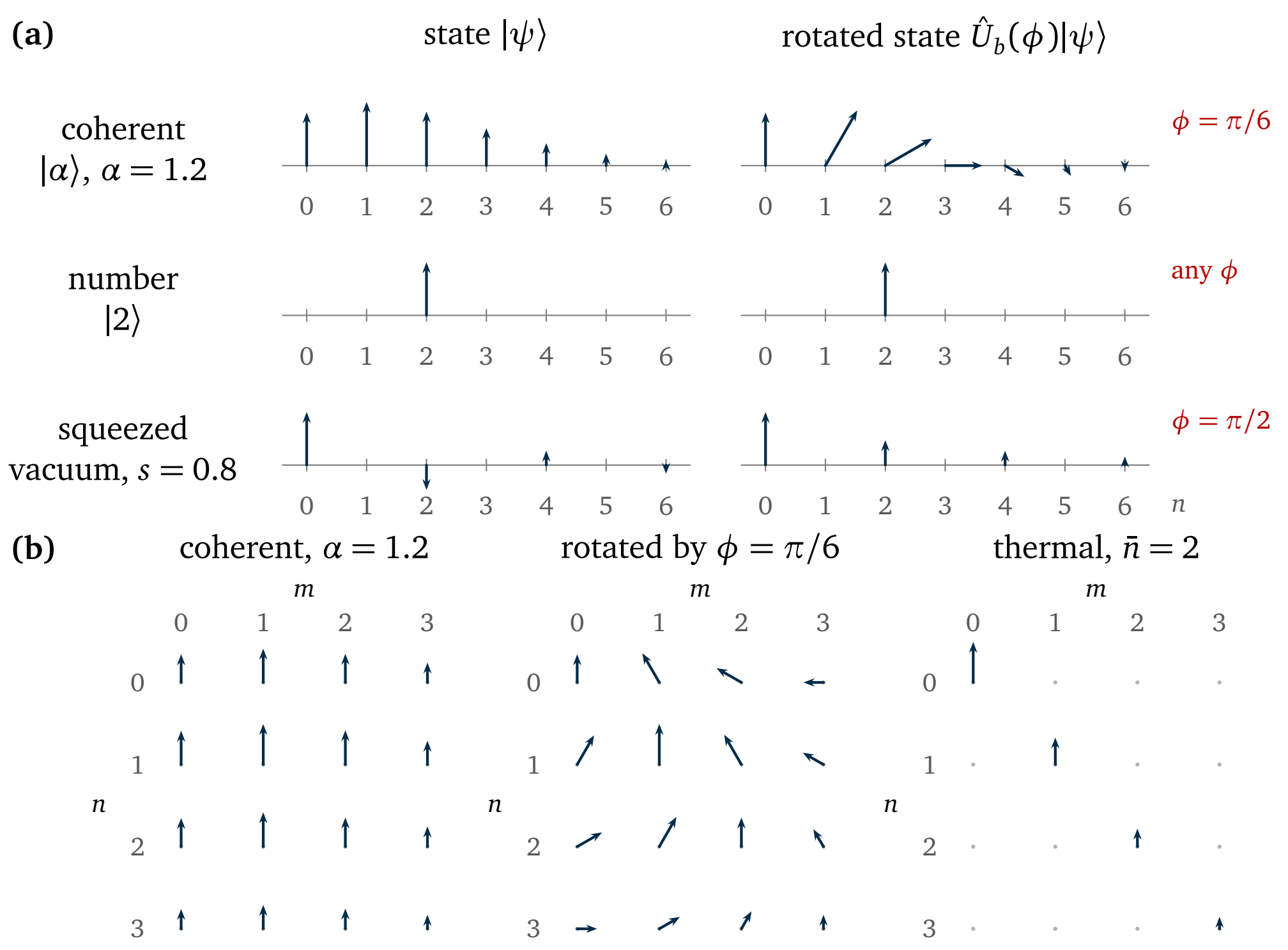

\paragraph{A picture of the criterion.} The criterion can be visualized with a simple diagram (Fig.~\ref{fig:arrows}). For a pure state, we expand $\ket{\psi} = \sum_n\psi_n\ket{n}$ in number states and divide all the amplitudes by the first one that is different from zero, $\psi_{n_0}$:
\begin{equation}
c_n = \frac{\psi_n}{\psi_{n_0}} , \qquad c_{n_0} = 1 .
\label{eq:cn}
\end{equation}
This removes the global phase of the state vector, which has no physical meaning. We then mark the points $n = 0,1,2,\ldots$ on a horizontal axis and draw at each point an arrow representing the complex number $c_n$: its length is $|c_n|$, and it is turned counterclockwise from the vertical by the angle $\arg c_n$, so that a positive amplitude points up. A rotation multiplies $\psi_n$ by $e^{-in\phi}$, and therefore $c_n$ by $e^{-i(n-n_0)\phi}$: the reference arrow at $n_0$ never moves, and the arrow at $n$ turns clockwise by $(n-n_0)\,\phi$.\footnote{The free evolution of the mode during a time $t$ is the rotation $\hat{U}_b(\omega t)$, up to an irrelevant global phase. The diagrams therefore also show how a state evolves in time: the arrows turn like the hands of a clock, the one at $n$ with angular velocity $(n-n_0)\,\omega$.} The diagram, and with it the state, is unchanged by every rotation if and only if it contains a single arrow, i.e., if the state is a number state.

For a mixed state we use a square grid instead, with an arrow representing $\bra{n}\hat{\rho}\ket{m}$ at the point $(n,m)$. By Eq.~\eqref{eq:rhonm}, the diagonal arrows point up and never move, while the arrow at $(n,m)$ turns clockwise by $(n-m)\,\phi$. A state has no phase if and only if its grid contains only diagonal arrows: this is Eq.~\eqref{eq:nophase} in pictures.

\subsubsection{Examples}
\label{sec:phaseexamples}

\paragraph{Number states and thermal states.} A number state only acquires a global phase factor, $\hat{U}_b(\phi)\ket{n}_b = e^{-in\phi}\ket{n}_b$. The ket changes, but the density operator $\ket{n}_b\bra{n}$ does not, and no measurement can distinguish the state from its rotated copies: its diagram has a single arrow (Fig.~\ref{fig:arrows}a, middle). A number state has no phase at all: its phase is completely undetermined, uniformly distributed between $0$ and $2\pi$~\cite{Loudon}. This holds in particular for a single photon, $\ket{1}_b$. The same is true of every mixture of number states, such as a thermal state, $\hat{\rho} = \sum_np_n\ket{n}_b\bra{n}$, which is diagonal in the number basis (Fig.~\ref{fig:arrows}b, right).

\paragraph{Coherent states.} From $\hat{U}_b(\phi)\ket{n}_b = e^{-in\phi}\ket{n}_b$ and the number-state expansion of a coherent state, $\ket{\alpha}_b = e^{-|\alpha|^2/2}\sum_n\alpha^n/\sqrt{n!}\,\ket{n}_b$, it follows that
\[
\hat{U}_b(\phi)\ket{\alpha}_b = e^{-|\alpha|^2/2}\sum_{n=0}^{\infty}\frac{(\alpha\,e^{-i\phi})^n}{\sqrt{n!}}\,\ket{n}_b = \ket{\alpha\,e^{-i\phi}}_b .
\]
A coherent state is turned into another coherent state, whose complex amplitude is rotated exactly as a classical amplitude. Its amplitudes $c_n = \alpha^n/\sqrt{n!}$ are all different from zero for $\alpha\ne0$, and so are its coherences, $\bra{n}\hat{\rho}\ket{m} = e^{-|\alpha|^2}\alpha^n\alpha^{*m}/\sqrt{n!\,m!}$. For real $\alpha$ all the arrows point up; after a rotation the arrow at $n$ points at the angle $n\phi$ clockwise from the vertical, and the diagram is clearly different (Fig.~\ref{fig:arrows}a, top). The coherent state has a phase, $\arg\alpha$. How sharply this phase is defined depends on the spread of the photon number; Appendix~\ref{app:sharp} shows that for a coherent state it is defined to within an angle of order $1/|\alpha|$, so that it is sharp only for $|\alpha|\gg1$.

\paragraph{The squeezed vacuum.} The squeezed vacuum $\hat{S}(s)\ket{0}$, with $\hat{S}(s) = \exp[s(\hat{b}^2-\hat{b}^{\dagger2})/2]$, is an intermediate case. The operators $\hat{b}^2$ and $\hat{b}^{\dagger2}$ change the photon number by two, so the squeezed vacuum contains only even photon numbers, and its coherences connect photon numbers that differ by $n - m = \pm2,\pm4,\ldots$. By Eq.~\eqref{eq:rhonm} they are all unchanged if and only if $e^{-2i\phi} = 1$, i.e., for $\phi = 0$ or $\phi = \pi$. The squeezed vacuum is invariant under a rotation by $\pi$, and only under it (Fig.~\ref{fig:arrows}a, bottom): its phase is defined only modulo $\pi$.

\medskip
The phase of a coherent state, $\arg\alpha$, is the phase of a laser beam. This is the phase the students had in mind, and a single photon does not have it.

\subsubsection{The mean intensity is blind to the phase of the state}
\label{sec:blind}

We can now compare the two phases. For a single-mode excitation, Eq.~\eqref{eq:central} contains the state of mode $\hat{b}$ only through the mean photon number $\bar{n}_b = \tr(\hat{N}_b\,\hat{\rho})$. This number does not change when the state is rotated, because $\hat{N}_b$ commutes with $\hat{U}_b$:
\[
\tr\!\left(\hat{N}_b\,\hat{U}_b\hat{\rho}\,\hat{U}_b^\dagger\right) = \tr\!\left(\hat{U}_b^\dagger\hat{N}_b\hat{U}_b\,\hat{\rho}\right) = \tr\!\left(\hat{N}_b\,\hat{\rho}\right) .
\]
The mean intensity, and with it the whole fringe pattern, is therefore the same for every rotated copy of the state, whether the state has a sharp phase (a coherent state with $|\alpha|\gg1$), a phase defined only modulo $\pi$ (a squeezed vacuum), or no phase at all (a number state).

The same holds for any state of the two modes, not only for single-mode excitations. The intensity operator~\eqref{eq:IKop} commutes with the total number of photons, $\hat{N} = \ad{1}\an{1} + \ad{2}\an{2}$. Indeed, using $[\an{k},\ad{m}\an{m}] = \delta_{km}\an{k}$ and $[\ad{j},\ad{m}\an{m}] = -\delta_{jm}\ad{j}$,
\begin{equation}
[\ad{j}\an{k},\hat{N}] = \sum_m\left(\ad{j}[\an{k},\ad{m}\an{m}] + [\ad{j},\ad{m}\an{m}]\an{k}\right) = \ad{j}\an{k} - \ad{j}\an{k} = 0 :
\label{eq:INcomm}
\end{equation}
each term of $\hat{I}$ destroys one photon and creates another, so that the number of photons is unchanged. Hence $\tr(e^{-i\phi\hat{N}}\hat{\rho}\,e^{i\phi\hat{N}}\hat{I}) = \tr(\hat{\rho}\,e^{i\phi\hat{N}}\hat{I}\,e^{-i\phi\hat{N}}) = \tr(\hat{\rho}\,\hat{I})$: the mean intensity is unchanged by a common phase rotation of the two modes, for any state whatsoever.

\subsubsection{A relative phase, set by the mode}
\label{sec:relphase}

What does shift the fringes is a \emph{relative} phase between the two interfering modes. A rotation of mode 2 alone, $\hat{U}_2(\phi) = e^{-i\phi\,\ad{2}\an{2}}$, produced for instance by a delay line in the second arm, changes the state into $\hat{U}_2\hat{\rho}\,\hat{U}_2^\dagger$. By the cyclic property of the trace and Eq.~\eqref{eq:Ubrot} applied to mode 2,
\begin{equation}
C_{12} \;\to\; \tr\!\left(\hat{U}_2\hat{\rho}\,\hat{U}_2^\dagger\,\ad{1}\an{2}\right) = \tr\!\left(\hat{\rho}\;\ad{1}\,\hat{U}_2^\dagger\an{2}\hat{U}_2\right) = C_{12}\,e^{-i\phi},
\label{eq:relrot}
\end{equation}
which shifts the fringes in Eq.~\eqref{eq:IQ}.

For a single-mode excitation such a rotation does not change the state of the excitation at all: it changes the mode that carries it. This needs a word of care, because the density operator itself does change, $\hat{U}_2\hat{\rho}\,\hat{U}_2^\dagger \neq \hat{\rho}$ in general, as the change of $C_{12}$ in Eq.~\eqref{eq:relrot} shows. What does not change are the numbers that specify the state: the old density operator, written in the number states of mode $\hat{b}$, and the new one, written in the number states of a rotated mode $\hat{b}'$ (constructed below), have the same matrix elements. Equation~\eqref{eq:Ubrot} applied to mode 2 gives $\hat{U}_2\,\ad{2}\,\hat{U}_2^\dagger = \ad{2}\,e^{-i\phi}$, and therefore
\[
\hat{b}'^{\dagger} \equiv \hat{U}_2\,\hat{b}^\dagger\,\hat{U}_2^\dagger = \cos\theta\,\ad{1} + e^{-i(\xi+\phi)}\sin\theta\,\ad{2} ,
\]
which is the creation operator of the mode~\eqref{eq:betapar} with $\xi$ replaced by $\xi + \phi$. Since $\hat{U}_2$ leaves the vacuum unchanged, $\hat{U}_2\ket{0} = \ket{0}$, it maps the number states of mode $\hat{b}$ onto those of mode $\hat{b}'$:
\[
\hat{U}_2\ket{n}_b = \frac{1}{\sqrt{n!}}\left(\hat{U}_2\,\hat{b}^\dagger\,\hat{U}_2^\dagger\right)^{n}\hat{U}_2\ket{0} = \frac{1}{\sqrt{n!}}\left(\hat{b}'^{\dagger}\right)^{n}\ket{0} = \ket{n}_{b'} .
\]
Hence, if $\hat{\rho} = \sum_{n,m}\rho_{nm}\ket{n}_b\bra{m}$, then
\[
\hat{U}_2\hat{\rho}\,\hat{U}_2^\dagger = \sum_{n,m}\rho_{nm}\ket{n}_{b'}\bra{m} ,
\]
with the same coefficients $\rho_{nm}$. The photon-number probabilities and the coherences of the excitation are carried over unchanged; what changes is the mode they refer to, $\hat{b}'$ instead of $\hat{b}$. For $0 < \theta < \pi/2$ and $\phi \neq 0 \pmod{2\pi}$, $\hat{b}'$ is a different mode from $\hat{b}$, so that the kets $\ket{n}_{b'}$ with $n \geq 1$ differ from the kets $\ket{n}_b$: this is why the operator $\hat{\rho}$ changes, even though the state of the excitation does not. The relative phase $\xi$ is a parameter of the mode: it is the phase of $\beta_2/\beta_1$ in Eq.~\eqref{eq:betapar}, which, together with the phase $\phi_{12}(\bm{r})$ of $\bm{u}_1^*\cdot\bm{u}_2$, makes up the position-dependent phase of the mode function $\bm{w}(\bm{r})$.

To see this concretely, write a single photon in mode $\hat{b}$ in the number basis of modes 1 and 2:
\begin{equation}
\ket{1}_b = \hat{b}^\dagger\ket{0} = \left(\cos\theta\,\ad{1} + e^{-i\xi}\sin\theta\,\ad{2}\right)\ket{0,0} = \cos\theta\,\ket{1,0} + e^{-i\xi}\sin\theta\,\ket{0,1} .
\label{eq:onephoton}
\end{equation}
The criterion~\eqref{eq:nophase}, applied to modes 1 and 2, tells us exactly which phase this state has. As always, we must examine the density operator, here $\hat{\rho} = \ket{1}_b\bra{1}$, not the ket. Both components of Eq.~\eqref{eq:onephoton} contain exactly one photon, so that $\hat{N}\ket{1}_b = \ket{1}_b$, where $\hat{N} = \ad{1}\an{1} + \ad{2}\an{2}$ is the total photon number. Therefore $\hat{N}\hat{\rho} = \hat{\rho}\hat{N} = \hat{\rho}$, i.e., $[\hat{N},\hat{\rho}] = 0$: a common rotation $e^{-i\phi\hat{N}}$ of the two modes leaves $\hat{\rho}$ unchanged, because it only multiplies the ket by the global phase factor $e^{-i\phi}$, which has no physical meaning. The density operator does not commute, instead, with the photon number of mode 2 alone, $\hat{N}_2 = \ad{2}\an{2}$. The states $\ket{1,0}$ and $\ket{0,1}$ contain $0$ and $1$ photons in mode 2, and the matrix element of $\hat{\rho}$ between them is the coherence
\[
\bra{1,0}\hat{\rho}\ket{0,1} = \langle1,0|1\rangle_b\;{}_b\langle1|0,1\rangle = \cos\theta\,\sin\theta\,e^{i\xi} .
\]
By Eq.~\eqref{eq:rhonm}, with $\hat{N}_2$ in place of $\hat{N}_b$, a rotation of mode 2 alone multiplies this coherence by $e^{-i(0-1)\phi} = e^{i\phi}$, i.e., it changes $\xi$ into $\xi + \phi$, as found above. For $0 < \theta < \pi/2$ the coherence is different from zero, and its phase is precisely the relative phase $\xi$ between the two components of the state, which describe the photon in the two arms of the interferometer. This phase is the same for every photon prepared by the setup, because it is fixed by the setup---slits, beam splitters, path lengths and polarizers---and not by the photon (Problem~\ref{prob:bs}). The fringes measure $\phi_{12}(\bm{r}) - \xi$, Eq.~\eqref{eq:Itheta}, and nothing else.

\begin{keybox}[Answer to the students' question]
The state of the photon has no definite phase, but its \emph{mode} does. The fringes reveal the phase of the mode, $\phi_{12}(\bm{r}) - \xi$, and the mean intensity is completely blind to the phase of the state.
\end{keybox}

\subsection{Randomizing the phase}
\label{sec:random}

The distinction becomes sharper if phases are randomized on purpose.

\paragraph{A common random phase.} Suppose first that the phase of a coherent single-mode excitation fluctuates randomly, as for a laser whose phase drifts, and that the light is then split into the two interfering beams. The state is then the average of $\ket{\alpha e^{i\phi}}_b\bra{\alpha e^{i\phi}}$ over a uniformly distributed phase $\phi$. Using the number-state expansion of the coherent state (Appendix~\ref{app:oscillator}),
\begin{equation}
\ket{\alpha e^{i\phi}}\bra{\alpha e^{i\phi}} = e^{-|\alpha|^2}\sum_{n,m=0}^{\infty}\frac{\alpha^n\alpha^{*m}}{\sqrt{n!\,m!}}\,e^{i(n-m)\phi}\,\ket{n}\bra{m},
\end{equation}
and $\int_0^{2\pi}e^{i(n-m)\phi}\,\dif\phi/2\pi = \delta_{nm}$, we obtain
\begin{equation}
\hat{\rho} = \int_0^{2\pi}\frac{\dif\phi}{2\pi}\,\ket{\alpha e^{i\phi}}_b\bra{\alpha e^{i\phi}} = \sum_{n=0}^{\infty}e^{-|\alpha|^2}\frac{|\alpha|^{2n}}{n!}\,\ket{n}_b\bra{n} ,
\label{eq:phaseavg}
\end{equation}
a Poisson mixture of number states of mode $\hat{b}$, whose phase is completely undetermined. Mode $\hat{c}$ is in the vacuum in every term, so the state is still a single-mode excitation. Let us compute its coherence matrix explicitly. Each term of the mixture is the two-mode state $\ket{n}_b\ket{0}_c$, with $n$ photons in mode $\hat{b}$ and none in mode $\hat{c}$. Writing $\an{j} = \beta_j^*\hat{b} + \zeta_j^*\hat{c}$, Eq.~\eqref{eq:ainv}, we have
\[
\ad{j}\an{k} = \beta_j\beta_k^*\,\hat{b}^\dagger\hat{b} + \beta_j\zeta_k^*\,\hat{b}^\dagger\hat{c} + \zeta_j\beta_k^*\,\hat{c}^\dagger\hat{b} + \zeta_j\zeta_k^*\,\hat{c}^\dagger\hat{c} .
\]
The second and fourth terms give zero when they act on $\ket{n}_b\ket{0}_c$, because $\hat{c}\ket{0}_c = 0$. The third has zero expectation value, because ${}_c\bra{0}\hat{c}^\dagger = (\hat{c}\ket{0}_c)^\dagger = 0$. Only the first term survives, and since $\hat{b}^\dagger\hat{b}\ket{n}_b = n\ket{n}_b$,
\begin{align*}
C_{jk} = \tr\!\left(\hat{\rho}\,\ad{j}\an{k}\right)
&= \sum_{n=0}^{\infty}e^{-|\alpha|^2}\frac{|\alpha|^{2n}}{n!}\;{}_b\bra{n}{}_c\bra{0}\,\ad{j}\an{k}\,\ket{n}_b\ket{0}_c
\\
&= \beta_j\beta_k^*\sum_{n=0}^{\infty}n\,e^{-|\alpha|^2}\frac{|\alpha|^{2n}}{n!} = |\alpha|^2\,\beta_j\beta_k^* .
\end{align*}
In the last step we used the mean of the Poisson distribution: with $x = |\alpha|^2$,
\[
\sum_{n=0}^{\infty}n\,e^{-x}\frac{x^n}{n!} = x\,e^{-x}\sum_{n=1}^{\infty}\frac{x^{n-1}}{(n-1)!} = x\,e^{-x}e^{x} = x .
\]
This is Eq.~\eqref{eq:Csingle} with $\bar{n}_b = |\alpha|^2$, so the fringe pattern is exactly that of the coherent state: randomizing the phase of the state has no effect at all. In terms of the two beams, $\ket{\alpha e^{i\phi}}_b = \ket{\alpha e^{i\phi}\beta_1^*}_1\ket{\alpha e^{i\phi}\beta_2^*}_2$ by Eq.~\eqref{eq:cohb} below: both beams acquire the same random phase, and their relative phase is unaffected. Indeed, for each coherent state $\langle\ad{j}\an{k}\rangle = (\alpha e^{i\phi}\beta_j^*)^*(\alpha e^{i\phi}\beta_k^*) = |\alpha|^2\beta_j\beta_k^*$: the random phase cancels even before the average is taken.

\paragraph{Independent random phases.} Suppose instead that the two beams come from independent sources, so that their phases $\phi_1$ and $\phi_2$ fluctuate independently. Averaging the product coherent state $\ket{\alpha_1e^{i\phi_1}}_1\ket{\alpha_2e^{i\phi_2}}_2$ over both phases, the same calculation applied to each mode gives
\begin{equation}
\hat{\rho} = \sum_{n_1,n_2=0}^{\infty}e^{-|\alpha_1|^2-|\alpha_2|^2}\,\frac{|\alpha_1|^{2n_1}|\alpha_2|^{2n_2}}{n_1!\,n_2!}\,\ket{n_1,n_2}\bra{n_1,n_2} ,
\label{eq:phaseavg2}
\end{equation}
a mixture of products of number states. Its coherence matrix follows as in the previous paragraph. Each term of the mixture is a number state $\ket{n_1,n_2}$ with probability
\[
p(n_1,n_2) = e^{-|\alpha_1|^2}\frac{|\alpha_1|^{2n_1}}{n_1!}\;e^{-|\alpha_2|^2}\frac{|\alpha_2|^{2n_2}}{n_2!} ,
\]
the product of two Poisson distributions. On the diagonal, $\ad{1}\an{1}\ket{n_1,n_2} = n_1\ket{n_1,n_2}$, so that
\[
C_{11} = \sum_{n_1,n_2}p(n_1,n_2)\,n_1 = |\alpha_1|^2 , \qquad\text{and similarly}\qquad C_{22} = |\alpha_2|^2 ,
\]
by the mean of the Poisson distribution, since the sum over $n_2$ of the second factor of $p(n_1,n_2)$ gives 1. Off the diagonal, $\ad{1}\an{2}$ moves one photon from mode 2 to mode 1,
\[
\ad{1}\an{2}\ket{n_1,n_2} = \sqrt{(n_1+1)\,n_2}\;\ket{n_1+1,n_2-1} ,
\]
which is orthogonal to $\ket{n_1,n_2}$. Hence $\bra{n_1,n_2}\ad{1}\an{2}\ket{n_1,n_2} = 0$ for every term, and
\[
C_{12} = C_{21}^* = \sum_{n_1,n_2}p(n_1,n_2)\bra{n_1,n_2}\ad{1}\an{2}\ket{n_1,n_2} = 0 .
\]
The coherence matrix is diagonal, $\gamma_{12} = 0$, and Eq.~\eqref{eq:IC} gives
\[
\langle\hat{I}(\bm{r})\rangle = \hbar\omega c\left(|\alpha_1|^2K_{11}(\bm{r}) + |\alpha_2|^2K_{22}(\bm{r})\right):
\]
the sum of the intensities of the two beams, with no interference term. There are no fringes, exactly as for the classical random amplitudes of Eq.~\eqref{eq:Irandom}. What destroys the fringes is the randomness of the \emph{relative} phase of the two modes, not that of the phase of the state. This is what happens with two independent lasers: their fringes are visible only over times shorter than the time on which their relative phase drifts~\cite{Paul1986,PfleegorMandel}.

\begin{keybox}[In a nutshell]
For a single-mode excitation, $\langle\hat{I}\rangle = \hbar\omega c\,\bar{n}_b|\bm{w}(\bm{r})|^2$: the pattern is the squared modulus of the excited mode function, whatever the state. Conversely, $|\gamma_{12}| = 1$ implies a single-mode excitation. The phase of the state is generated by $\hat{N}$, which commutes with $\hat{I}$: the mean intensity is blind to it. Fringes are produced by the phase of the mode. Randomizing the phase of the state leaves the fringes intact; randomizing the relative phase of the two modes destroys them.
\end{keybox}
\section{Examples}
\label{sec:states}

In this section we apply the general results to the most common states of light. For each state we compute the coherence matrix, and in particular $C_{12}$, and read off the fringes from Eqs.~\eqref{eq:IQ} and \eqref{eq:central}. The results are collected in Table~\ref{tab:states}.

\subsection{Coherent states}
\label{sec:coherentstates}

For $\ket{\psi} = \ket{\alpha_1}_1\ket{\alpha_2}_2$, with $\alpha_j = |\alpha_j|e^{i\varphi_j}$, Sec.~\ref{sec:coherent_classical} gives $C_{jk} = \alpha_j^*\alpha_k$. Hence $\bar{n}_j = |\alpha_j|^2$, $C_{12} = |\alpha_1||\alpha_2|e^{i(\varphi_2-\varphi_1)}$, $|\gamma_{12}| = 1$ and $\psi = \varphi_2-\varphi_1$, and Eq.~\eqref{eq:IQ} reproduces the classical result \eqref{eq:I2} with $a_j\to\alpha_j$. As a check of Eq.~\eqref{eq:Cfock}, insert the number-state amplitudes of the product coherent state, $\Psi_{n_1n_2} = e^{-(|\alpha_1|^2+|\alpha_2|^2)/2}\,\alpha_1^{n_1}\alpha_2^{n_2}/\sqrt{n_1!\,n_2!}$. Then
\begin{equation}
\Psi^*_{m_1+1,m_2-1}\Psi_{m_1,m_2}\sqrt{(m_1+1)\,m_2} = e^{-|\alpha_1|^2-|\alpha_2|^2}\,\frac{|\alpha_1|^{2m_1}}{m_1!}\,\frac{|\alpha_2|^{2(m_2-1)}}{(m_2-1)!}\;\alpha_1^*\alpha_2 ,
\end{equation}
and the double sum gives $e^{-|\alpha_1|^2-|\alpha_2|^2}\,e^{|\alpha_1|^2}\,e^{|\alpha_2|^2}\,\alpha_1^*\alpha_2 = \alpha_1^*\alpha_2$, as expected.

\subsection{Products of number states}
\label{sec:numberstates}

For a product of number states $\ket{n_1}_1\ket{n_2}_2 = \ket{n_1,n_2}$, the operator $\ad{1}\an{2}$ produces a multiple of $\ket{n_1+1,n_2-1}$, which is orthogonal to $\ket{n_1,n_2}$. Hence $C_{12} = 0$, while $\bar{n}_j = n_j$, and
\begin{equation}
\langle\hat{I}(\bm{r})\rangle = \hbar\omega c\left[n_1K_{11}(\bm{r}) + n_2K_{22}(\bm{r})\right].
\label{eq:IFock}
\end{equation}
There are no fringes, $\mathcal{V} = 0$, whatever the mode functions: the intensities of the two beams add, exactly as for the classical random amplitudes of Eq.~\eqref{eq:Irandom}. The same is true of any mixture of products of number states, such as Eq.~\eqref{eq:phaseavg2}.

This result is sometimes attributed to the undetermined phase of number states. \emph{That cannot be the reason}: the single photon \eqref{eq:onephoton} is a number state of mode $\hat{b}$, with an equally undetermined phase, and it does produce fringes. The reason is that no photon is shared between the two modes, $C_{12} = 0$; equivalently, the relative phase of the two modes is completely random. The absence of fringes concerns only the mean signal: correlations between the signals at two points can display interference even when the mean signal does not, as in the experiments of Pfleegor and Mandel with independent laser beams~\cite{PfleegorMandel,Paul1986} (Problem~\ref{prob:g2}).

\subsection{One, \texorpdfstring{$n$}{n}, and two photons in mode \texorpdfstring{$\hat{b}$}{b}}
\label{sec:nphotons}

The $n$-photon state of mode $\hat{b}$ is $\ket{n}_b = (\hat{b}^\dagger)^n\ket{0}/\sqrt{n!}$, with $\hat{b}^\dagger = \cos\theta\,\ad{1} + e^{-i\xi}\sin\theta\,\ad{2}$ from Eq.~\eqref{eq:betapar}. Since $\ad{1}$ and $\ad{2}$ commute, the binomial theorem applies:
\begin{equation}
(\hat{b}^\dagger)^n = \sum_{m=0}^{n}\binom{n}{m}\cos^m\theta\left(e^{-i\xi}\sin\theta\right)^{n-m}(\ad{1})^m(\ad{2})^{n-m} .
\end{equation}
Using (Appendix~\ref{app:oscillator})
\[
(\ad{1})^m(\ad{2})^{n-m}\ket{0,0} = \sqrt{m!\,(n-m)!}\,\ket{m,n-m}
\quad\text{and}\quad
\binom{n}{m}\frac{\sqrt{m!\,(n-m)!}}{\sqrt{n!}} = \binom{n}{m}^{1/2},
\]
we obtain
\begin{equation}
\ket{n}_b = \frac{(\hat{b}^\dagger)^n}{\sqrt{n!}}\ket{0} = \sum_{m=0}^{n}\binom{n}{m}^{1/2}\cos^m\theta\,\left(e^{-i\xi}\sin\theta\right)^{n-m}\,\ket{m,n-m} .
\label{eq:nb}
\end{equation}
The photon numbers in modes 1 and 2 are therefore binomially distributed, with probabilities $\binom{n}{m}\cos^{2m}\theta\,\sin^{2(n-m)}\theta$ and means $n\cos^2\theta$ and $n\sin^2\theta$: each photon goes to mode 1 with probability $\cos^2\theta$ and to mode 2 with probability $\sin^2\theta$, independently of the others. By Eq.~\eqref{eq:Csingle}, $C_{12} = n\,\beta_1\beta_2^* = n\sin\theta\cos\theta\,e^{-i\xi}$ and $|\gamma_{12}| = 1$.

For $n = 1$, Eq.~\eqref{eq:nb} is the single-photon state \eqref{eq:onephoton}, shared between the two modes. According to Eq.~\eqref{eq:central}, it produces the full fringe pattern $|\bm{w}(\bm{r})|^2$, which is built up photon by photon in a sequence of single-photon experiments~\cite{Walls,Grangier1986,Galvez2005,Aspden2016}: this is Dirac's dictum in field-theoretic form. For $n = 2$, $\theta = \pi/4$ and $\xi = 0$,
\begin{equation}
\ket{2}_b = \tfrac{1}{2}\ket{2,0} + \tfrac{1}{\sqrt{2}}\ket{1,1} + \tfrac{1}{2}\ket{0,2} ,
\label{eq:twophoton}
\end{equation}
with photon-number probabilities $(1/4, 1/2, 1/4)$ and $\bar{n}_1 = \bar{n}_2 = 1$. With Eq.~\eqref{eq:Cfock}, the only pairs of components that differ by one photon moved from mode 2 to mode 1 are $(\ket{1,1}\to\ket{2,0})$ and $(\ket{0,2}\to\ket{1,1})$, and
\begin{equation}
C_{12} = \underbrace{\tfrac{1}{2}\cdot\tfrac{1}{\sqrt{2}}\cdot\sqrt{2\cdot1}}_{m_1=1,\,m_2=1} + \underbrace{\tfrac{1}{\sqrt{2}}\cdot\tfrac{1}{2}\cdot\sqrt{1\cdot2}}_{m_1=0,\,m_2=2} = \tfrac{1}{2} + \tfrac{1}{2} = 1 ,
\end{equation}
in agreement with $C_{12} = n\sin\theta\cos\theta = 1$. Thus $\langle\hat{I}\rangle = \hbar\omega c\,[K_{11} + K_{22} + 2|K_{12}|\cos\phi_{12}]$ and $\mathcal{V} = 2|K_{12}|/(K_{11}+K_{22})$. The coherence comes from the chain of components $\ket{2,0}\leftrightarrow\ket{1,1}\leftrightarrow\ket{0,2}$, each differing from the next by one photon moved between the modes.

\subsection{Squeezed and thermal states}
\label{sec:squeezedthermal}

\paragraph{Squeezed vacuum.} The squeezed vacuum of mode $\hat{b}$ is $\hat{S}(s)\ket{0}$, with real squeezing parameter $s$ and
\[
\hat{S}(s) = \exp\!\left[\frac{s}{2}\left(\hat{b}^2 - \hat{b}^{\dagger2}\right)\right].
\] Expanding the exponential, $\hat{S}(s)$ creates and destroys photons of mode $\hat{b}$ only, in pairs, so the state contains only even numbers of photons, all in mode $\hat{b}$: it is a single-mode excitation. Its mean photon number follows from the transformation $\hat{S}^\dagger\hat{b}\,\hat{S} = \hat{b}\cosh s - \hat{b}^\dagger\sinh s$ (Appendix~\ref{app:oscillator}):
\begin{align}
\bar{n}_b &= \bra{0}\hat{S}^\dagger\hat{b}^\dagger\hat{S}\,\hat{S}^\dagger\hat{b}\,\hat{S}\ket{0}
\nonumber\\
&= \bra{0}\left(\hat{b}^\dagger\cosh s - \hat{b}\sinh s\right)\left(\hat{b}\cosh s - \hat{b}^\dagger\sinh s\right)\ket{0} = \sinh^2s ,
\label{eq:nsq}
\end{align}
because only the term $\hat{b}\hat{b}^\dagger\sinh^2s$ has a nonzero vacuum expectation value, equal to $\sinh^2 s$. Equation~\eqref{eq:central} gives the intensity $\hbar\omega c\,\sinh^2s\,|\bm{w}(\bm{r})|^2$, with the visibility \eqref{eq:vistheta}.

\paragraph{Thermal state.} A thermal state of mode $\hat{b}$,
\begin{equation}
\hat{\rho} = \sum_{n=0}^{\infty}\frac{\bar{n}^n}{(1+\bar{n})^{n+1}}\,\ket{n}_b\bra{n},
\label{eq:thermal}
\end{equation}
is a mixture of number states of mode $\hat{b}$, hence a single-mode excitation with $\bar{n}_b = \bar{n}$. Physically, chaotic light, once it has been filtered into a single spatial mode---for instance by a pinhole small enough to make it spatially coherent---produces fringes of the same visibility as laser light.

For both states, as for number states, the mean field vanishes, $\langle\hat{\bm{E}}^+\rangle = 0$, because $\langle\hat{b}\rangle = 0$. Fringes in the mean intensity do not require a nonzero mean field, nor a well-defined phase of the state; they require only that all photons occupy the same mode.

\subsection{Coherent states revisited; a two-photon state without fringes}
\label{sec:cohb}

\paragraph{Product coherent states are single-mode excitations.} The coherent state of mode $\hat{b}$ is $\ket{\alpha}_b = \hat{D}_b(\alpha)\ket{0}$, with the displacement operator $\hat{D}_b(\alpha) = \exp(\alpha\hat{b}^\dagger - \alpha^*\hat{b})$ (Appendix~\ref{app:oscillator}). Since $\hat{b}^\dagger = \sum_j\beta_j^*\ad{j}$ and $\hat{b} = \sum_j\beta_j\an{j}$,
\begin{equation}
\alpha\hat{b}^\dagger - \alpha^*\hat{b} = \sum_{j=1}^{2}\left(\alpha\beta_j^*\,\ad{j} - \alpha^*\beta_j\,\an{j}\right).
\end{equation}
The two terms of the sum act on different modes and commute, so the exponential of their sum is the product of their exponentials:
\begin{align}
\hat{D}_b(\alpha) &= \prod_{j=1}^{2}\exp\!\left(\alpha\beta_j^*\,\ad{j} - \alpha^*\beta_j\,\an{j}\right) = \hat{D}_1(\alpha\beta_1^*)\,\hat{D}_2(\alpha\beta_2^*),
\nonumber\\
\ket{\alpha}_b &= \hat{D}_b(\alpha)\ket{0,0} = \ket{\alpha\beta_1^*}_1\,\ket{\alpha\beta_2^*}_2 .
\label{eq:cohb}
\end{align}
Conversely, every product $\ket{\alpha_1}_1\ket{\alpha_2}_2$ is the coherent state $\ket{\alpha}_b$ with $\alpha = (|\alpha_1|^2+|\alpha_2|^2)^{1/2}$ and $\beta_j^* = \alpha_j/\alpha$~\cite{Walls}. This is why two independent, phase-stable lasers produce fringes, whereas two independent number states do not.

\paragraph{Same photon numbers, no fringes.} A state with the same mean photon numbers as Eq.~\eqref{eq:twophoton}, $\bar{n}_1 = \bar{n}_2 = 1$, but no fringes is~\cite{Walls}
\begin{equation}
\ket{\psi} = \tfrac{1}{\sqrt{2}}\left(\ket{2,0} + \ket{0,2}\right).
\label{eq:NOON}
\end{equation}
Its two components differ by \emph{two} photons moved between the modes. The operator $\ad{1}\an{2}$, which moves a single photon, has no matrix element between them: $\ad{1}\an{2}\ket{2,0} = 0$ and $\ad{1}\an{2}\ket{0,2} = \sqrt{2}\ket{1,1}$, which is orthogonal to both components. Hence $C_{12} = 0$, and the intensity reduces to $\hbar\omega c\,(K_{11}+K_{22})$.

\begin{table}[t]
\centering
\caption{First-order coherence of the states discussed in Secs.~\ref{sec:phase} and \ref{sec:states}, with $\hat{b} = \cos\theta\,\an{1} + e^{i\xi}\sin\theta\,\an{2}$. The first five states are single-mode excitations: they have $|\gamma_{12}| = 1$, and their fringe pattern is given by Eq.~\eqref{eq:central}, whatever the phase properties of the state. The last three have no first-order coherence between the two modes and give no fringes; Eq.~\eqref{eq:phaseavg2} is the product coherent state with independently randomized phases.}
\label{tab:states}
\begin{tabular}{lcccc}
\toprule
State & $\bar{n}_1$ & $\bar{n}_2$ & $C_{12} = \langle\ad{1}\an{2}\rangle$ & $|\gamma_{12}|$ \\
\midrule
$\ket{\alpha_1}_1\ket{\alpha_2}_2$ & $|\alpha_1|^2$ & $|\alpha_2|^2$ & $\alpha_1^*\alpha_2$ & 1 \\
$\ket{\alpha}_b$, or Eq.~\eqref{eq:phaseavg} & $|\alpha|^2\cos^2\theta$ & $|\alpha|^2\sin^2\theta$ & $|\alpha|^2\sin\theta\cos\theta\,e^{-i\xi}$ & 1 \\
$\ket{n}_b$ & $n\cos^2\theta$ & $n\sin^2\theta$ & $n\sin\theta\cos\theta\,e^{-i\xi}$ & 1 \\
Squeezed vacuum of $\hat{b}$ & $\sinh^2\! s\cos^2\theta$ & $\sinh^2\! s\sin^2\theta$ & $\sinh^2\! s\sin\theta\cos\theta\,e^{-i\xi}$ & 1 \\
Thermal state of $\hat{b}$ & $\bar{n}\cos^2\theta$ & $\bar{n}\sin^2\theta$ & $\bar{n}\sin\theta\cos\theta\,e^{-i\xi}$ & 1 \\
\midrule
$\ket{n_1}_1\ket{n_2}_2$ & $n_1$ & $n_2$ & 0 & 0 \\
Eq.~\eqref{eq:phaseavg2} & $|\alpha_1|^2$ & $|\alpha_2|^2$ & 0 & 0 \\
$(\ket{2,0} + \ket{0,2})/\sqrt{2}$ & 1 & 1 & 0 & 0 \\
\bottomrule
\end{tabular}
\end{table}

\begin{keybox}[In a nutshell]
Coherent, number, squeezed and thermal states of a single mode $\hat{b}$ all give the same fringes, of maximal visibility. Products of number states, independently phase-randomized coherent states, and the state $(\ket{2,0}+\ket{0,2})/\sqrt{2}$ give no fringes, because no single photon is shared coherently between the two modes ($C_{12} = 0$). Whether fringes appear is decided by $C_{12} = \langle\ad{1}\an{2}\rangle$. This depends on the state, but only through its coherences between components that differ by one photon moved from one mode to the other, such as $\ket{n_1+1,n_2-1}$ and $\ket{n_1,n_2}$. These coherences are not changed by a common phase rotation: the fringes are decided by the relative phase of the two modes, not by the phase of the state, as shown in Secs.~\ref{sec:blind} and~\ref{sec:relphase}.
\end{keybox}
\section{Problems}
\label{sec:problems}

The first two problems are warm-up exercises on Secs.~\ref{sec:modes}--\ref{sec:energy}; the others extend Secs.~\ref{sec:intensity}--\ref{sec:states}. Solutions are given in Appendix~\ref{app:solutions}.

\begin{enumerate}
\item\label{prob:shells} \emph{Plane waves of a given frequency.} List all wave vectors $\bm{k}_l$ of the box with $(\omega/\omega_0)^2 = 5$, and check that there are 24 of them. Show that there is no field with $(\omega/\omega_0)^2 = 7$. \emph{Hint:} the square of an integer leaves remainder 0, 1 or 4 when divided by 8.

\item\label{prob:halfenergy} \emph{Electric and magnetic energy.} Using Eqs.~\eqref{eq:WE} and \eqref{eq:WB}, show that the cycle averages of the electric and magnetic energies of the field \eqref{eq:Afield} are equal, $\langle W_E\rangle_\ca = \langle W_B\rangle_\ca = W/2$. Then show that, for a single standing wave $\bm{u} = \sqrt{2}\,\hat{\bm{x}}\sin(kz)/L^{3/2}$, the electric energy oscillates between $0$ and $W$. Why does this not contradict energy conservation?

\item\label{prob:pancharatnam} \emph{Pancharatnam phase.} In the geometry of Fig.~\ref{fig:geometry}, let $\hat{\bm{p}}_j = \hat{\bm{x}}\times\hat{\bm{k}}_j$ and take beam 1 circularly polarized, $\ue_1 = (\hat{\bm{x}} + i\hat{\bm{p}}_1)/\sqrt{2}$, and beam 2 linearly polarized at $45^\circ$, $\ue_2 = (\hat{\bm{x}} + \hat{\bm{p}}_2)/\sqrt{2}$. Show that $\ue_1^*\cdot\ue_2 = (1 - i\cos 2\alpha)/2$. Conclude that, for $|a_1| = |a_2|$, the visibility is $\frac{1}{2}(1+\cos^2 2\alpha)^{1/2}$ and that the fringes are displaced, with respect to those of two $\hat{\bm{x}}$-polarized beams, by the fraction $\arctan(\cos 2\alpha)/2\pi$ of a period---one eighth of a period for small $\alpha$---although the amplitudes $a_1$ and $a_2$ are unchanged.

\item\label{prob:helicity} \emph{Circular polarizations.} In the same geometry, let both beams be circularly polarized, $\ue_j = (\hat{\bm{x}} + i\sigma_j\hat{\bm{p}}_j)/\sqrt{2}$ with helicities $\sigma_j = \pm 1$. Show that $|\mu_{12}| = |1 + \sigma_1\sigma_2\cos 2\alpha|/2$, i.e., $\cos^2\alpha$ for equal and $\sin^2\alpha$ for opposite helicities. Discuss the limits $\alpha \to 0$, where one recovers a Fresnel--Arago law, and $\alpha \to \pi/2$, where the two waves counterpropagate.

\item\label{prob:bs} \emph{The beam splitter defines the mode.} A lossless beam splitter maps two input modes $\hat{c}_{\rm in}$ and $\hat{c}'_{\rm in}$ onto the two output modes according to $\an{1} = \tau\hat{c}_{\rm in} + \kappa\hat{c}'_{\rm in}$ and $\an{2} = \kappa\hat{c}_{\rm in} + \tau\hat{c}'_{\rm in}$, with $|\tau|^2 + |\kappa|^2 = 1$ and $\tau\kappa^* + \kappa\tau^* = 0$. Show that if $\hat{c}'_{\rm in}$ is in the vacuum and $\hat{c}_{\rm in}$ is in an arbitrary state, the output is a single-mode excitation of $\hat{b} = \hat{c}_{\rm in} = \tau^*\an{1} + \kappa^*\an{2}$, with mode function $\bm{w} = \tau\bm{u}_1 + \kappa\bm{u}_2$. Conclude that the relative phase $\arg(\kappa/\tau)$, and hence the position of the fringes, is fixed by the beam splitter and not by the state of the light.

\item\label{prob:squeezed} \emph{One-mode and two-mode squeezing.} Show that the squeezed vacuum of mode $\hat{b}$ has $C_{jk} = \sinh^2 s\,\beta_j\beta_k^*$. Then consider the two-mode squeezed vacuum
\begin{equation*}
\exp[s(\an{1}\an{2} - \ad{1}\ad{2})]\,\ket{0,0} = \frac{1}{\cosh s}\sum_{n=0}^{\infty}(-\tanh s)^n\,\ket{n,n} ,
\end{equation*}
produced, for instance, by parametric down-conversion. Show that the mean photon numbers are $\bar{n}_1 = \bar{n}_2 = \sinh^2 s$, but $C_{12} = 0$. Thus, although the photon numbers in the two modes are perfectly correlated, the mean intensity shows no fringes.

\item\label{prob:g2} \emph{Intensity correlations.} For the state $\ket{1,1}$ and the two $\hat{\bm{x}}$-polarized plane waves of Fig.~\ref{fig:geometry}, show that the mean intensity has no fringes, but that the correlation function $\langle\hat{E}_x^-(\bm{r})\,\hat{E}_x^-(\bm{r}')\,\hat{E}_x^+(\bm{r}')\,\hat{E}_x^+(\bm{r})\rangle$, which is proportional to the probability of joint detections at $\bm{r}$ and $\bm{r}'$, is proportional to $1 + \cos[(\bm{k}_1 - \bm{k}_2)\cdot(\bm{r} - \bm{r}')]$: the joint detection probability shows fringes of full contrast as a function of the separation of the two detectors. Compare with the experiments of Pfleegor and Mandel~\cite{PfleegorMandel}.
\end{enumerate}

\section{Conclusions}
\label{sec:conclusions}

We have answered a question that students ask when they first meet photon-number states: if the phase of a photon is completely undetermined, how can single photons build up interference fringes? Two different phases are involved. The phase of the quantum state of a mode---undetermined for a number state, sharp for an intense coherent state---does not enter the mean photodetection signal at all, because the intensity operator conserves the number of photons. The fringes are produced by the phase of the mode: the position-dependent phase of the vector mode function that the photons occupy, which is fixed by the optical setup and includes the polarization of the light.

To make this explicit we have built, from Maxwell's equations, a formalism that treats classical and quantized fields alike. The field is expanded in transverse vector modes, and nonorthogonal modes are reduced to orthonormal ones by diagonalizing their Gram matrix. The intensity is a quadratic form whose kernel, the intensity matrix $K_{jk}(\bm{r}) = \bm{u}_j^*(\bm{r})\cdot\bm{u}_k(\bm{r})$, contains the modes, and whose coefficients---the classical products $a_j^*a_k$ or the elements $\langle\ad{j}\an{k}\rangle$ of the coherence matrix---contain the state; with the prefactor $2\varepsilon_0 c$, the classical and quantum intensities coincide for coherent states. For every single-mode excitation---a number, coherent, squeezed or thermal state of one mode---the pattern is the squared modulus of the excited mode function, Eq.~\eqref{eq:central}, and its visibility is the largest allowed by the geometry; conversely, full first-order coherence implies single-mode excitation. Randomizing the phase of the state leaves the fringes unchanged, whereas randomizing the relative phase of the two modes, as for independent sources or for products of number states, destroys them. Polarization, being part of the mode, can reduce the contrast, suppress the fringes, or interchange bright and dark fringes without any change in the state of the light.

The same separation between modes and states extends to higher-order correlations, which are sensitive to the state beyond its coherence matrix and are the natural next step: Problem~\ref{prob:g2} gives a first example, and the Hong--Ou--Mandel effect and the intensity interferometry of Hanbury Brown and Twiss are discussed in the textbooks cited above~\cite{Loudon,MandelWolf}.

\section*{Acknowledgements}
The author thanks the students whose question motivated these notes.

\paragraph{Funding information}
This work had no specific funding.
\begin{appendix}
\numberwithin{equation}{section}

\section{Mathematical tools}
\label{app:math}

\subsection{Orthonormality of the plane waves in the box}
\label{app:orthogonality}

For an integer $n$,
\begin{equation}
\int_0^L e^{2\pi inx/L}\,\dif x =
\begin{cases}
L, & n = 0,\\[2pt]
\dfrac{L}{2\pi in}\left(e^{2\pi in} - 1\right) = 0, & n\ne0,
\end{cases}
\;\;=\; L\,\delta_{n0},
\label{eq:int1d}
\end{equation}
because $e^{2\pi in} = 1$. In three dimensions the integrand of Eq.~\eqref{eq:orthonormal_pw} factorizes into three one-dimensional integrals:
\begin{align}
\int_{L^3}\frac{e^{i(\bm{k}_{l'}-\bm{k}_l)\cdot\bm{r}}}{L^3}\,\dif^3r
&= \frac{1}{L^3}\int_0^Le^{2\pi i(l_1'-l_1)x/L}\dif x\int_0^Le^{2\pi i(l_2'-l_2)y/L}\dif y\int_0^Le^{2\pi i(l_3'-l_3)z/L}\dif z
\nonumber\\
&= \delta_{l_1l_1'}\delta_{l_2l_2'}\delta_{l_3l_3'} = \delta_{ll'} .
\label{eq:int3d}
\end{align}
In the same way, $\int_{L^3}e^{i(\bm{k}_l+\bm{k}_{l'})\cdot\bm{r}}\,\dif^3r/L^3 = \delta_{l',-l}$, which was used in Sec.~\ref{sec:Delta}.

\subsection{Laplace's equation in the periodic box}
\label{app:laplace}

Let $\Phi(\bm{r})$ be periodic in the box and satisfy $\nabla^2\Phi = 0$. Expanding it in a Fourier series, $\Phi = \sum_l\Phi_l\,e^{i\bm{k}_l\cdot\bm{r}}/L^{3/2}$, and using Eq.~\eqref{eq:lap_pw}, we get $\sum_l|\bm{k}_l|^2\Phi_l\,e^{i\bm{k}_l\cdot\bm{r}} = 0$. By uniqueness of the Fourier expansion, $|\bm{k}_l|^2\Phi_l = 0$ for every $l$, so $\Phi_l = 0$ for every $l \ne (0,0,0)$. Hence $\Phi = \Phi_{(0,0,0)}/L^{3/2}$ is a constant.

\subsection{Vector identities and integrals over the box}
\label{app:vector}

We use index notation, with summation over repeated indices, $\partial_i = \partial/\partial x_i$, the Levi-Civita symbol $\epsilon_{ijk}$, so that $(\bm{a}\times\bm{b})_i = \epsilon_{ijk}a_jb_k$ and $(\nabla\times\bm{a})_i = \epsilon_{ijk}\partial_ja_k$, and the identity $\epsilon_{kij}\epsilon_{klm} = \delta_{il}\delta_{jm} - \delta_{im}\delta_{jl}$.

\paragraph{Curl of a curl.}
\begin{equation}
[\nabla\times(\nabla\times\bm{A})]_i = \epsilon_{ijk}\partial_j\,\epsilon_{klm}\partial_lA_m = (\delta_{il}\delta_{jm} - \delta_{im}\delta_{jl})\,\partial_j\partial_lA_m = \partial_i(\partial_mA_m) - \partial_j\partial_jA_i ,
\end{equation}
i.e., $\nabla\times(\nabla\times\bm{A}) = \nabla(\nabla\cdot\bm{A}) - \nabla^2\bm{A}$.

\paragraph{Divergence of a cross product.} By the product rule,
\begin{equation}
\nabla\cdot(\bm{F}\times\bm{G}) = \partial_i(\epsilon_{ijk}F_jG_k) = G_k\,\epsilon_{kij}\partial_iF_j - F_j\,\epsilon_{jik}\partial_iG_k = \bm{G}\cdot(\nabla\times\bm{F}) - \bm{F}\cdot(\nabla\times\bm{G}),
\end{equation}
where we used $\epsilon_{ijk} = \epsilon_{kij} = -\epsilon_{jik}$.
With $\bm{F} = \bm{b}$ and $\bm{G} = \nabla\times\bm{a}$ this gives
\[
\nabla\cdot\left[\bm{b}\times(\nabla\times\bm{a})\right] = (\nabla\times\bm{a})\cdot(\nabla\times\bm{b}) - \bm{b}\cdot\left[\nabla\times(\nabla\times\bm{a})\right],
\]
which is Eq.~\eqref{eq:vecid}.

\paragraph{Integral of a divergence.} If $\bm{F}$ is periodic in the box, then
\begin{align}
\int_{L^3}\nabla\cdot\bm{F}\,\dif^3r &= \int_{L^3}\left(\frac{\partial F_x}{\partial x} + \frac{\partial F_y}{\partial y} + \frac{\partial F_z}{\partial z}\right)\dif x\,\dif y\,\dif z
\nonumber\\
&= \int_0^L\!\!\int_0^L\left[F_x(L,y,z) - F_x(0,y,z)\right]\dif y\,\dif z + \cdots = 0 ,
\end{align}
where we integrated $\partial F_x/\partial x$ over $x$, and the dots stand for the analogous terms with $F_y$ and $F_z$. The result vanishes because each square bracket vanishes by periodicity. This is the divergence theorem applied to the cube: the fluxes through opposite faces cancel.

\subsection{Hermitian matrices}
\label{app:hermitian}

An $N\times N$ complex matrix $\Gm$ is Hermitian if $\Gm^\dagger = \Gm$, i.e., $G_{kj} = G_{jk}^*$. We recall the facts used in the text.
\begin{enumerate}
\item The eigenvalues of a Hermitian matrix are real, and its eigenvectors can be chosen orthonormal. Collecting the normalized eigenvectors as the columns of a matrix $U$, one has $U^\dagger U = UU^\dagger = \Id$ ($U$ is unitary) and $\Gm = UDU^\dagger$, with $D$ the diagonal matrix of the eigenvalues $g_l$. In components, $G_{jk} = \sum_lU_{jl}\,g_l\,U_{kl}^*$.
\item The quadratic form of $\Gm$ in the basis of eigenvectors is $\sum_{jk}z_j^*G_{jk}z_k = \sum_lg_l|w_l|^2$, with $w_l = \sum_kU_{kl}^*z_k$. Hence $\Gm$ is positive semidefinite (the form is $\ge0$ for all $z$) if and only if all $g_l\ge0$, and positive definite (the form is $>0$ for all $z\ne0$) if and only if all $g_l>0$.
\item If all $g_l\ne0$, the inverse is $\Gm^{-1} = UD^{-1}U^\dagger$, since $UDU^\dagger\,UD^{-1}U^\dagger = \Id$.
\item The determinant is the product of the eigenvalues, $\det\Gm = \prod_lg_l$. For a positive-semidefinite $2\times2$ matrix, therefore, $G_{11}G_{22} - |G_{12}|^2 \ge 0$.
\end{enumerate}

\subsection{The Cauchy--Schwarz inequality}
\label{app:cauchy}

For complex vectors $\bm{a}$, $\bm{b}$ of $\mathbb{C}^n$ (we need $n = 3$),
\begin{equation}
|\bm{a}^*\cdot\bm{b}|^2 \le |\bm{a}|^2\,|\bm{b}|^2 ,
\label{eq:cs}
\end{equation}
with equality if and only if $\bm{a}$ and $\bm{b}$ are parallel, i.e., one is a complex multiple of the other. \emph{Proof:} if $\bm{b} = 0$ there is nothing to prove. Otherwise let $\lambda = (\bm{b}^*\cdot\bm{a})/|\bm{b}|^2$; then
\begin{equation}
0 \le |\bm{a} - \lambda\bm{b}|^2 = |\bm{a}|^2 - \lambda^*\,(\bm{b}^*\cdot\bm{a}) - \lambda\,(\bm{a}^*\cdot\bm{b}) + |\lambda|^2|\bm{b}|^2 = |\bm{a}|^2 - \frac{|\bm{a}^*\cdot\bm{b}|^2}{|\bm{b}|^2},
\end{equation}
which is Eq.~\eqref{eq:cs}. Equality holds if and only if $\bm{a} = \lambda\bm{b}$. Equivalently, the $2\times2$ Gram matrix of $\bm{a}$ and $\bm{b}$ is positive semidefinite, and its determinant is nonnegative.

\section{Number states and coherent states}
\label{app:oscillator}

\paragraph{Number states.} For a single mode with $[\hat{a},\hat{a}^\dagger] = 1$, the number states $\ket{n}$, $n = 0,1,2,\ldots$, are the eigenstates of $\hat{N} = \hat{a}^\dagger\hat{a}$, $\hat{N}\ket{n} = n\ket{n}$. They are orthonormal, $\langle n|m\rangle = \delta_{nm}$, and
\begin{equation}
\hat{a}\ket{n} = \sqrt{n}\,\ket{n-1}, \qquad \hat{a}^\dagger\ket{n} = \sqrt{n+1}\,\ket{n+1}, \qquad \ket{n} = \frac{(\hat{a}^\dagger)^n}{\sqrt{n!}}\ket{0} .
\label{eq:fock}
\end{equation}
For several modes,
\[
\ket{n_1,n_2} = \ket{n_1}_1\ket{n_2}_2 = \frac{(\ad{1})^{n_1}(\ad{2})^{n_2}}{\sqrt{n_1!\,n_2!}}\ket{0,0} ,
\]
and operators of different modes act on their own factor only. For example,
\[
\ad{1}\an{2}\ket{m_1,m_2} = \sqrt{(m_1+1)\,m_2}\,\ket{m_1+1,m_2-1} .
\]

\paragraph{Phase rotation.} Since $\hat{N}\ket{n} = n\ket{n}$, $e^{-i\phi\hat{N}}\ket{n} = e^{-in\phi}\ket{n}$: a number state only acquires a global phase.

\paragraph{Coherent states.} For a complex number $\alpha$, the coherent state is
\begin{equation}
\ket{\alpha} = e^{-|\alpha|^2/2}\sum_{n=0}^{\infty}\frac{\alpha^n}{\sqrt{n!}}\,\ket{n} .
\label{eq:coh}
\end{equation}
Its main properties follow from this expansion.
\begin{enumerate}
\item It is normalized: $\langle\alpha|\alpha\rangle = e^{-|\alpha|^2}\sum_n|\alpha|^{2n}/n! = 1$.
\item It is an eigenstate of $\hat{a}$. Using $\hat{a}\ket{n} = \sqrt{n}\ket{n-1}$ and $\sqrt{n}/\sqrt{n!} = 1/\sqrt{(n-1)!}$, and setting $m = n-1$,
\[
\hat{a}\ket{\alpha} = e^{-|\alpha|^2/2}\sum_{n=1}^{\infty}\frac{\alpha^n}{\sqrt{(n-1)!}}\,\ket{n-1} = \alpha\,e^{-|\alpha|^2/2}\sum_{m=0}^{\infty}\frac{\alpha^m}{\sqrt{m!}}\,\ket{m} = \alpha\ket{\alpha} .
\]
Taking the Hermitian conjugate, $\bra{\alpha}\hat{a}^\dagger = \alpha^*\bra{\alpha}$.
\item The photon-number distribution is Poissonian, $|\langle n|\alpha\rangle|^2 = e^{-|\alpha|^2}|\alpha|^{2n}/n!$, with mean $\langle\hat{N}\rangle = \bra{\alpha}\hat{a}^\dagger\hat{a}\ket{\alpha} = |\alpha|^2$.
\item Two coherent states are not orthogonal:
\[
\langle\beta|\alpha\rangle = e^{-(|\alpha|^2+|\beta|^2)/2}\sum_{n=0}^{\infty}\frac{(\beta^*\alpha)^n}{n!} = e^{-(|\alpha|^2+|\beta|^2)/2+\beta^*\alpha},
\]
so $|\langle\beta|\alpha\rangle|^2 = e^{-|\alpha|^2-|\beta|^2+\beta^*\alpha+\beta\alpha^*} = e^{-|\alpha-\beta|^2}$.
\item Phase rotation: since $e^{-i\phi\hat{N}}\ket{n} = e^{-in\phi}\ket{n}$,
\[
e^{-i\phi\hat{N}}\ket{\alpha} = e^{-|\alpha|^2/2}\sum_{n=0}^{\infty}\frac{(\alpha e^{-i\phi})^n}{\sqrt{n!}}\,\ket{n} = \ket{\alpha e^{-i\phi}} .
\]
\item Displacement operator. If $[\hat{X},\hat{Y}]$ commutes with $\hat{X}$ and $\hat{Y}$, then (a special case of the Baker--Campbell--Hausdorff formula)
\[
e^{\hat{X}+\hat{Y}} = e^{\hat{X}}e^{\hat{Y}}e^{-[\hat{X},\hat{Y}]/2} .
\] With $\hat{X} = \alpha\hat{a}^\dagger$ and $\hat{Y} = -\alpha^*\hat{a}$ we have $[\hat{X},\hat{Y}] = -|\alpha|^2[\hat{a}^\dagger,\hat{a}] = |\alpha|^2$, so
\begin{equation}
\hat{D}(\alpha) \equiv e^{\alpha\hat{a}^\dagger - \alpha^*\hat{a}} = e^{-|\alpha|^2/2}\,e^{\alpha\hat{a}^\dagger}\,e^{-\alpha^*\hat{a}},
\qquad
\hat{D}(\alpha)\ket{0} = e^{-|\alpha|^2/2}\,e^{\alpha\hat{a}^\dagger}\ket{0} = \ket{\alpha},
\end{equation}
where we used $e^{-\alpha^*\hat{a}}\ket{0} = \ket{0}$ and $e^{\alpha\hat{a}^\dagger}\ket{0} = \sum_n\alpha^n(\hat{a}^\dagger)^n/n!\,\ket{0} = \sum_n\alpha^n/\sqrt{n!}\,\ket{n}$.
\end{enumerate}

\paragraph{Squeezing.} Let $\hat{S}(s) = e^{s\hat{G}}$ with $\hat{G} = (\hat{a}^2 - \hat{a}^{\dagger2})/2$, which satisfies $\hat{G}^\dagger = -\hat{G}$, so that $\hat{S}$ is unitary. Define $\hat{a}(s) = \hat{S}^\dagger\hat{a}\,\hat{S} = e^{-s\hat{G}}\hat{a}\,e^{s\hat{G}}$. Then $\dif\hat{a}(s)/\dif s = e^{-s\hat{G}}[\hat{a},\hat{G}]e^{s\hat{G}}$, and $[\hat{a},\hat{G}] = -\frac{1}{2}[\hat{a},\hat{a}^{\dagger2}] = -\hat{a}^\dagger$. Similarly $[\hat{a}^\dagger,\hat{G}] = \frac{1}{2}[\hat{a}^\dagger,\hat{a}^2] = -\hat{a}$. Hence $\dif\hat{a}(s)/\dif s = -\hat{a}^\dagger(s)$ and $\dif\hat{a}^\dagger(s)/\dif s = -\hat{a}(s)$, whose solution with $\hat{a}(0) = \hat{a}$ is
\begin{equation}
\hat{S}^\dagger(s)\,\hat{a}\,\hat{S}(s) = \hat{a}\cosh s - \hat{a}^\dagger\sinh s .
\end{equation}

\paragraph{Thermal states.} A thermal state of a mode is $\hat{\rho} = \sum_np_n\ket{n}\bra{n}$ with $p_n = \bar{n}^n/(1+\bar{n})^{n+1}$. With $x = \bar{n}/(1+\bar{n})$ we have $1-x = 1/(1+\bar{n})$ and $p_n = (1-x)\,x^n$, a geometric distribution. Using $\sum_nx^n = 1/(1-x)$ and $\sum_nn\,x^n = x/(1-x)^2$,
\[
\sum_{n=0}^{\infty}p_n = 1 , \qquad \langle\hat{N}\rangle = \sum_{n=0}^{\infty}n\,p_n = \frac{x}{1-x} = \bar{n} .
\]

\section{A mode that is not a plane wave: Young's two pinholes}
\label{app:pinholes}

In Secs.~\ref{sec:planewaves} and~\ref{sec:polarization} the two modes were plane waves. Here we work out an example in which they are not: the two waves radiated by the pinholes of Young's experiment. The example shows that the intensity matrix and the local overlap $\mu_{12}$ are computed in exactly the same way for any modes, and that polarization is part of the phase of the mode also when the modes are not plane waves. Throughout, $\hat{\bm{e}}$ is a real unit vector (linear polarization), $k = \omega/c$, and the observation point $\bm{r}$ is at a distance $R_j = |\bm{r} - \bm{r}_j|$ from the pinhole $j$, located at $\bm{r}_j$. We write $\bm{R}_j = \bm{r} - \bm{r}_j$ and $\hat{\bm{R}}_j = \bm{R}_j/R_j$.

\subsection{From a scalar spherical wave to a transverse mode}
\label{app:spherical}

\paragraph{The scalar spherical wave.} The outgoing spherical wave $f(R) = e^{ikR}/R$ satisfies the Helmholtz equation everywhere except at $R = 0$. For a function of $R$ alone the Laplacian is $\nabla^2f = R^{-1}\,\dif^2(Rf)/\dif R^2$, so that
\[
\nabla^2\frac{e^{ikR}}{R} = \frac{1}{R}\,\frac{\dif^2}{\dif R^2}\,e^{ikR} = -k^2\,\frac{e^{ikR}}{R} \qquad (R\ne0) .
\]
We shall also need its first two derivatives. From $f = e^{ikR}/R$,
\begin{equation}
f' = \left(ik - \frac{1}{R}\right)f , \qquad
f'' = \frac{1}{R^2}\,f + \left(ik - \frac{1}{R}\right)f' = \left(-k^2 - \frac{2ik}{R} + \frac{2}{R^2}\right)f ,
\label{eq:fprime}
\end{equation}
where the prime denotes $\dif/\dif R$.

\paragraph{A spherical wave with a fixed polarization is not transverse.} A simple model, used for instance by Walls~\cite{Walls}, treats each pinhole as the source of the scalar wave $f(R_j)$, multiplied by the polarization $\hat{\bm{e}}$ of the incident light. Since $\nabla R = \hat{\bm{R}}$, the gradient of a function of $R$ is $\nabla f(R) = f'(R)\,\hat{\bm{R}}$, and
\begin{equation}
\nabla\cdot\left(\hat{\bm{e}}\,\frac{e^{ikR}}{R}\right) = \hat{\bm{e}}\cdot\nabla f = (\hat{\bm{e}}\cdot\hat{\bm{R}})\left(ik - \frac{1}{R}\right)\frac{e^{ikR}}{R} .
\label{eq:divscalar}
\end{equation}
The divergence vanishes only in the directions perpendicular to $\hat{\bm{e}}$, so this field violates the transversality condition in Eq.~\eqref{eq:helmholtz}. Physically, the field has a component $(\hat{\bm{e}}\cdot\hat{\bm{R}})\,f$ along the direction of propagation $\hat{\bm{R}}$ that decreases as slowly as the field itself, as $1/R$. This is not allowed: far from the pinhole the spherical wave looks locally like a plane wave travelling along $\hat{\bm{R}}$, and a plane wave is transverse.

\paragraph{The transverse mode.} A transverse field that satisfies the Helmholtz equation is obtained by applying the operator $\nabla\times\nabla\times$ to $\hat{\bm{e}}\,f$. The result is divergence-free, because the divergence of a curl vanishes, and it still satisfies the Helmholtz equation, because $\nabla\times\nabla\times$ and $\nabla^2 + k^2$ are both built from derivatives with constant coefficients and therefore commute. Physically, this is the electric field radiated by an oscillating electric dipole oriented along $\hat{\bm{e}}$ and placed at the pinhole. To compute it, we use $\nabla\times(\nabla\times\bm{a}) = \nabla(\nabla\cdot\bm{a}) - \nabla^2\bm{a}$ (Appendix~\ref{app:vector}), Eq.~\eqref{eq:divscalar}, and $\nabla^2f = -k^2f$:
\[
\nabla\times\nabla\times(\hat{\bm{e}}f) = \nabla\big[(\hat{\bm{e}}\cdot\hat{\bm{R}})\,f'\big] + k^2\,\hat{\bm{e}}\,f .
\]
The gradient of $\hat{\bm{e}}\cdot\hat{\bm{R}} = (\hat{\bm{e}}\cdot\bm{R})/R$ follows from $\nabla(\hat{\bm{e}}\cdot\bm{R}) = \hat{\bm{e}}$ and $\nabla(1/R) = -\hat{\bm{R}}/R^2$:
\[
\nabla(\hat{\bm{e}}\cdot\hat{\bm{R}}) = \frac{\hat{\bm{e}}}{R} - \frac{(\hat{\bm{e}}\cdot\bm{R})\,\hat{\bm{R}}}{R^2} = \frac{\hat{\bm{e}} - (\hat{\bm{e}}\cdot\hat{\bm{R}})\hat{\bm{R}}}{R} .
\]
With the product rule and $\nabla f' = f''\hat{\bm{R}}$,
\[
\nabla\big[(\hat{\bm{e}}\cdot\hat{\bm{R}})\,f'\big] = \frac{f'}{R}\left[\hat{\bm{e}} - (\hat{\bm{e}}\cdot\hat{\bm{R}})\hat{\bm{R}}\right] + (\hat{\bm{e}}\cdot\hat{\bm{R}})\,f''\,\hat{\bm{R}} .
\]
Inserting the derivatives \eqref{eq:fprime} and writing $k^2\hat{\bm{e}} = k^2[\hat{\bm{e}} - (\hat{\bm{e}}\cdot\hat{\bm{R}})\hat{\bm{R}}] + k^2(\hat{\bm{e}}\cdot\hat{\bm{R}})\hat{\bm{R}}$, the terms proportional to $k^2(\hat{\bm{e}}\cdot\hat{\bm{R}})\hat{\bm{R}}$ cancel, and the remaining ones combine into
\begin{multline}
\nabla\times\nabla\times\left(\hat{\bm{e}}\,\frac{e^{ikR}}{R}\right) = k^2\,\frac{e^{ikR}}{R}\,\Big\{\left[\hat{\bm{e}} - (\hat{\bm{e}}\cdot\hat{\bm{R}})\hat{\bm{R}}\right]
+ \left(\frac{i}{kR} - \frac{1}{(kR)^2}\right)\left[\hat{\bm{e}} - 3(\hat{\bm{e}}\cdot\hat{\bm{R}})\hat{\bm{R}}\right]\Big\} .
\label{eq:curlcurl}
\end{multline}
The first term is the \emph{far field}; the second, the \emph{near field}, is smaller by a factor of order $1/(kR)$. For visible light, $\lambda = 500$~nm, at a distance $R = 1$~mm from the pinhole, $kR \approx 1.3\times10^4$. Keeping only the far field, we take as mode functions
\begin{equation}
\bm{u}_j(\bm{r}) = \mathcal{N}\,\bm{t}_j\,\frac{e^{ikR_j}}{R_j}, \qquad \bm{t}_j \equiv \hat{\bm{e}} - (\hat{\bm{e}}\cdot\hat{\bm{R}}_j)\,\hat{\bm{R}}_j \qquad (j = 1,2),
\label{eq:dipolemode}
\end{equation}
where $\mathcal{N}$ is a normalization constant that plays no role below. The vector $\bm{t}_j$ is the projection of $\hat{\bm{e}}$ onto the plane perpendicular to the ray $\hat{\bm{R}}_j$: in the far zone the field oscillates perpendicularly to its local direction of propagation. Only the near field in Eq.~\eqref{eq:curlcurl} has a component along $\hat{\bm{R}}$, coming from the term $-3(\hat{\bm{e}}\cdot\hat{\bm{R}})\hat{\bm{R}}$; it decreases as $1/R^2$ and is negligible where the fringes are observed.\footnote{A subwavelength hole in a perfectly conducting screen radiates instead as a magnetic dipole parallel to the screen~\cite{Bethe1944}. In the plane $x = 0$ considered below, this gives the same polarization directions, and hence the same overlap \eqref{eq:muTheta}, but a different envelope.}

Three remarks are in order. First, the modes of Sec.~\ref{sec:modes} are superpositions of the plane waves of a periodic box. In a large box, the field radiated by a pinhole is such a superposition of plane waves with $|\bm{k}_l| = k$, and we only need its form in the region beyond the screen where the fringes are observed. Second, the two modes \eqref{eq:dipolemode} are not exactly orthogonal. This does not matter here: the classical intensity \eqref{eq:IK} and the visibility \eqref{eq:visclass} follow from $I = 2\varepsilon_0c\,|\bm{E}^+|^2$ with $\bm{E}^+\propto\sum_ja_j\bm{u}_j$, for any mode functions. Third, Eq.~\eqref{eq:dipolemode} is a model of an ideal pinhole; it does not describe the details of a real aperture.

\subsection{The intensity matrix}
\label{app:Kpinholes}

With real vectors $\bm{t}_j$, the elements $K_{jk} = \bm{u}_j^*\cdot\bm{u}_k$ of the intensity matrix are
\begin{equation}
K_{12}(\bm{r}) = \frac{|\mathcal{N}|^2}{R_1R_2}\,(\bm{t}_1\cdot\bm{t}_2)\,e^{ik(R_2-R_1)}, \qquad
K_{jj}(\bm{r}) = \frac{|\mathcal{N}|^2}{R_j^2}\,|\bm{t}_j|^2 .
\label{eq:Kyoung}
\end{equation}
Writing $s_j = \hat{\bm{e}}\cdot\hat{\bm{R}}_j$ and using $\hat{\bm{e}}\cdot\hat{\bm{e}} = \hat{\bm{R}}_j\cdot\hat{\bm{R}}_j = 1$,
\[
\bm{t}_1\cdot\bm{t}_2 = 1 - s_1^2 - s_2^2 + s_1s_2\,(\hat{\bm{R}}_1\cdot\hat{\bm{R}}_2), \qquad
|\bm{t}_j|^2 = 1 - 2s_j^2 + s_j^2 = 1 - s_j^2 .
\]
The local overlap \eqref{eq:mu} is therefore
\begin{equation}
\mu_{12}(\bm{r}) = \frac{\bm{t}_1\cdot\bm{t}_2}{|\bm{t}_1|\,|\bm{t}_2|}\;e^{ik(R_2-R_1)} .
\label{eq:mupinholes}
\end{equation}
Its phase contains $k(R_2 - R_1)$, the familiar path difference of the scalar theory: the bright fringes lie on the surfaces $R_2 - R_1 = \text{const}$, which are hyperboloids of revolution with the two pinholes as foci. The real prefactor $\bm{t}_1\cdot\bm{t}_2/(|\bm{t}_1||\bm{t}_2|)$, which lies between $-1$ and $1$, comes from the polarization: its modulus multiplies the visibility, and where it is negative it adds $\pi$ to the phase of the mode. Moreover, the envelopes $K_{jj}\propto(1 - s_j^2)/R_j^2$ carry the angular distribution of dipole radiation, which the scalar spherical wave lacks.

\subsection{Observation in the plane of the pinholes}
\label{app:planepinholes}

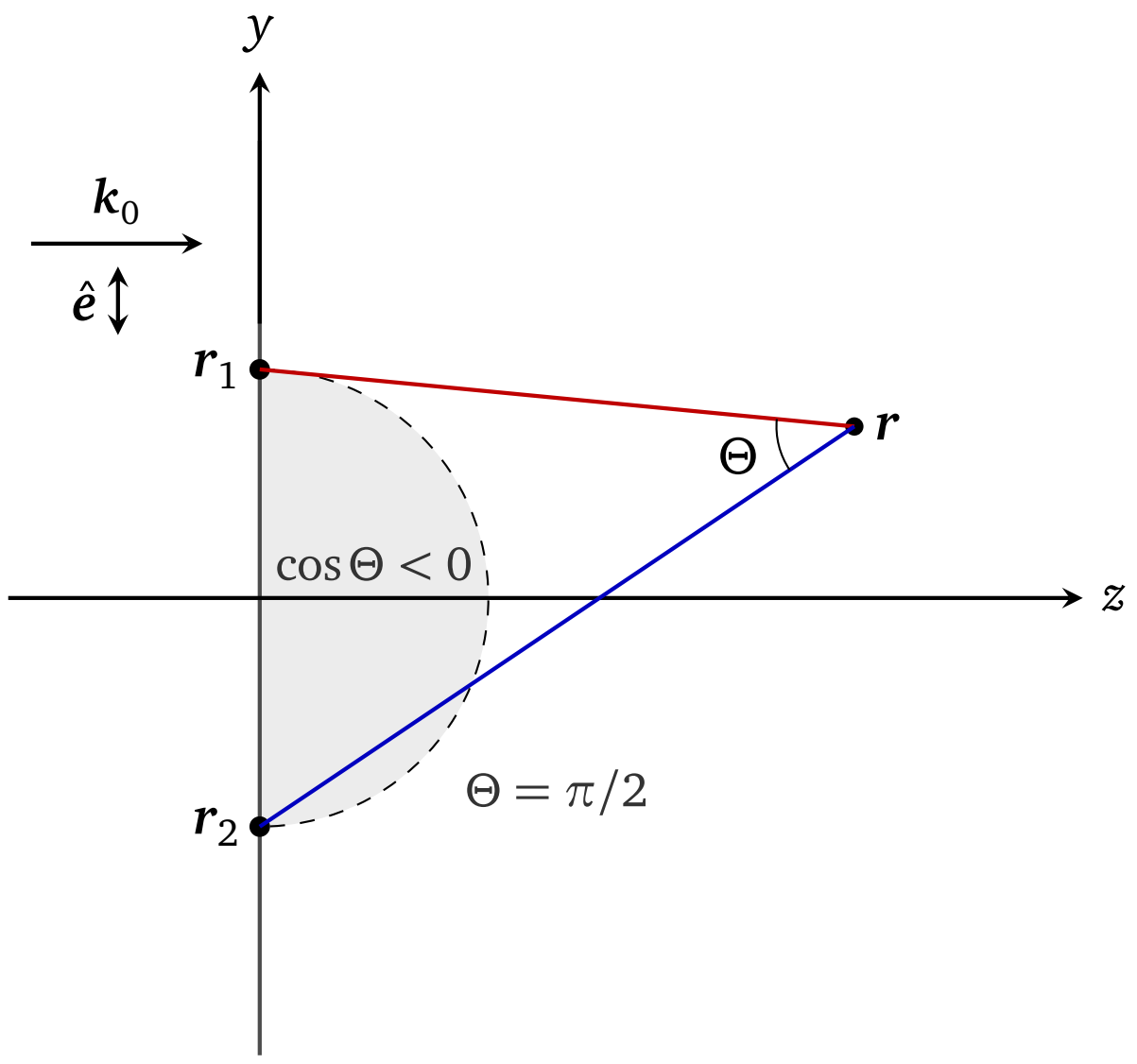
\begin{figure}[t]
\centering
\begin{tikzpicture}[scale=1.6, >=stealth]
  \fill[gray!15] (0,0) -- (0,1) arc (90:-90:1) -- cycle;
  \draw[gray!60!black, thick] (0,-2.0) -- (0,2.0);
  \draw[->, thick] (-1.1,0) -- (3.6,0) node[right] {$z$};
  \draw[->, thick] (0,1.2) -- (0,2.3) node[above] {$y$};
  \draw[dashed] (0,1) arc (90:-90:1);
  \fill (0,1) circle (0.045) node[left] {$\bm{r}_1$};
  \fill (0,-1) circle (0.045) node[left] {$\bm{r}_2$};
  \coordinate (P) at (2.6,0.75);
  \fill (P) circle (0.04) node[right] {$\bm{r}$};
  \draw[red!75!black, thick] (0,1) -- (P);
  \draw[blue!75!black, thick] (0,-1) -- (P);
  \draw (2.2616,0.7825) arc (174.51:213.94:0.34);
  \node at (2.096,0.622) {$\Theta$};
  \node[gray!40!black] at (0.5,0.15) {\small $\cos\Theta<0$};
  \node[gray!40!black] at (1.3,-0.85) {\small $\Theta=\pi/2$};
  \draw[->, thick] (-1.0,1.55) -- (-0.25,1.55) node[midway, above] {\small $\bm{k}_0$};
  \draw[<->, thick] (-0.62,1.15) -- (-0.62,1.45);
  \node[left] at (-0.62,1.3) {$\hat{\bm{e}}$};
\end{tikzpicture}
\caption{Young's experiment with two pinholes at $\bm{r}_{1,2} = \pm(d/2)\hat{\bm{y}}$ in an opaque screen (vertical line), illuminated from the left by a plane wave with wave vector $\bm{k}_0$ and polarization $\hat{\bm{e}}$. At the observation point $\bm{r}$ the rays from the two pinholes make the angle $\Theta$. For in-plane polarization, $\hat{\bm{e}} = \hat{\bm{y}}$, the local overlap of the two mode functions is $\cos\Theta$, Eq.~\eqref{eq:muTheta}: it vanishes on the dashed half-circle with diameter $d$ (Thales's theorem) and is negative inside it (shaded), where bright and dark fringes are interchanged.}
\label{fig:pinholes}
\end{figure}

The result is especially transparent for points in the plane $x = 0$ that contains the axis of the experiment and the two pinholes, placed at $\bm{r}_{1,2} = \pm(d/2)\,\hat{\bm{y}}$ in a screen at $z = 0$ (Fig.~\ref{fig:pinholes}). We observe at $\bm{r} = y\,\hat{\bm{y}} + z\,\hat{\bm{z}}$ with $z > 0$, and write the directions of the rays as
\begin{equation}
\hat{\bm{R}}_j = \hat{\bm{y}}\sin\psi_j + \hat{\bm{z}}\cos\psi_j , \qquad |\psi_j| < \pi/2 ,
\label{eq:rays}
\end{equation}
exactly as the directions \eqref{eq:pw} of the two plane waves. The angle between the two rays that reach $\bm{r}$ is $\Theta = \psi_2 - \psi_1$ (up to sign), with $\cos\Theta = \hat{\bm{R}}_1\cdot\hat{\bm{R}}_2$.

\paragraph{Polarization perpendicular to the plane, $\hat{\bm{e}} = \hat{\bm{x}}$.} Then $s_j = \hat{\bm{x}}\cdot\hat{\bm{R}}_j = 0$ and $\bm{t}_j = \hat{\bm{x}}$: both mode functions are parallel to $\hat{\bm{x}}$, and $\mu_{12} = e^{ik(R_2-R_1)}$ has unit modulus. In this plane the scalar theory is exact, with envelopes $K_{jj} = |\mathcal{N}|^2/R_j^2$.

\paragraph{Polarization in the plane, $\hat{\bm{e}} = \hat{\bm{y}}$.} Now $s_j = \sin\psi_j$, and
\[
\bm{t}_j = \hat{\bm{y}} - \sin\psi_j\left(\hat{\bm{y}}\sin\psi_j + \hat{\bm{z}}\cos\psi_j\right) = \cos\psi_j\left(\hat{\bm{y}}\cos\psi_j - \hat{\bm{z}}\sin\psi_j\right),
\]
where we used $1 - \sin^2\psi_j = \cos^2\psi_j$. Since $\cos\psi_j > 0$ for $z > 0$, we have $|\bm{t}_j| = \cos\psi_j$, and $\bm{t}_j/|\bm{t}_j| = \hat{\bm{y}}\cos\psi_j - \hat{\bm{z}}\sin\psi_j$ is, up to a sign, the in-plane polarization vector \eqref{eq:ppol} of a plane wave travelling along $\hat{\bm{R}}_j$. Hence, as in Eq.~\eqref{eq:cos2a},
\[
\frac{\bm{t}_1\cdot\bm{t}_2}{|\bm{t}_1|\,|\bm{t}_2|} = \cos\psi_1\cos\psi_2 + \sin\psi_1\sin\psi_2 = \cos(\psi_1 - \psi_2) = \hat{\bm{R}}_1\cdot\hat{\bm{R}}_2 ,
\]
and Eq.~\eqref{eq:mupinholes} becomes
\begin{equation}
\mu_{12}(\bm{r}) = \cos\Theta(\bm{r})\;e^{ik(R_2-R_1)} .
\label{eq:muTheta}
\end{equation}
This is the pinhole analog of the factor $\cos2\alpha$ in Eq.~\eqref{eq:Ip}. The reason is simple: near the observation point, the two spherical waves look like two plane waves travelling along $\hat{\bm{R}}_1$ and $\hat{\bm{R}}_2$, which cross at the angle $\Theta$, and the plane-wave result \eqref{eq:cos2a} applies locally with $2\alpha\to\Theta$. With $a_j = |a_j|e^{i\varphi_j}$, $K_{jj} = |\mathcal{N}|^2\cos^2\psi_j/R_j^2$ and Eq.~\eqref{eq:I2}, the intensity in the plane $x = 0$ is
\begin{multline}
I(\bm{r}) = \hbar\omega c\,|\mathcal{N}|^2\left\{\frac{|a_1|^2\cos^2\psi_1}{R_1^2} + \frac{|a_2|^2\cos^2\psi_2}{R_2^2}\right.
\\
\left. + \frac{2|a_1||a_2|\cos\psi_1\cos\psi_2}{R_1R_2}\,\cos\Theta\,\cos\!\left[k(R_2 - R_1) + \varphi_2 - \varphi_1\right]\right\}.
\label{eq:Ipinholes}
\end{multline}
The visibility is the ratio of the fringe amplitude to the background, Eq.~\eqref{eq:visdef}:
\begin{equation}
\mathcal{V}(\bm{r}) = \frac{2|a_1||a_2|\,\cos\psi_1\cos\psi_2\,|\cos\Theta|/(R_1R_2)}{|a_1|^2\cos^2\psi_1/R_1^2 + |a_2|^2\cos^2\psi_2/R_2^2}
= |\cos\Theta(\bm{r})|\;\frac{2\sqrt{I_1(\bm{r})\,I_2(\bm{r})}}{I_1(\bm{r}) + I_2(\bm{r})} ,
\label{eq:Vpinholes}
\end{equation}
where $I_j = \hbar\omega c\,|\mathcal{N}|^2|a_j|^2\cos^2\psi_j/R_j^2$ is the intensity that pinhole $j$ would produce alone. This is Eq.~\eqref{eq:visclass} with $|\mu_{12}| = |\cos\Theta|$: the visibility is the polarization factor $|\cos\Theta|$ times the balance of the two beams.

\paragraph{Where the fringes disappear: Thales's theorem.} The angle $\Theta$ is the angle at $\bm{r}$ of the triangle with vertices $\bm{r}_1$, $\bm{r}_2$ and $\bm{r}$, whose side opposite to $\bm{r}$ is the segment of length $d$ that joins the pinholes. By the law of cosines, $d^2 = R_1^2 + R_2^2 - 2R_1R_2\cos\Theta$. With $R_{1,2}^2 = (y\mp d/2)^2 + z^2$ we have $R_1^2 + R_2^2 = 2(y^2 + z^2) + d^2/2$, and therefore
\begin{equation}
\cos\Theta = \frac{R_1^2 + R_2^2 - d^2}{2R_1R_2} = \frac{\rho^2 - d^2/4}{R_1R_2}, \qquad \rho^2 \equiv y^2 + z^2 ,
\label{eq:cosTheta}
\end{equation}
where $\rho$ is the distance of $\bm{r}$ from the midpoint between the pinholes. 
Hence $\cos\Theta = 0$ on the half-circle $\rho = d/2$, whose diameter is the segment joining the pinholes: this is Thales's theorem, the angle inscribed in a semicircle is a right angle, which Eq.~\eqref{eq:cosTheta} proves directly.
On this half-circle the two fields are perpendicular, and the fringes disappear for in-plane polarization, although both beams reach $\bm{r}$ and are perfectly coherent. Inside it, $\rho < d/2$, $\cos\Theta < 0$, and bright and dark fringes are interchanged; outside it, $\cos\Theta > 0$.

\paragraph{The paraxial limit.} Far from the pinholes, $\rho\gg d$, Eq.~\eqref{eq:cosTheta} gives $\cos\Theta\simeq1$. On the axis, for instance, $y = 0$ and $R_1 = R_2 = (z^2 + d^2/4)^{1/2}$, so that
\[
\cos\Theta = \frac{z^2 - d^2/4}{z^2 + d^2/4} \simeq 1 - \frac{d^2}{2z^2} \qquad (z\gg d).
\]
For pinholes $d = 1$~mm apart and a screen at $z = 1$~m, $1 - \cos\Theta \approx 5\times10^{-7}$: the scalar description of the experiment is excellent, and it is recovered as the paraxial limit of the vector one. The vector structure of the modes matters close to the pinholes, within a distance of the order of $d$. For $d\gg\lambda$ the far-field approximation \eqref{eq:dipolemode} is still valid there, except within a few wavelengths of each pinhole.

\section{How sharp is the phase of a state?}
\label{app:sharp}

In Sec.~\ref{sec:criterion} we have seen that a state has a phase if and only if it contains coherences between different photon numbers. A state may have a phase and yet be hardly changed by small rotations. In this appendix we quantify how sharply the phase of a state is defined, and we show that the answer is set by the spread of its photon number.

\subsection{The overlap of rotated states}
\label{app:overlap}

To measure how much a rotation changes a pure state $\ket{\psi}$, we compare it with the rotated state $\ket{\psi(\phi)} = \hat{U}_b(\phi)\ket{\psi}$. The squared overlap $|\langle\psi|\psi'\rangle|^2$ of two states is the probability that a measurement asking ``is the system in the state $\ket{\psi}$?'' answers yes when the system is in the state $\ket{\psi'}$. It equals 1 if the two states coincide, and 0 if they are orthogonal, i.e., perfectly distinguishable. For small rotations,
\begin{equation}
|\langle\psi|\psi(\phi)\rangle|^2 = 1 - \phi^2\,(\Delta N_b)^2 + O(\phi^4) ,
\label{eq:fidelity}
\end{equation}
where $(\Delta N_b)^2 = \langle\hat{N}_b^2\rangle - \langle\hat{N}_b\rangle^2$ is the variance of the photon number. To derive this result, we expand the exponential to second order:
\[
\langle\psi|\psi(\phi)\rangle = \bra{\psi}e^{-i\phi\hat{N}_b}\ket{\psi} = 1 - i\phi\,\langle\hat{N}_b\rangle - \frac{\phi^2}{2}\,\langle\hat{N}_b^2\rangle + O(\phi^3) .
\]
The real part is $1 - \phi^2\langle\hat{N}_b^2\rangle/2 + O(\phi^4)$, because the odd powers of $-i\phi$ are imaginary, and the imaginary part is $-\phi\langle\hat{N}_b\rangle + O(\phi^3)$. Therefore
\[
|\langle\psi|\psi(\phi)\rangle|^2 = \left(1 - \frac{\phi^2}{2}\langle\hat{N}_b^2\rangle\right)^2 + \phi^2\langle\hat{N}_b\rangle^2 + O(\phi^4) = 1 - \phi^2\left(\langle\hat{N}_b^2\rangle - \langle\hat{N}_b\rangle^2\right) + O(\phi^4) .
\]
Equation~\eqref{eq:fidelity} holds only for $\phi\,\Delta N_b\ll1$. It describes how the overlap \emph{starts} to decrease, not its value at larger angles: taken literally, it would vanish at $\phi = 1/\Delta N_b$ and become negative beyond, which is impossible. What it tells us is the angular scale on which the state changes: the overlap begins to drop appreciably when $\phi$ becomes comparable with $1/\Delta N_b$. An exact inequality, the analogue of the Mandelstam--Tamm time--energy uncertainty relation~\cite{MandelstamTamm}, makes this precise (the proof is given in Appendix~\ref{app:MT} below):
\begin{equation}
|\langle\psi|\psi(\phi)\rangle|^2 \ge \cos^2(\phi\,\Delta N_b) \qquad\text{for } 0\le\phi\,\Delta N_b\le\pi/2 .
\label{eq:MTbound}
\end{equation}
Its right-hand side has the same expansion as Eq.~\eqref{eq:fidelity}, but it shows that at $\phi = 1/\Delta N_b$ the overlap is still at least $\cos^21\approx0.29$, and that a rotation by at least $\pi/(2\Delta N_b)$ is needed to reach an orthogonal state. In short, the spread of the photon number sets how sharply the phase is defined: the larger $\Delta N_b$, the smaller the rotation that produces a distinguishable state.

\subsection{Coherent states}
\label{app:sharpcoh}

\begin{figure}[t]
\centering
\begin{tikzpicture}[scale=0.95,>=Stealth]
  \begin{scope}
  \fill[scipostblue!25] (40:1.6) circle (0.38);
  \draw[scipostblue] (40:1.6) circle (0.38);
  \draw[scipostblue,dashed] (-8:1.6) circle (0.38);
  \draw[gray] (0,0) -- (40:1.6);
  \draw[->,thick,red!70!black] (36:2.25) arc (36:-4:2.25);
  \node[red!70!black] at (16:2.6) {$\phi$};
  \draw[->] (-0.4,0) -- (2.9,0) node[right] {\small$\Real\,\alpha$};
  \draw[->] (0,-1.1) -- (0,1.9) node[above] {\small$\Imag\,\alpha$};
  \node[align=center,font=\small] at (1.2,-1.6) {coherent state:\\ rotated into a new state};
  \end{scope}
  \begin{scope}[xshift=6.3cm]
  \fill[scipostblue!25,even odd rule] (0,0) circle (1.5) (0,0) circle (0.95);
  \draw[scipostblue] (0,0) circle (1.5) (0,0) circle (0.95);
  \draw[->,thick,red!70!black] (62:1.85) arc (62:22:1.85);
  \node[red!70!black] at (42:2.2) {$\phi$};
  \draw[->] (-1.9,0) -- (2.2,0) node[right] {\small$\Real\,\alpha$};
  \draw[->] (0,-1.75) -- (0,1.95) node[above] {\small$\Imag\,\alpha$};
  \node[align=center,font=\small] at (0.2,-2.3) {number state:\\ invariant under rotations};
  \end{scope}
\end{tikzpicture}
\caption{Schematic phase-space pictures of a coherent state and of a number state. A coherent state is represented by a disk displaced from the origin; the rotation $\hat{U}_b(\phi)$ moves it to a new position, so the state has a phase, $\arg\alpha$. A number state is represented by a ring centered at the origin, which a rotation leaves unchanged: its phase is completely undetermined.}
\label{fig:phasespace}
\end{figure}
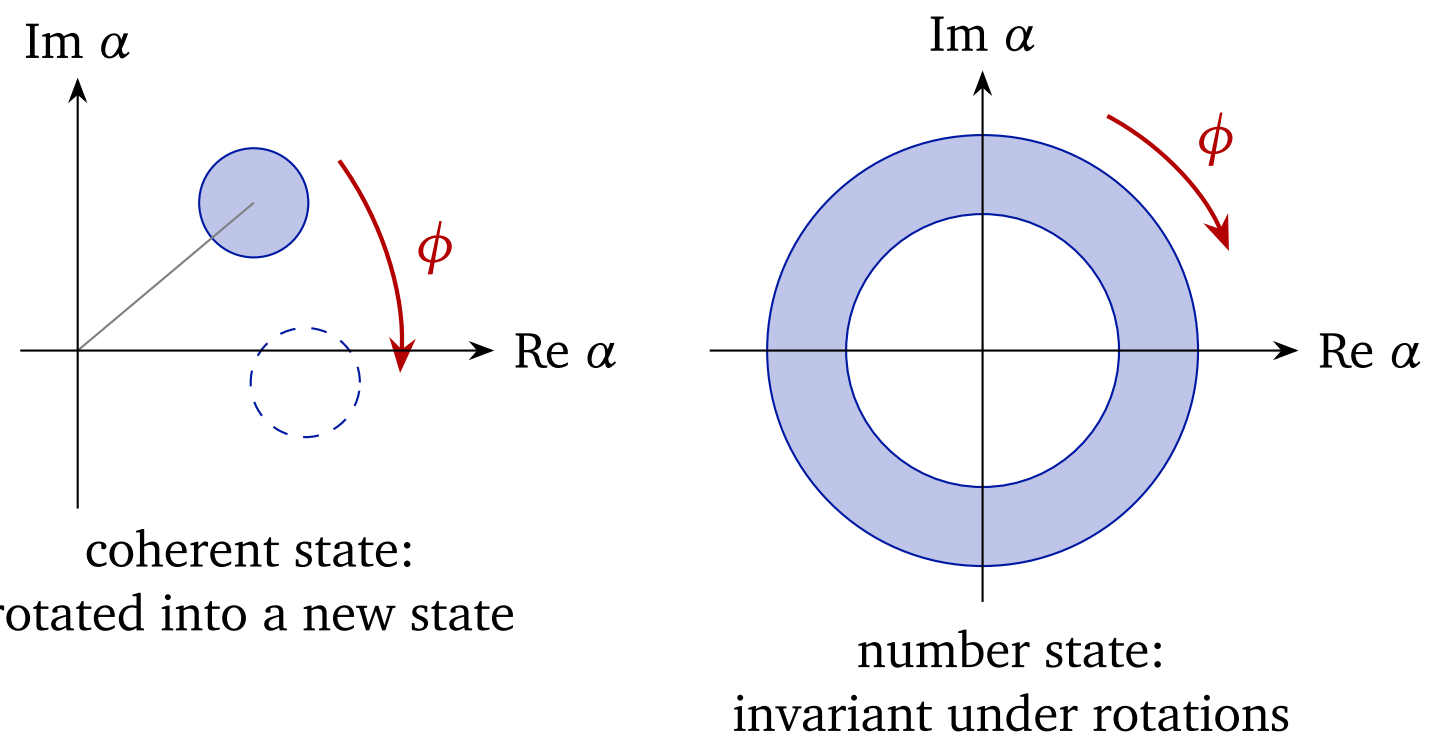

For a number state $\Delta N_b = 0$, and the overlap is exactly 1 for every $\phi$, as we already know from Sec.~\ref{sec:phaseexamples}. For a coherent state,
\[
\langle\hat{N}_b^2\rangle = \bra{\alpha}\hat{b}^\dagger\hat{b}\,\hat{b}^\dagger\hat{b}\ket{\alpha} = \bra{\alpha}\hat{b}^\dagger\left(\hat{b}^\dagger\hat{b} + 1\right)\hat{b}\ket{\alpha} = |\alpha|^4 + |\alpha|^2 ,
\]
so that $(\Delta N_b)^2 = |\alpha|^2$. In this case the overlap is known exactly. Using $|\langle\beta|\alpha\rangle|^2 = e^{-|\alpha-\beta|^2}$ (Appendix~\ref{app:oscillator}) and $|1-e^{-i\phi}|^2 = 2-2\cos\phi = 4\sin^2(\phi/2)$, we find
\[
|\langle\alpha|\alpha e^{-i\phi}\rangle|^2 = e^{-|\alpha|^2|1-e^{-i\phi}|^2} = e^{-4|\alpha|^2\sin^2(\phi/2)} \approx e^{-|\alpha|^2\phi^2}
\quad\text{for } |\phi|\ll1 ,
\]
whose expansion, $1 - |\alpha|^2\phi^2$, agrees with Eq.~\eqref{eq:fidelity}. For $|\alpha|\gg1$ the overlap equals $e^{-1}\approx0.37$ at $\phi = 1/|\alpha|$ and $e^{-4}\approx0.02$ at $\phi = 2/|\alpha|$, in agreement with the bound~\eqref{eq:MTbound}: a rotation by a few times $1/|\alpha|$ already produces a state that can be distinguished from the original one, and the phase $\arg\alpha$ is defined to within an angle of order $1/|\alpha|\ll1$. In the picture of Fig.~\ref{fig:phasespace}, the disk has a size of order 1 and lies at distance $|\alpha|$ from the origin; a rotation by $\phi$ moves it by about $|\alpha|\phi$, and the rotated disk no longer overlaps the original one when $|\alpha|\phi\gtrsim1$. For $|\alpha|$ of order 1 the phase is defined only to within about a radian. For $|\alpha|\ll1$ the overlap stays close to 1 for every $\phi$ (its smallest value, at $\phi = \pi$, is $e^{-4|\alpha|^2}\approx1-4|\alpha|^2$): no rotation produces a clearly different state, and the phase is poorly defined. For the vacuum, $\alpha = 0$, it is not defined at all.

A sharp phase requires a broad photon-number distribution, and a sharp photon number destroys the phase. This is the content of the number--phase uncertainty relation, usually written heuristically as $\Delta N_b\,\Delta\phi\gtrsim1/2$, with near equality for coherent states with $|\alpha|\gg1$. It cannot easily be made rigorous, because there is no Hermitian operator for the phase.

\subsection{Proof of the Mandelstam--Tamm bound}
\label{app:MT}

Here we prove the inequality~\eqref{eq:MTbound}. The argument is the one used by Mandelstam and Tamm for the time--energy uncertainty relation~\cite{MandelstamTamm}, with the photon-number operator in place of $\hat{H}/\hbar$ and the angle $\phi$ in place of the time. To lighten the notation we write $\hat{N}$ for $\hat{N}_b$ and $\Delta N$ for $\Delta N_b$, which we assume to be finite.

\paragraph{Step 1: the Robertson uncertainty relation.} Let $\hat{A}$ and $\hat{B}$ be Hermitian operators and $\ket{\chi}$ a normalized state, and write $\langle\cdot\rangle = \bra{\chi}\cdot\ket{\chi}$. We define the deviations $\delta\hat{A} = \hat{A} - \langle\hat{A}\rangle$ and $\delta\hat{B} = \hat{B} - \langle\hat{B}\rangle$, and the spreads $(\Delta A)^2 = \langle\delta\hat{A}^2\rangle$ and $(\Delta B)^2 = \langle\delta\hat{B}^2\rangle$. Numbers commute with everything, so $[\hat{A},\hat{B}] = [\delta\hat{A},\delta\hat{B}]$. Moreover, $(\delta\hat{A}\,\delta\hat{B})^\dagger = \delta\hat{B}\,\delta\hat{A}$, so that $\langle\delta\hat{B}\,\delta\hat{A}\rangle = \langle\delta\hat{A}\,\delta\hat{B}\rangle^*$. Hence
\[
\langle[\hat{A},\hat{B}]\rangle = \langle\delta\hat{A}\,\delta\hat{B}\rangle - \langle\delta\hat{A}\,\delta\hat{B}\rangle^* = 2i\,\Imag\langle\delta\hat{A}\,\delta\hat{B}\rangle .
\]
The number $\langle\delta\hat{A}\,\delta\hat{B}\rangle$ is the scalar product of the vectors $\delta\hat{A}\ket{\chi}$ and $\delta\hat{B}\ket{\chi}$, whose norms are $\Delta A$ and $\Delta B$. The proof of the Cauchy--Schwarz inequality in Appendix~\ref{app:cauchy} uses only the properties of the scalar product, so it holds in the space of quantum states as well. Therefore
\begin{equation}
|\langle[\hat{A},\hat{B}]\rangle| \le 2\,|\langle\delta\hat{A}\,\delta\hat{B}\rangle| \le 2\,\Delta A\,\Delta B .
\label{eq:robertson}
\end{equation}

\paragraph{Step 2: a differential inequality.} Let $Q(\phi) = |\langle\psi|\psi(\phi)\rangle|^2$, with $\ket{\psi(\phi)} = e^{-i\phi\hat{N}}\ket{\psi}$. Using the projector $\hat{P} = \ket{\psi}\bra{\psi}$, we write
\[
Q(\phi) = \langle\psi(\phi)|\psi\rangle\langle\psi|\psi(\phi)\rangle = \bra{\psi(\phi)}\hat{P}\ket{\psi(\phi)} = \bra{\psi}e^{i\phi\hat{N}}\,\hat{P}\,e^{-i\phi\hat{N}}\ket{\psi} .
\]
Differentiating, and using that $\hat{N}$ commutes with $e^{\pm i\phi\hat{N}}$,
\[
\frac{\dif Q}{\dif\phi} = \bra{\psi}e^{i\phi\hat{N}}\left(i\hat{N}\hat{P} - i\hat{P}\hat{N}\right)e^{-i\phi\hat{N}}\ket{\psi} = \bra{\psi(\phi)}\,i[\hat{N},\hat{P}]\,\ket{\psi(\phi)} .
\]
We now apply Eq.~\eqref{eq:robertson} in the state $\ket{\chi} = \ket{\psi(\phi)}$, with $\hat{A} = \hat{N}$ and $\hat{B} = \hat{P}$. The spread of $\hat{N}$ in $\ket{\psi(\phi)}$ is the same as in $\ket{\psi}$, because $\hat{N}$ commutes with $e^{-i\phi\hat{N}}$; it is therefore $\Delta N$. The spread of $\hat{P}$ follows from $\hat{P}^2 = \hat{P}$ and $\bra{\psi(\phi)}\hat{P}\ket{\psi(\phi)} = Q$: $(\Delta P)^2 = \langle\hat{P}^2\rangle - \langle\hat{P}\rangle^2 = Q - Q^2$. Hence
\begin{equation}
\left|\frac{\dif Q}{\dif\phi}\right| \le 2\,\Delta N\,\sqrt{Q\,(1-Q)} .
\label{eq:Qineq}
\end{equation}

\paragraph{Step 3: integration.} Since $0\le Q\le1$, we can write $Q(\phi) = \cos^2\theta(\phi)$ with $\theta = \arccos\sqrt{Q}$, a continuous function with values in $[0,\pi/2]$ and $\theta(0) = 0$, because $Q(0) = 1$. Where $0 < Q < 1$, $\theta$ is differentiable, with
\[
\frac{\dif Q}{\dif\phi} = -2\cos\theta\,\sin\theta\,\frac{\dif\theta}{\dif\phi} , \qquad \sqrt{Q\,(1-Q)} = \cos\theta\,\sin\theta ,
\]
and Eq.~\eqref{eq:Qineq} becomes simply
\[
\left|\frac{\dif\theta}{\dif\phi}\right| \le \Delta N .
\]
The angle $\theta$ starts from zero and can grow at most at the rate $\Delta N$, so that
\begin{equation}
\theta(\phi) \le \Delta N\,\phi \qquad (\phi\ge0) .
\label{eq:thetabound}
\end{equation}
If $\Delta N\,\phi\le\pi/2$, both $\theta(\phi)$ and $\Delta N\,\phi$ lie in $[0,\pi/2]$, where $\cos^2$ is a decreasing function. Therefore $Q(\phi) = \cos^2\theta(\phi) \ge \cos^2(\Delta N\,\phi)$, which is Eq.~\eqref{eq:MTbound}. Since $Q(-\phi) = Q(\phi)$, the same bound holds for negative angles, with $|\phi|$ in place of $\phi$.

\paragraph{A technical remark.} The derivative of $\theta$ may not exist at angles where $Q = 1$ or $Q = 0$, but Eq.~\eqref{eq:thetabound} still holds. Take $\phi_1 > 0$ with $\theta(\phi_1) > 0$, and let $\phi_0$ be the largest angle in $[0,\phi_1)$ at which $\theta = 0$. If $\theta < \pi/2$ everywhere in $(\phi_0,\phi_1)$, the mean value theorem gives $\theta(\phi_1) = \theta(\phi_1) - \theta(\phi_0) \le \Delta N\,(\phi_1-\phi_0) \le \Delta N\,\phi_1$. If instead $\theta$ reaches $\pi/2$ inside $(\phi_0,\phi_1)$, the same argument applied up to the first such angle $\phi_2$ gives $\pi/2 \le \Delta N\,\phi_2 \le \Delta N\,\phi_1$, and then $\theta(\phi_1) \le \pi/2 \le \Delta N\,\phi_1$.

\paragraph{The bound cannot be improved.} For the superposition $\ket{\psi} = (\ket{n} + \ket{m})/\sqrt{2}$ of two number states, with $n > m$,
\[
\langle\psi|\psi(\phi)\rangle = \frac{e^{-in\phi} + e^{-im\phi}}{2} = e^{-i(n+m)\phi/2}\cos\frac{(n-m)\phi}{2} ,
\qquad
Q(\phi) = \cos^2\frac{(n-m)\phi}{2} .
\]
The photon number takes the values $n$ and $m$ with probability $1/2$ each, so that
\[
\langle\hat{N}\rangle = \frac{n+m}{2}, \qquad \langle\hat{N}^2\rangle = \frac{n^2+m^2}{2}, \qquad (\Delta N)^2 = \frac{n^2+m^2}{2} - \frac{(n+m)^2}{4} = \frac{(n-m)^2}{4} .
\]
Hence $Q(\phi) = \cos^2(\Delta N\,\phi)$: the bound is attained, and this state becomes orthogonal to its rotated copy at $\phi = \pi/(2\Delta N)$, the smallest angle allowed. A coherent state, in contrast, never becomes exactly orthogonal to its rotated copies, since $e^{-4|\alpha|^2\sin^2(\phi/2)} > 0$.

\section{Solutions of the problems}
\label{app:solutions}

\paragraph{Problem~\ref{prob:shells}.} The integers whose squares add up to 5 are, up to order and signs, $(2,1,0)$. There are 3 positions for the zero, 2 ways of placing $2$ and $1$ in the remaining positions, and $2\times2$ choices of signs: $3\times2\times4 = 24$ wave vectors, e.g.\ $(2\pi/L)(2,1,0)$, $(2\pi/L)(0,-1,2)$, and so on. For $m = 7$: an even integer $2p$ has square $4p^2$, which leaves remainder 0 or 4 when divided by 8; an odd integer $2p+1$ has square $4p(p+1)+1$, which leaves remainder 1 because $p(p+1)$ is even. The sum of three numbers taken from $\{0,1,4\}$ can leave the remainders $0,1,2,3,4,5,6$ when divided by 8, but never 7. Hence $l_1^2+l_2^2+l_3^2 = 7$ has no solution, and there is no field with $(\omega/\omega_0)^2 = 7$.

\paragraph{Problem~\ref{prob:halfenergy}.} The cycle average of $e^{\mp2i\omega t}$ vanishes, Eq.~\eqref{eq:ca_zero}, so Eqs.~\eqref{eq:WE} and \eqref{eq:WB} give $\langle W_E\rangle_\ca = \langle W_B\rangle_\ca = \varepsilon_0\omega^2\sum_{jk}A_j^*G_{jk}A_k = W/2$. The standing wave $\bm{u} = \sqrt{2}\,\hat{\bm{x}}\sin(kz)/L^{3/2}$, with $k = 2\pi l_3/L$, is a normalized mode, being the superposition $(\bm{u}_+ - \bm{u}_-)/(i\sqrt{2})$ of the plane waves $\bm{u}_\pm = \hat{\bm{x}}e^{\pm ikz}/L^{3/2}$. Since $\bm{u}$ is real, $\Delta = \int_{L^3}\bm{u}\cdot\bm{u}\,\dif^3r = (\bm{u},\bm{u}) = 1$, and with $A = |A|e^{i\varphi}$ Eq.~\eqref{eq:WE} gives
\begin{equation}
W_E = \varepsilon_0\omega^2|A|^2 - \varepsilon_0\omega^2|A|^2\cos(2\omega t - 2\varphi) ,
\end{equation}
which oscillates between $0$ and $2\varepsilon_0\omega^2|A|^2 = W$. The magnetic energy $W_B = W - W_E$ oscillates in antiphase: in a standing wave the energy sloshes back and forth between the electric and the magnetic field, and only the total is conserved.

\paragraph{Problem~\ref{prob:pancharatnam}.} From Eq.~\eqref{eq:ppol}, $\hat{\bm{p}}_j$ is the in-plane polarization vector, so $\hat{\bm{x}}\cdot\hat{\bm{p}}_j = 0$ and $\hat{\bm{p}}_1\cdot\hat{\bm{p}}_2 = \cos2\alpha$ by Eq.~\eqref{eq:cos2a}. Then
\begin{equation}
\ue_1^*\cdot\ue_2 = \tfrac{1}{2}(\hat{\bm{x}} - i\hat{\bm{p}}_1)\cdot(\hat{\bm{x}} + \hat{\bm{p}}_2) = \tfrac{1}{2}\left(1 - i\cos2\alpha\right).
\end{equation}
For $|a_1| = |a_2|$ and $K_{11} = K_{22} = 1/L^3$ the balance factor in Eq.~\eqref{eq:visclass} is 1, so
\[
\mathcal{V} = |\mu_{12}| = |\ue_1^*\cdot\ue_2| = \tfrac{1}{2}\left(1+\cos^22\alpha\right)^{1/2} .
\]
The phase of $\mu_{12}$ is $\phi_{12} = -2ky\sin\alpha - \arctan(\cos2\alpha)$, so by Eq.~\eqref{eq:brightsurf} the bright fringes lie at
\[
y_m = m\,\Delta y + \frac{\Delta y}{2\pi}\left[\varphi_2-\varphi_1 - \arctan(\cos2\alpha)\right] .
\]
Compared with the fringes of two $\hat{\bm{x}}$-polarized beams, they are displaced toward negative $y$ by the fraction $\arctan(\cos2\alpha)/2\pi$ of a period. For $\alpha\to0$, $\arctan1 = \pi/4$ and the displacement is one eighth of a period.

\paragraph{Problem~\ref{prob:helicity}.} With $\hat{\bm{x}}\cdot\hat{\bm{p}}_j = 0$ and $\hat{\bm{p}}_1\cdot\hat{\bm{p}}_2 = \cos2\alpha$,
\begin{equation}
\ue_1^*\cdot\ue_2 = \tfrac{1}{2}(\hat{\bm{x}} - i\sigma_1\hat{\bm{p}}_1)\cdot(\hat{\bm{x}} + i\sigma_2\hat{\bm{p}}_2) = \tfrac{1}{2}\left(1 + \sigma_1\sigma_2\cos2\alpha\right),
\end{equation}
so $|\mu_{12}| = |1+\sigma_1\sigma_2\cos2\alpha|/2$, which equals $(1+\cos2\alpha)/2 = \cos^2\alpha$ for $\sigma_1 = \sigma_2$ and $(1-\cos2\alpha)/2 = \sin^2\alpha$ for $\sigma_1 = -\sigma_2$. For $\alpha\to0$ the beams are collinear: equal helicities give full contrast, and opposite helicities, which are orthogonal polarizations, give no fringes, in accordance with the Fresnel--Arago laws. For $\alpha\to\pi/2$ the beams counterpropagate along $\pm\hat{\bm{y}}$, with $\hat{\bm{p}}_1 = \hat{\bm{z}}$ and $\hat{\bm{p}}_2 = -\hat{\bm{z}}$: now \emph{equal} helicities give no fringes and opposite helicities give full contrast. The reason is that helicity is defined with respect to the direction of propagation: for counterpropagating beams, equal helicities correspond to electric fields rotating in opposite senses in the laboratory, i.e., to orthogonal polarizations.

\paragraph{Problem~\ref{prob:bs}.} In matrix form $(\an{1},\an{2})^T = \mathsf{T}\,(\hat{c}_{\rm in},\hat{c}'_{\rm in})^T$ with $\mathsf{T} = \begin{pmatrix}\tau&\kappa\\\kappa&\tau\end{pmatrix}$. The two conditions make $\mathsf{T}$ unitary: $(\mathsf{T}^\dagger\mathsf{T})_{11} = |\tau|^2+|\kappa|^2 = 1$ and $(\mathsf{T}^\dagger\mathsf{T})_{12} = \tau^*\kappa + \kappa^*\tau = 0$. Hence $(\hat{c}_{\rm in},\hat{c}'_{\rm in})^T = \mathsf{T}^\dagger(\an{1},\an{2})^T$, i.e., $\hat{c}_{\rm in} = \tau^*\an{1} + \kappa^*\an{2}$, and $\hat{c}'_{\rm in} = \kappa^*\an{1} + \tau^*\an{2}$ is the orthogonal mode. If $\hat{c}'_{\rm in}$ is in the vacuum, the state is a single-mode excitation of $\hat{b} = \hat{c}_{\rm in}$, with $\beta_1 = \tau^*$, $\beta_2 = \kappa^*$, and by Eq.~\eqref{eq:w} $\bm{w} = \beta_1^*\bm{u}_1 + \beta_2^*\bm{u}_2 = \tau\bm{u}_1 + \kappa\bm{u}_2$. The relative phase of the two components of $\bm{w}$ is $\arg(\kappa/\tau)$, a property of the beam splitter; the state of $\hat{c}_{\rm in}$ only fixes $\bar{n}_b$ in Eq.~\eqref{eq:central}.

\paragraph{Problem~\ref{prob:squeezed}.} The squeezed vacuum of $\hat{b}$ is a single-mode excitation with $\bar{n}_b = \sinh^2s$, Eq.~\eqref{eq:nsq}, so Eq.~\eqref{eq:Csingle} gives $C_{jk} = \sinh^2s\,\beta_j\beta_k^*$. For the two-mode squeezed vacuum, the probability of $\ket{n,n}$ is $p_n = \tanh^{2n}s/\cosh^2s$, so with $x = \tanh^2s$ and $\sum_nnx^n = x/(1-x)^2$,
\begin{equation}
\bar{n}_1 = \bar{n}_2 = \sum_np_n\,n = \frac{1}{\cosh^2s}\,\frac{\tanh^2s}{(1-\tanh^2s)^2} = \frac{\tanh^2s\,\cosh^4s}{\cosh^2s} = \sinh^2s ,
\end{equation}
where we used $1-\tanh^2s = 1/\cosh^2s$. The operator $\ad{1}\an{2}$ maps $\ket{n,n}$ onto a multiple of $\ket{n+1,n-1}$, which is orthogonal to every $\ket{m,m}$; hence $C_{12} = 0$, and Eq.~\eqref{eq:IQ} shows no fringes.

\paragraph{Problem~\ref{prob:g2}.} The state $\ket{1,1}$ is a product of number states, so $C_{12} = 0$ and the mean intensity \eqref{eq:IFock} shows no fringes. For the correlation function, write $\hat{E}_x^+(\bm{r}) = \mathcal{E}\sum_j\an{j}\,e^{i\bm{k}_j\cdot\bm{r}}$, with $\mathcal{E} = i\sqrt{\hbar\omega/2\varepsilon_0L^3}$ and the time dependence omitted. Since $\an{1}\an{1}\ket{1,1} = \an{2}\an{2}\ket{1,1} = 0$ and $\an{1}\an{2}\ket{1,1} = \an{2}\an{1}\ket{1,1} = \ket{0,0}$,
\begin{equation}
\hat{E}_x^+(\bm{r}')\,\hat{E}_x^+(\bm{r})\ket{1,1} = \mathcal{E}^2\left[e^{i(\bm{k}_1\cdot\bm{r}'+\bm{k}_2\cdot\bm{r})} + e^{i(\bm{k}_2\cdot\bm{r}'+\bm{k}_1\cdot\bm{r})}\right]\ket{0,0} .
\end{equation}
The correlation function is the squared norm of this vector:
\begin{multline}
\langle\hat{E}_x^-(\bm{r})\hat{E}_x^-(\bm{r}')\hat{E}_x^+(\bm{r}')\hat{E}_x^+(\bm{r})\rangle = |\mathcal{E}|^4\,\big|e^{i(\bm{k}_1-\bm{k}_2)\cdot(\bm{r}'-\bm{r})} + 1\big|^2
\\
= 2|\mathcal{E}|^4\left\{1 + \cos\!\left[(\bm{k}_1-\bm{k}_2)\cdot(\bm{r}-\bm{r}')\right]\right\}.
\end{multline}
The joint detection probability oscillates with full contrast as a function of $\bm{r}-\bm{r}'$, with the same period $\Delta y$ as the one-photon fringes of Eq.~\eqref{eq:Isym}, although each detector alone sees a uniform intensity. Pfleegor and Mandel observed such correlations with two independent lasers at very low intensity~\cite{PfleegorMandel}.

\end{appendix}

\nolinenumbers

\end{document}